\documentclass{aa}
\usepackage[colorlinks=true,     linkcolor=blue, citecolor=blue, filecolor=blue, urlcolor=blue]{hyperref}

\usepackage{natbib}
\bibpunct{(}{)}{;}{a}{}{,}

\usepackage{siunitx}
\usepackage{float}
\usepackage{gensymb}
\usepackage{makecell}
\usepackage{enumitem}
\setlist{nolistsep}
\usepackage{multirow}
\usepackage{adjustbox}
\usepackage{booktabs}
\usepackage{mathtools}
\usepackage{tikz}
\usetikzlibrary{calc}
\usepackage{algorithm,algpseudocode}
\usepackage{graphicx}
\usepackage{txfonts}
\usepackage[utf8]{inputenc}
\UseRawInputEncoding

\usetikzlibrary{shapes.geometric, arrows}
\tikzstyle{ML} = [trapezium, trapezium left angle=70, trapezium right angle=110, minimum width=1cm, minimum height=1cm, text centered, draw=black, fill=red!30]
\tikzstyle{discard} = [rectangle, rounded corners, minimum width=1cm, minimum height=1cm,text centered, draw=black, fill=orange!30]
\tikzstyle{kept} = [rectangle, rounded corners, minimum width=2cm, minimum height=1cm,text centered, draw=black, fill=blue!30]
\tikzstyle{invisible} = [rectangle]
\tikzstyle{decision} = [diamond, minimum width=0.1cm, minimum height=0.1cm, text centered, draw=black, fill=yellow!30]
\tikzstyle{class} = [circle, text centered, draw=black]
\tikzstyle{arrow} = [thick,->,>=stealth]
\tikzstyle{line}=[draw] % here
\begin{document}

% \title{An Example Article using \aastex v6.2\footnote{Released on January, 8th, 2018}}
\title{Machine learning applications in studies of the physical properties of active galactic nuclei based on photometric observations}
%\subtitle{Estimation of obscuration, redshift, luminosities, black hole mass, and Eddington ratio \LEt{ AA  prefers titles to be no more than three lines. I\ suggest removing the subtitle.***}}

\titlerunning{Machine learning applications for AGN studies}
\authorrunning{Mechbal et al}

   \author{Sarah Mechbal\inst{\ref{desy}}
          \and Markus Ackermann\inst{\ref{desy}}
          \and Marek Kowalski\inst{\ref{desy}}}
    \institute{DESY, Platanenallee 6, D-15738, Zeuthen, Germany\\
     \email{sarah.mechbal@desy.de} \label{desy}}

   \date{Received November 15, 2022; accepted November 16, 2022}

  \abstract
   {We investigate the physical nature of active galactic nuclei (AGNs) using machine learning (ML) tools.}
   {We show that the redshift, $z$,  bolometric luminosity, $L_{\rm Bol}$,  central mass of the supermassive black hole (SMBH), $M_{\rm BH}$,  Eddington ratio, $\lambda_{\rm Edd}$, and AGN class (obscured or unobscured) can be reconstructed through multi-wavelength photometric observations only.}
   {We trained a random forest regressor (RFR) ML-model on \num{7616} spectroscopically observed AGNs from the SPIDERS-AGN survey, which had previously been cross-matched with soft X-ray observations (from ROSAT or XMM), WISE mid-infrared photometry, and optical photometry from SDSS \textit{ugriz} filters. We built a catalog of \num{21050} AGNs that were subsequently reconstructed with the trained RFR; for \num{9687} sources, we found archival redshift measurements. All AGNs were classified as either type 1 or type 2 using a random forest classifier (RFC) algorithm on a subset of known sources. All known photometric measurement uncertainties were incorporated via a simulation-based approach.}
   {We present the reconstructed catalog of \num{21050} AGNs with redshifts ranging from $ 0 < z < 2.5$. We determined $z$ estimations for \num{11363} new sources, with both accuracy and outlier rates within 2\%. The distinction between type 1 or type 2 AGNs could be identified with respective efficiencies of 94\% and 89\%. The estimated obscuration level, a proxy for AGN classification, of all sources is given in the dataset. The $L_{\rm Bol}$, $M_{\rm BH}$, and $\lambda_{\rm Edd}$ values are given for \num{21050} new sources with their estimated error. These results have been made publicly available.}
   {The release of this catalog will advance AGN studies by presenting key parameters of the accretion history of 6 dex in luminosity over a wide range of $z$. Similar applications of ML techniques using photometric data only will be essential in the future, with large datasets from eROSITA, JSWT, and the VRO poised to be released in the next decade.}

   \maketitle


\section{Introduction} \label{sec:intro}
\indent Active galactic nuclei (AGNs)  are known to be the most luminous sources in the Universe \citep{burbidge_nuclear_1958, minkowski_new_1960, matthews_optical_1963, schmidt_3c_1963}. These systems consist of a central supermassive black hole (SMBH), around which an accretion disk is formed \citep{lynden-bell_galactic_1969, rees_black_1984}. Although much is yet to be learned about the feedback mechanisms linking SMBH growth and the evolution of their host galaxies \citep{ferrarese_fundamental_2000,gebhardt_relationship_2000,yu_observational_2002}, we already know that their mass, $M_{\rm BH}$, scales with a number of galaxy properties, such as the stellar velocity dispersion, $\sigma$, bulge mass, and luminosity (see a review in \cite{kormendy_coevolution_2013}). Furthermore, the extreme energetics of these objects also makes them a favored source of cosmic ray acceleration \citep{mannheim_high-energy_1995, halzen_neutrino_1997,murase_neutrinos_2022, abbasi_search_2022}, as underlined by the recent discovery of neutrinos originating from the AGN NGC 1068 with IceCube \citep{icecube_collaboration_evidence_2022}. \newline
\indent Collecting the physical parameters of SMBHs (e.g., redshifts, $z$, black hole mass, $M_{\rm BH}$, and Eddington luminosity and ratio, $L_{\rm Edd}$ and $\lambda_{\rm Edd}$) from a complete and unbiased sample of AGNs stands as a necessary step in  studying their accretion history \citep{soltan_masses_1982,magorrian_demography_1998,schulze_low_2010}. However, spectroscopic techniques are almost always needed to measure these variables and although the number of spectroscopically observed AGNs has undoubtedly grown in the last decade \citep{plotkin_large_2008, kochanek_ages_2012, menzel_spectroscopic_2016, dwelly_spiders_2017, comparat_final_2020}, the discrepancy between photometrically identified AGNs and those followed up with spectroscopic surveys remains large \citep{blanton_sloan_2017, comparat_final_2020}. Fortunately, AGNs have been well covered by multi-wavelength surveys \citep{elvis_atlas_1994, edelson_multiwavelength_1996}: X-ray telescopes have observed the extragalactic sky \citep{brandt_deep_2005}, revealing a pattern of stars, black holes in binary systems, and AGNs, while infrared (IR) telescopes have allowed us to distinguish the latter from the large stellar population \citep{stern_midinfrared_2005, assef_mid-infrared_2013}. \newline
Enlarging the sample and sky coverage of AGN observations with reliably estimated physical parameters is particularly important for multimessenger astronomy, where signals from individual sources are often weak \citep{aartsen_icecube_2020, abbasi_constraining_2023}. This limitation can be overcome by searching for correlations between a messenger (e.g., neutrinos or cosmic rays) and a population of AGNs instead. However, the power of correlation searches increases with the sky coverage of the counterpart observations and, additionally, this requires a model for the expected production of the messenger in question for each object included in the correlation study \citep{achterberg_selection_2006,aartsen_contribution_2017, abbasi_search_2022}. Such models usually depend on the physical parameters of the specific AGN.

\indent Machine learning (ML) techniques have been applied in recent years to characterize AGNs and galaxies in general, for  classification tasks, redshift determinations \citep{cunha_photometric_2022, dainotti_predicting_2021, fotopoulou_cpz_2018, sadeh_annz2_2016}, and reconstructions of their physical properties. \cite{li_identification_2021}, \cite{clarke_identifying_2020}, \cite{fotopoulou_cpz_2018}, and \cite{khramtsov_northern_2020} all trained ML classifiers using multi-wavelength photometric datasets to distinguish galaxies, quasars, and stars from one another. 
While \cite{ucci_inferring_2017,rhea_machine-learning_2021} used spectral measurements of galaxies and ML algorithms to infer the physical properties of the host galaxies, \cite{bonjean_star_2019} and \cite{simet_comparison_2021} used photometric measurements only and ML regression to determine star formation rates, while the $M_{\rm \star}$, parameters have usually been reconstructed from spectroscopy observations or template fitting methods. We follow this latter approach in this paper.

We report on a novel attempt to employ ML regression tasks to reconstruct the fundamental parameters of AGNs. We trained a ML algorithm to estimate $z$, $L_{\rm x}$, and $L_{\rm Bol}$, along with the soft X-ray (SXR) and bolometric luminosities, as well as $M_{\rm BH}$, $L_{\rm Edd}$, and $\lambda_{\rm Edd}$ on \num{21050} AGNs, all observed in the IR, optical, and X-ray bands photometrically -- but not spectroscopically. To train the model, we used the recent SPIDERS-AGN spectroscopic survey \citep{clerc_spiders_2016,dwelly_spiders_2017}, which has compiled and released a sample of $\sim$ 7600 type 1 AGNs \citep{coffey_sdss-ivspiders_2019}. In addition, we also trained a ML classifier to identify type 2 (or obscured) from type 1 (unobscured) AGNs. 

The structure of the paper is as follows: we detail the catalogs used to expand and build both the AGN training sample and the data sample that is to be reconstructed in Sect. \ref{sec:data} and we describe the procedures to select AGNs from stellar, galactic, and blazar populations. We detail in Sect. \ref{sec:pseudo-sets} how the errors on the input parameters have been incorporated to generate pseudo-sets for the classification, training, and reconstruction of AGNs, and underline the advantages of the simulation-based method, before classifying the unlabeled sources using ML tools in Sect. \ref{sec:classification}. In Sect. \ref{sec:training}, we describe the details of the main ML regression task built to parametrize the core of AGNs. We also discuss our comparison of several models and the final results from the spectroscopic parameter predictions.  In Sect. \ref{sec:results}, we present in the largest catalog (to date) of AGN physical properties, stemming from the ML reconstruction of \num{21050} sources, including \num{11363} new $z$ measurements, and $L_{\rm Bol}$, $M_{\rm BH}$, and $\lambda_{\rm Edd}$ values for all. The limits of  the type 2 AGN reconstruction are also discussed. We turn to future prospects following the release of this dataset in Sect. \ref{sec:summary} and discuss the role ML tools will play with the advent of future missions. The catalog columns are described in Appendix \ref{appendix:catalogue column}.

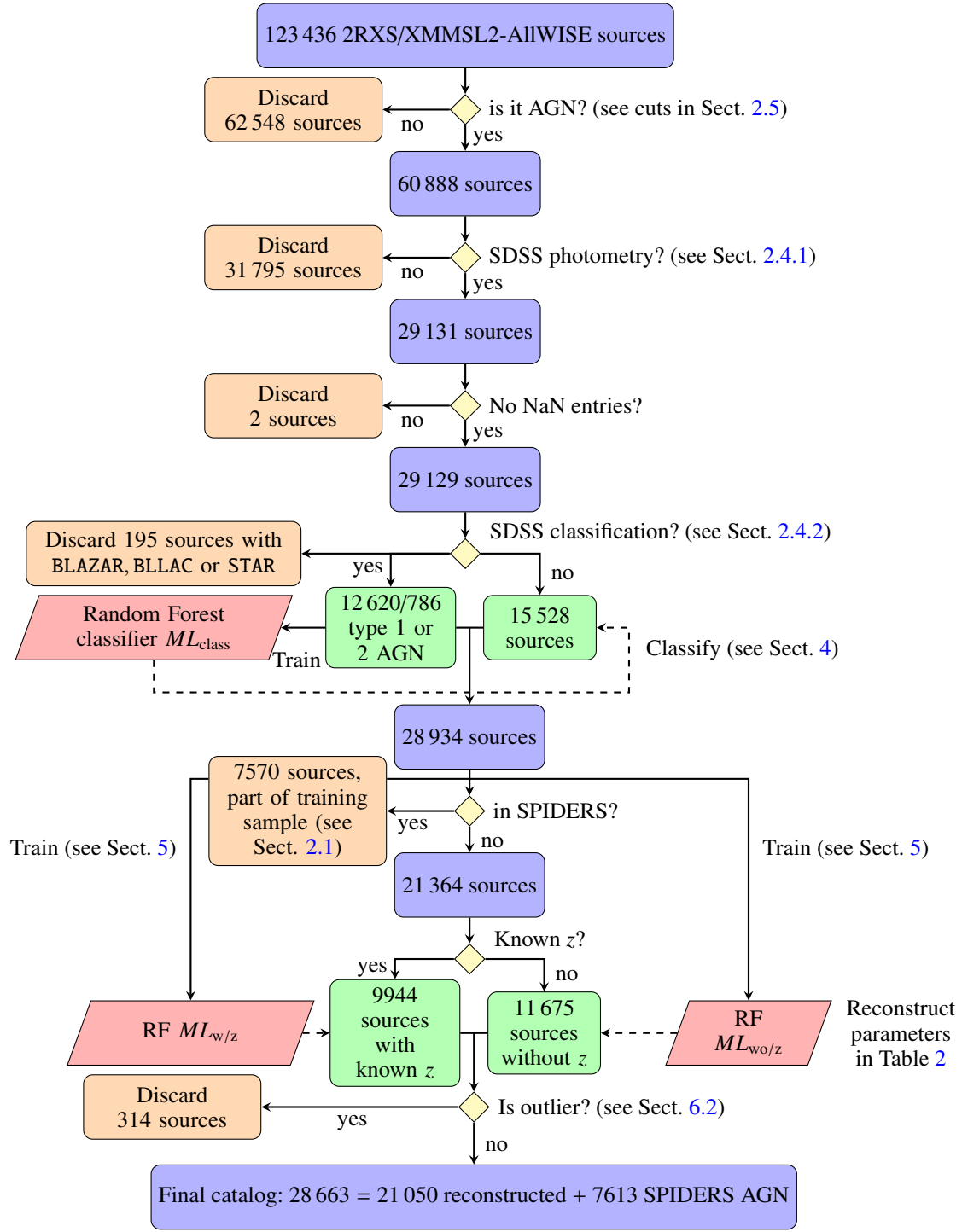
\begin{figure*}
\centering
\begin{tikzpicture}[node distance=1.15cm]

% FIRST STEP
\node (step) [kept] {\num{123436} 2RXS/XMMSL2-AllWISE sources};

% is AGN ? 
\node (dec00) [decision, below of=step, label=0:is it AGN? (see cuts in Sect. \ref{subsec:agn_selection})] {};
\node (discard00) [discard, text width=2.5cm,left of=dec00, xshift=-1.5cm] {Discard \num{62548} sources};
\node (step00) [kept, below of=dec00] {\num{60888} sources} {};

\draw [arrow] (step) -- (dec00);
\draw [arrow] (dec00) -- node[anchor=east, xshift=0.25cm, yshift=-0.25cm] {no} (discard00);
\draw [arrow] (dec00) -- node[anchor=south, yshift=-0.25cm ,xshift=0.35cm] {yes} (step00);

% SDSS photometry? 
\node (dec1) [decision, below of=step00, label=0:SDSS photometry? (see Sect. \ref{subsubsec:photometry})] {};
\node (discard1) [discard, text width=2.5cm,left of=dec1, xshift=-1.5cm] {Discard \num{31795} sources};
\node (step1) [kept, below of=dec1] {\num{29131} sources} {};

\draw [arrow] (step00) -- (dec1);
\draw [arrow] (dec1) -- node[anchor=east, xshift=0.25cm, yshift=-0.25cm] {no} (discard1);
\draw [arrow] (dec1) -- node[anchor=south, yshift=-0.25cm ,xshift=0.35cm] {yes} (step1);

% No NaN in WISE columns?  
\node (dec22) [decision, below of=step1, label=0:No NaN entries?] {};
\node (discard22) [discard, text width=2.5cm,left of=dec22, xshift=-1.5cm] {Discard \num{2} sources};
\node (step22) [kept, below of=dec22] {\num{29129} sources};

\draw [arrow] (step1) -- (dec22);
\draw [arrow] (dec22) -- node[anchor=east,xshift=0.25cm, yshift=-0.25cm] {no} (discard22);
\draw [arrow] (dec22) -- node[anchor=south, yshift=-0.25cm ,xshift=0.35cm] {yes} (step22);

% SDSS spectroscopy?
\node (dec3) [decision, below of=step22, label={[label distance=0.05cm]10:SDSS classification? (see Sect. \ref{subsubsec:spectroscopy_and_redshift})}] {};
\node (discard3bis) [discard, text width=4.cm,left of=dec3, xshift=-3.5cm] {Discard \num{195} sources with $\tt{BLAZAR, BLLAC}$ or $\tt{STAR}$};

\node (aux3) [invisible, below of=dec3]  {};
\node (aux) []  {};
\node (aux_last) []  {};

\node (step3) [kept,left of=aux3, text width=1.75cm, align=center, fill=green!30] {\num{12620}/\num{786} type 1 or 2 AGN};
\node (discard3) [discard, text width=1.5cm,right of=aux3, fill=green!30] {\num{15528} sources};

\draw [arrow] (step22) -- (dec3);
\draw[arrow] (dec3) -| node [ near end, xshift=-0.35cm] {yes} (step3);
\draw[arrow] (dec3) -| node [ near end, xshift=0.35cm] {no} (discard3);
\draw[arrow] (dec3) -- (discard3bis);

% Classifier NN 
\node (nn1) [ML, left of=step3,  text width=3.0cm,xshift=-2.5cm] {Random Forest classifier $ML_{\rm class}$};
\path (aux) (step3) -- (discard3) coordinate[midway] (aux);

\draw [arrow] (step3) -- node[yshift=-0.5cm, xshift=-0.10cm] {Train} (nn1);
\coordinate (right d3) at ($(discard3.east)+(0.5,0cm)$);
\coordinate (Below nn1) at ($(nn1.south) + (0,-0.5cm)$);
\draw[dashed, arrow]  (nn1.south) -- (Below nn1)  -|  (right d3)  -- (discard3.east) node [yshift=-0.35cm, xshift=2.25cm] {Classify (see Sect. \ref{sec:classification})};
\node  (step4)  [kept, below of=aux, yshift=-0.55cm] (step4) {\num{28934} sources};
\draw [line, thick] (discard3) -- (step3);
\draw [arrow] (aux) -- (step4);

%% is SPIDERS?
\node (dec34) [decision, below of=step4, label=0:in SPIDERS?] {};
\node (discard34) [discard, text width=2.5cm,left of=dec34, xshift=-1.5cm] {\num{7570} sources, part of training sample (see Sect. \ref{subsec:spiders})};
\node (step34) [kept, below of=dec34] {\num{21364} sources};

\draw [arrow] (step4) -- (dec34);
\draw [arrow] (dec34) -- node[anchor=east, xshift=0.25cm, yshift=-0.25cm] {yes} (discard34);
\draw [arrow] (dec34) -- node[anchor=south, yshift=-0.25cm ,xshift=0.35cm] {no} (step34);

%% z information? 
\node (dec4) [decision, below of=step34, label={[label distance=0.05cm]10:Known $z$?}] {};
\node (aux4) [invisible, below of=dec4]  {};
\node (step5) [kept,left of=aux4, text width=1.5cm, fill=green!30] {\num{9944} sources with known $z$};
\node (discard5) [discard, text width=1.5cm,right of=aux4, fill=green!30] {\num{11675} sources without $z$};

\draw [arrow] (step34) -- (dec4);
\draw[arrow] (dec4) -| node [ near end, xshift=-0.35cm] {yes} (step5);
\draw[arrow] (dec4) -| node [ near end, xshift=0.35cm] {no} (discard5);

% Final ML-algorithm

\node (nn_w_z) [ML, left of=step5,  text width=1.5cm,  xshift=-2.0cm] {RF $ML_{\rm w/z}$};
\node (nn_wo_z) [ML, right of=discard5,  text width=1.5cm,  xshift=2.0cm] {RF $ML_{\rm wo/z}$};

%\draw [arrow] (discard34) -- node[yshift=-0.5cm, xshift=-0.75cm] {Train} (nn_w_z);

\coordinate (north nn_wo_z) at ($(nn_wo_z.north)+(0,0.5cm)$);
\coordinate (north nn_w_z) at ($(nn_w_z.north)+(0,0.5cm)$);
\coordinate (northeast discard34) at ($(discard34.east) + (0,0.5cm)$);
\coordinate (northwest discard34) at ($(discard34.west) + (0.,0.5cm)$);
\coordinate (right discard34) at ($(discard34.east) + (3,0.5cm)$);
\coordinate (left discard34) at ($(northwest discard34) + (-0.25,0cm)$);

\draw[arrow]  (northwest discard34) -- (left discard34)  -|  (north nn_w_z)  -- (nn_w_z.north) node [yshift=2.35cm, xshift=-1.5cm] {Train (see Sect. \ref{sec:training})};

\draw[arrow]  (northeast discard34) -- (right discard34)  -|  (north nn_wo_z)  -- (nn_wo_z.north) node [yshift=2.35cm, xshift=1.5cm] {Train (see Sect. \ref{sec:training})};
\draw [dashed, arrow] (nn_w_z) --  (step5);
\draw [dashed, arrow] (nn_wo_z) --  (discard5) node [xshift=5.5cm,  align=center] {Reconstruct\\parameters\\in Table \ref{tab:outputs}};

% Final released catalogue
\path (aux_last) (step5) -- (discard5) coordinate[midway] (aux_last);

\node (dec_last) [decision, below of=aux_last, label={[yshift=-0.35cm, label distance=0.05cm]10:Is outlier? (see Sect. \ref{outliers})}] {};
\node (discard_last) [discard, text width=2.5cm,left of=dec_last,  xshift=-3.5cm] {Discard \num{314} sources};
\node  (step_last)  [kept, below of=dec_last, yshift=-0.25cm] (step_last) {Final catalog: $\num{28663} = \num{21050}$ reconstructed + \num{7613} SPIDERS AGN};
\draw [line, thick] (discard5) -- (step5);

\draw [arrow] (aux_last) -- (dec_last);
\draw [arrow] (dec_last) -- node[anchor=south, yshift=-0.25cm ,xshift=0.35cm] {no} (step_last);
\draw [arrow] (dec_last) -- node[anchor=east, xshift=0.25cm, yshift=-0.25cm] {yes} (discard_last);

\end{tikzpicture}
\caption{Flowchart of the analysis.The starting point 2RXS/XMMSL2-AllWISE catalog released in \cite{salvato_finding_2018}, leading to the catalog of \num{21050} reconstructed AGN sources presented in this work.}
\label{fig:flowchart}
\end{figure*}

\section{Data} \label{sec:data}

\begin{table*}
\centering
%\resizebox{\hsize}{!}{
\begin{tabular}{|cccccc|}
\hline
 Observation type & Instrument & Spectral band  & Input provided & 	$N_{\rm AGN}$\footnote{After all selections} & Reference \\   
\hline 
Soft X-ray & \textit{ROSAT} & 0.1-2.4 keV   & \makecell{\\X-ray flux and err.} & \num{19896} &\cite{boller_second_2016} \\
\cline{2-3}
 & \textit{XMM-Newton} & 0.2-12 keV & & \num{1468} &  \cite{saxton_first_2008}  \\ 
\hline
Mid-Infrared photometry & WISE  & 3.4 - 22 \si{\micro\metre} & \makecell{W1,W2,W3,W4 \\mag. and err.} & \num{21050} & \cite{cutri_vizier_2021}  \\
\cline{2-5}
% &	NEOWISE &  & & \citep{necker_jannisnetimewise_2022} &   W1, W2 excess variance    \\
\hline
Optical photometry & SDSS I-IV &  3543-9134 \si{\angstrom} & \textit{ugriz} mag. and err. & \num{21050} & \cite{blanton_sloan_2017}   \\
 & \textit{Gaia} & 330–1050 \si{\nm} &  Flux and err. &  & \cite{arenou_gaia_2017}  \\
\hline
Optical spectroscopy & SDSS I-III & 380-920 nm  & \makecell{Classification} & \num{9944} & \cite{dwelly_spiders_2017} \\
\cline{2-2}
 & see Ref. &  & redshift  & & \cite{veron-cetty_catalogue_2010}  \\
\hline
\end{tabular}
%}
\caption{Catalogues and their references used to build the multiwavelength inputs to the machine learning algorithm.}
\label{tab:inputs}
\end{table*}

\begin{table}
\resizebox{0.45\textwidth}{!}
{\begin{tabular}{|c|c|c|c|c|}
\hline
Target parameter & Notes & Training range & Error on the estimate   \\   
\hline
$z$ & Redshift & [0.008,2.5] & - \\
log $L_{\rm X}$ & X-ray luminosity & [40.8,45.9] & $\sim$0.05 dex\\
log $L_{\rm bol}$ & Bolometric luminosity  & [42.9,47.6] & $\sim$0.02 dex\\
log $M_{\rm BH}/M_{\rm \odot}$ & Black hole mass & [6.2,10.4] & $\sim$0.02 dex \\
log $\lambda_{\rm Edd}$ & Eddington ratio & [-3.2,0.54] & $\sim$0.01 dex \\
\hline
\end{tabular}}
\caption {Target variables, their domain range and error estimate from the SPIDERS catalogue \citep{coffey_sdss-ivspiders_2019}. The variables are presented in the order of their sequential prediction by the ML algorithm, as executed by the chain regressor.}
\label{tab:outputs}
\end{table}

Here, we provide a high-level summary of the different datasets used to build the reconstructed catalog and augment the training dataset. A detailed account of the multiwavelength observations used is given in Sections \ref{subsec:spiders} to \ref{subsec:agn_selection}. Table \ref{tab:inputs} summarizes all input features used for the training and reconstruction of the ML regressor, along with their catalog of origin. The analysis sequence to create the full reconstructed catalog is described in Fig. \ref{fig:flowchart}. The target parameters for the ML regression are presented in Table \ref{tab:outputs}. 

The starting point for building the reconstructed catalog is the 2RXS/XMMSL2-AllWISE dataset released in \cite{salvato_finding_2018}. Those \num{123436} AGNs have been soft X-ray- and infrared-selected and already cross-matched, using the 2RXS \citep{boller_second_2016} and XMMSL2 \footnote{\url{https://www.cosmos.esa.int/web/xmm-newton/xmmsl2-ug}} catalogs from \textit{ROSAT} (0.1 - 2.4 keV) and \textit{XMM-Newton} (0.2 - 12 keV)  for X-ray flux and error, respectively. In the infrared, the AllWISE catalog \citep{cutri_vizier_2021} provided observations at 3.4, 4.6, 12 and 22 \si{\micro\meter} (W1, W2, W3, and W4 bands, respectively), with their respective errors. Cuts are performed on the X-ray flux, and the IR color-color bands to remove non-AGN sources, leaving \num{60888} sources (see Sect. \ref{subsec:agn_selection}). For optical photometry and spectroscopy information, we cross-matched the remaining sources with data from the Sloan Digital Sky Survey (SDSS): \num{21050} sources were found to have been  photometrically observed in the instrument specific \textit{ugriz} (3543-9134 \si{\angstrom}) filters \citep{lyke_sloan_2020}, while a subset of \num{9944} sources were also observed spectroscopically (380-920 nm band): for these, the redshift, $z,$ and the classification $\tt{CLASS\_BEST}$ were recorded, and \num{195} sources classified by SDSS as blazars, BL Lacertae (BLLACs), or stars were removed. We required for  all sources to have SDSS photometry observations. Gaia's DR1 catalog \citep{arenou_gaia_2017} was also cross-matched in \cite{salvato_finding_2018}: the white-light G-band (330–1050 nm) mean flux and mean flux error was also recorded for \num{14887} cross-matched sources, expressed in photo-electrons s$^{-1}$. For the remaining \num{6097} without Gaia observations, we handled the null entries by creating synthetic values of the flux and error. The same method was applied to fill \num{21}, \num{89}, \num{1931} and \num{7156} missing W1, W2, W3, and W4 error entries. This was done for features found to have a minimal importance in the regression task, while aiming to maximize the number of sources in the catalog, to avoid the indiscriminate removal of any point with a missing parameter. Appendix \ref{appendix:null entries} describes the handling of null entries and its effect on the reconstruction in more detail.

For the training dataset, we used \num{7616} type 1 AGN sources from the SPIDERS-AGN catalog \citep{coffey_sdss-ivspiders_2019}, a completed SDSS-IV spectroscopic survey. All AGNs  that were all observed photometrically in the X-ray, IR, and optical bands prior to the survey, and the input catalog was augmented with the datasets listed above. The released dataset gives key AGN parameters derived from the spectral lines: redshift, $z$, bolometric luminosity, $L_{\rm Bol}$, black hole mass estimate, $M_{\rm BH}$, and Eddington ratio, $\lambda_{\rm Edd}$. These variables constitute the target parameters for the ML regression.

For matters pertaining to machine learning methods, a few terms also need be clearly defined:
\begin{itemize}
\item by \textit{training} and \textit{testing} sample, we mean the dataset of \num{7616} sources built from the SPIDERS catalog used to train the ML model tasked to learn the correlations between photometric and spectroscopic parameters. The performance of the ML model is assessed by comparing the true and predicted values of target parameters (also referred in the ML literature as the "validation" set),
\item we refer to the AGN catalog we  built for as the call the "reconstructed" or "full" dataset, which we used to make estimations on the target parameters using the previously trained ML model.
\end{itemize}

\subsection{The SPIDERS-AGN catalog} \label{subsec:spiders}

SPectroscopic IDentification of ERosita Sources (SPIDERS) is a completed SDSS-IV \citep{blanton_sloan_2017} 5128.9 deg$^{2}$ survey over the SDSS footprint. The AGN sources were originally pre-selected based on the 1RXS and XMMSL1 \citep{saxton_first_2008} catalogs, which were then later updated once the 2RXS \citep{boller_second_2016} and XMMSL2 were released. The details of the mission targeting and summary are documented in \cite{dwelly_spiders_2017} and \cite{comparat_final_2020},  respectively. The spectroscopic data was made available in the 16th SDSS data release (DR16) \citep{ahumada_16th_2020} as a catalog of type 1 AGNs containing X-ray fluxes, optical spectral and photometric measurements, black hole estimates, and other derived quantities\footnote{\url{https://data.sdss.org/datamodel/files/SPIDERS_ANALYSIS/spiders_quasar_bhmass.html}}. We refer the reader to \cite{coffey_sdss-ivspiders_2019} for a detailed description of the dataset and to \cite{wolf_exploring_2020} for a principal component analysis (PCA) of type 1 AGN properties. \\
The survey probed the brightest X-ray sources in the sky, at the higher end of the luminosity distribution with $41 < \rm log_{\rm 10}(L_{\rm X}/\rm erg s^{-1}) < 46$  for a mean redshift $\overline{z}$ = 0.47. The bolometric luminosity, $L_{\rm Bol}$, was also derived from the monochromatic luminosity, $L_{\rm 3000 \si{\angstrom}}$  and  $L_{\rm 5100 \si{\angstrom}}$, using bolometric corrections.  
Fitting the H$\beta$ and MgII emission lines, the SPIDERS-AGN study derived the $M_{\rm BH}$, $L_{\rm Edd} = 1.26 \times 10^{38} (\frac{M_{\rm BH}}{M_{\rm \odot}})$ erg s$^{-1}$, and $\lambda_{\rm Edd} = L_{\rm Bol}/L_{\rm Edd}$, with $M_{\rm \odot}$ being the solar mass.
%4\pi GM_{\rm BH}m_{\rm p}/\sigma_{\rm T}$,
For \num{2337} sources, both the H$\beta$ and MgII lines were observed, and two estimates of $M_{\rm BH}$ and derived quantities were provided: in such cases, we selected the values with the smallest associated error $\sigma_{\rm M_{\rm BH}}$, which has a typical value of $\sim$ 0.02 dex. Table \ref{tab:outputs} presents a list of the key properties found in the SPIDERS-AGN catalog, with their respective range and median estimate error.  Out of \num{7670} AGNs,  there are \num{7616} with complete spectroscopic information, which we used as the basis of our training sample and we expanded on it using several other astronomical catalogs.

\subsection{X-ray data} \label{subsec:x-ray}

X-ray band observations are some of the most effective data samples for identifying AGNs: emission is believed to come from above the accretion disk; from there, photon scatter onto the hot corona gas and emit X-rays via inverse Compton. Although binary systems such as accreting neutron stars and stellar-mass black holes are also X-ray emitters, AGNs are generally more luminous by an order of magnitude  ($L_{\rm X} > 10^{42}$ erg s$^{-1}$) \citep{hickox_obscured_2018}. 

The \textit{ROSAT} telescope \citep{trumper_rosat_1982} performed the first all-sky survey (RASS) between 1990 and 1991 in the 0.1-2.4 keV band. Two catalogs, one for faint and another for bright sources, were then released \citep{voges_rosat_2000}. The data were reprocessed decades later, leading to a second data release, the 2RXS catalog, comprising $\sim$\num{135000} sources \citep{boller_second_2016}. \
XMMSL2 is the second catalog of X-ray sources found in slew data taken by the \textit{XMM-Newton} European Photon Imaging Camera pn (EPIC-pn) in three bands: 0.2–12 keV (B8), 0.2–2 keV (B7), and 2–12 keV (B6). The B8 band is the most complete and the one of interest. \
The starting point of our reconstructed catalog building is the work of \cite{salvato_finding_2018}: \num{106573} X-ray sources from 2RXS and \num{17665} sources from XMMSL2 (with $\mid b \mid>15^{\circ}$) were cross-matched with their AllWISE \citep{wright_wide-field_2010} and \textit{Gaia} \citep{gaia_collaboration_gaia_2018} counterparts using a newly developed Bayesian algorithm to overcome the large positional uncertainties of the X-ray observations. Two catalogs were subsequently released: 2RXS-AllWISE and XMMSL2-AllWISE\footnote{\url{https://www.mpe.mpg.de/XraySurveys/2RXS_XMMSL2}}.

To combine the two X-ray datasets, several steps must be taken to match the different response functions and detection range of the instruments. We first converted the ROSAT fluxes from the original 0.1-2.4 keV into the classical soft X-ray band 0.5-2 keV \citep{dwelly_spiders_2017} as follows: 
\begin{equation}
 \label{eq:flux_conversion}
\frac{F[E'_{\rm max}:E_{\rm min}^{\prime}]}{F[E_{\rm max}:E_{\rm min}]} = \frac{E_{\rm max}^{\prime 2-\Gamma} - E_{\rm min}^{\prime 2-\Gamma}}{E_{\rm max}^{2-\Gamma} - E_{\rm min}^{2-\Gamma}},
\end{equation}
where $\Gamma$ = 1.7 for 2RXS sources.

\begin{figure}
\centering
\includegraphics[width=0.45\textwidth]{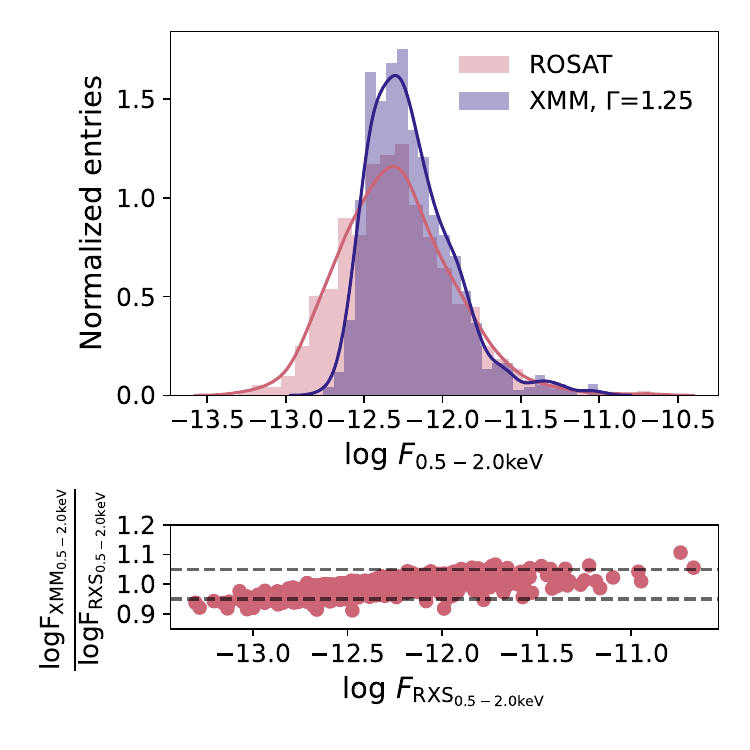}
\caption{\textit{Top}: 2RXS and XMMSL2 fluxes for the $\sim$ 1000 SPIDERS-AGN sources observed with both instruments. The X-ray fluxes were converted to the soft X-ray band 0.5-2.0 keV using $\Gamma$=1.7 for 2RXS and $\Gamma$=1.25 for XMMSL2, value chosen to match the peaks of the two distributions. \textit{Bottom}: Ratio of the converted flux logs as a function of the 2RXS fluxes for the same sources shown in the above panel.The dashed lines represent the $\pm$ 5\% level on the ratio, within which 94\% of the converted fluxes are. }
\label{fig:Xray_conversion}
\end{figure}

In \cite{dwelly_spiders_2017}, $\Gamma$ = 2.4 was chosen for XMMSL2 fluxes: however, the 2RXS/XMMSL2 datasets were kept separate. About a thousand sources from the SPIDERS AGN catalog have been observed with both instruments: we use these as a control group for matching the XMM to the converted RXS fluxes, by varying the $\Gamma$ power-law index of Eq. \ref{eq:flux_conversion} in order to match the peaks of the two X-ray flux distributions (as shown in Fig. \ref{fig:Xray_conversion}). Choosing $\Gamma_{\rm XMM}$=1.25, 94\% of sources present in both datasets have a flux ratio $\dfrac {\rm log F_{\rm XMM_{\rm 0.5-2.0 keV}}}{\rm log F_{\rm RXS_{\rm 0.5-2.0 keV}}}$ within 5\% of one another.
In the converted SRX band, the distribution of X-ray fluxes is contained between $10^{-14}< F_{\rm 0.5-2keV} < 10^{-9}$ erg cm$^{-2}$s$^{-1}$. From the X-ray catalogs, we only kept the X-ray fluxes and corresponding errors in the converted 0.5-2 keV band as as input to the ML-model. Whereas the hardness ratio and/or column density would have offered valuable information on the class of AGNs, both being a known proxy for the obscuration level of accretion disks, this parameter was neither complete, nor very accurate, in the case of ROSAT observations; thus, it had to be dropped. 

\subsection{Infrared observations} \label{subsec:infrared}

In the mid-infrared (MIR, 3-30 \si{\micro\meter}), AGNs are bright. The dusty torus is responsible for this thermal emission, as it absorbs shorter-wavelength photons from the accretion disk and re-emits them in the MIR. Although star-forming galaxies are also bright in this band, their SED is cooler and can be distinguished from those of AGNs \citep{padovani_active_2017}. \
The Wide-field Infrared Survey Explorer (WISE) \citep{wright_wide-field_2010} is a satellite launched in 2009. The missions was then later extended under a new appellation, NEOWISE \citep{mainzer_preliminary_2011}. The combination of WISE and NEOWISE data was made available to the public with the release of the AllWISE catalog \citep{cutri_vizier_2021}. The WISE survey scanned the sky at 3.4, 4.6, 12, and 22 \si{\micro\meter} (the bands designated as W1, W2, W3, and W4, respectively), at a depth at which the majority of the resolved 2RXS and XMMSL2 populations are to be detected \citep{salvato_finding_2018}. In addition to the 4 MIR magnitudes and their associated errors, we explicitly record the relative magnitudes W1-W2, W2-W3, and W3-W4. These values are readily available from \cite{salvato_finding_2018}, as previously mentioned. 

\subsection{Optical data} \label{subsec:optical}
\subsubsection{Photometry} \label{subsubsec:photometry}

\paragraph{SDSS:}
The Sloan Digital Sky Survey (SDSS) has in the course of its runs observed over \num{700000} quasars in the optical band, most of them in broadband photometry in the instrument specific \textit{ugriz} (3543-9134 \si{\angstrom}) filters \citep{lyke_sloan_2020}. To pair the AGN sources from our training and unknown sets with their SDSS observations, the $\tt{astroquery}$ software tool \citep{ginsburg_astroquery_2019} was used: we cross-matched the best AllWISE counterpart to the X-ray sources in our training and reconstructed samples with an optical counterpart from the DR17 photometric catalog, setting a maximum radius of \num{5} arcsec. For the matched sources, we add as features the SDSS PSF magnitudes $\tt{psfMag}$ and their associated error $\tt{psfMagErr}$ for the five \textit{ugriz} bands. These values are most appropriate when studying the photometry of distant quasars. Logically, all \num{7616} SPIDERS sources have SDSS photometry counterparts, however, \num{47739} (\num{5985}) 2RXS (XMMSL2) sources have been observed photometrically, which we added as a requirement (see Fig. \ref{fig:flowchart}).
\paragraph{\textit{Gaia:}}
\cite{salvato_finding_2018} also cross-matched the 2RXS/XMMSL2 sources to the first release catalog of the \textit{Gaia} mission \citep{arenou_gaia_2017}. The astrometric instrument performs broadband photometry in Gaia’s white-light G-band (330–1050 nm): we kept the mean flux and mean flux error for all sources, expressed in photo-electrons s$^{-1}$.

\subsubsection{Spectroscopy and redshift} \label{subsubsec:spectroscopy_and_redshift}
Prior to the start of the SPIDERS mission, X-ray+AllWISE AGN targets were cross-matched with the already observed SDSS I-II-III runs \citep{dwelly_spiders_2017}: $\sim$\num{12000} ROSAT and $\sim$\num{1500} XMM–Newton sources were found to already have been observed with spectroscopy. After performing a visual inspection of the optical spectra, two value-added catalogs (VAC) were released \footnote{\url{https://data.sdss.org/datamodel/files/SPIDERS_ANALYSIS}}.
These included redshift measurements, along with object main and sub-classifications. Matching sources from the reconstructed catalog by their X-ray name, we found \num{21287} (\num{2540}) 2RXS (XMMSL2) sources previously observed spectroscopically: of these, \num{11242} (\num{1250}) contain redshift information (but no black hole mass or Eddington ratio). As we explain in Sect. \ref{subsec:performance}, this latter variable greatly improves the ML predictions. Furthermore, we will make use of the sub-sample of sources for which we have an AGN classification to correlate multiwavelength observations with the AGN obscuration level (see Sect. \ref{sec:classification}).
Following the classification scheme of \cite{comparat_final_2020}, we mark AGNs as either type 1\footnote{$\tt{CLASS\_BEST}$==“$\tt{BALQSO}$”, “$\tt{QSO\_BAL}$, “$\tt{QSO}$”, “$\tt{BLAGN}$”} or 2\footnote{$\tt{CLASS\_BEST}$==“$\tt{NLAGN}$”, “$\tt{GALAXY}$”}, following the SDSS pipeline automated classification \citep{bolton_spectral_2012}.

Additional spectroscopic classification and redshift measurements are found in the VERONCAT catalog \citep{veron-cetty_catalogue_2010}, a collection of some $\sim$ \num{150000} quasars from multiple surveys. The X-ray positions of sources were then cross-matched with the optical or radio positions given by VERONCAT, using a maximum matching radius of \num{60} arcsec, a value taken from a past study \citep{abbasi_search_2022}. This allowed us to get additional AGN classification and spectral class features. We collected some $\sim$ \num{9000} redshifts from VERONCAT, in addition to those already found from previous SDSS surveys. To verify the accuracy of the cross-matching, we compare the redshift entries from the \num{6163} SPIDERS sources that are already present in VERONCAT, $\Delta_{\rm z} = \mid z_{\rm SPIDERS} - z_{\rm VERONCAT}\mid$. We found that 98$\%$ of the sources have $\Delta_{\rm z} < $ 0.01, confirming the adequacy of the cross-matching radius used.

\subsection{AGN selection} \label{subsec:agn_selection}
\begin{figure}
\centering
\includegraphics[width=0.35\textwidth]{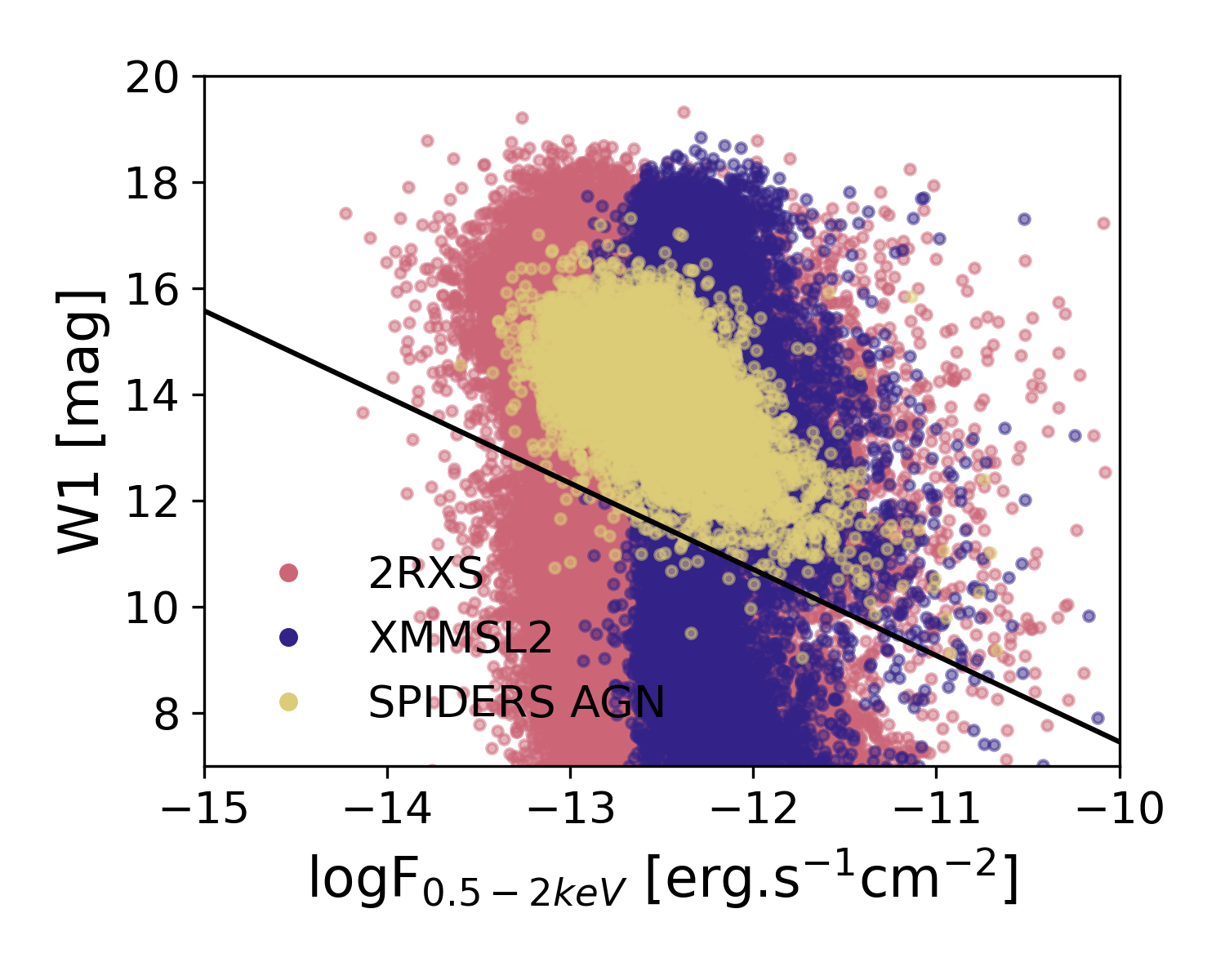}
\includegraphics[width=0.35\textwidth]{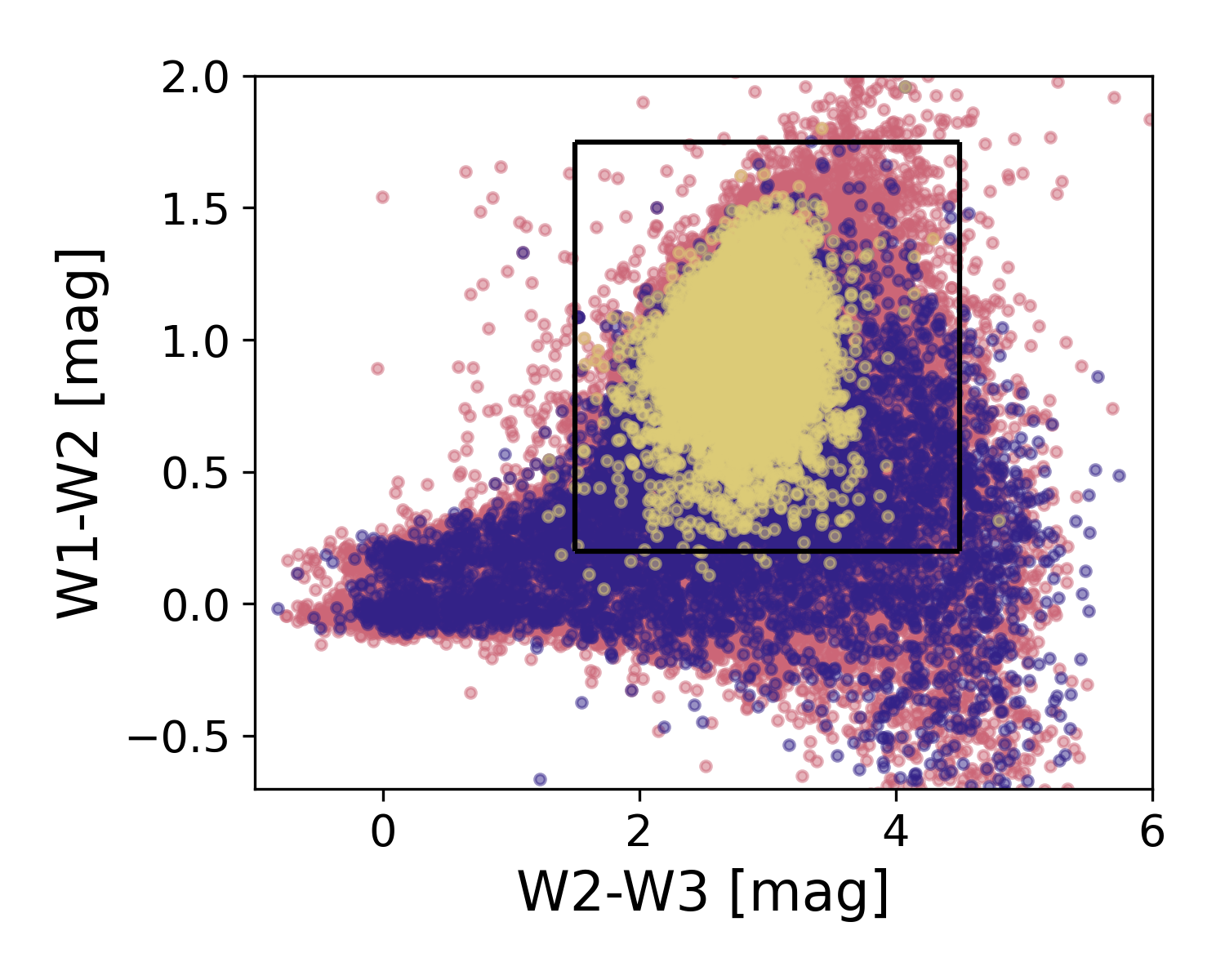}
\includegraphics[width=0.35\textwidth]{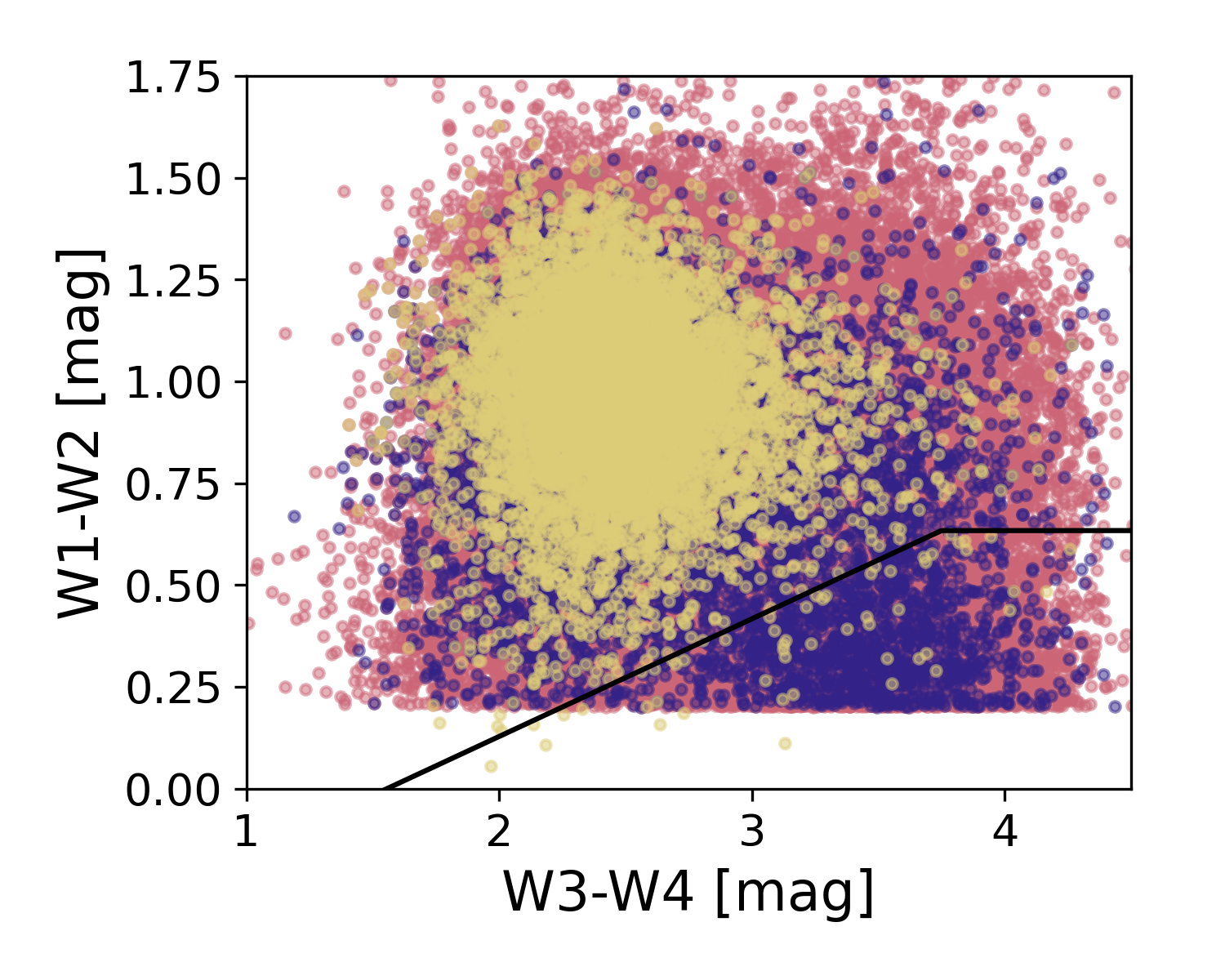}
\caption{\textit{Top}: Distribution of sources in the W1 band vs soft X-ray flux parameter space for the ALLWISE counterparts to 2RXS (pink) and XMMSL2 (blue). The confirmed SPIDERS AGN are represented in yellow. The cut defined in Eq. \ref{eq:xray_cut} is also shown. W1–W2 magnitude plotted against the W2-W3 (\textit{middle}) and W3-W4 (\textit{bottom}). The black lines show the cuts applied based on the SPIDERS AGN position.}
\label{fig:IR_color_color}
\end{figure}

The multiwavelength data collected in the previous sections can then be used to select AGNs from a larger sample comprising of blazars, galaxies, and stars using X-ray and IR colors observations.
Following the source characterization methods already established in \cite{salvato_finding_2018} and the references therein, we proceeded to a first selection of AGNs in the X-ray/MIR plane (see top panel of Fig. \ref{fig:IR_color_color}). An empirical relationship was found, separating AGNs from stars and galaxies.
\begin{equation}\label{eq:xray_cut}
W1 \geq \num{1.625} \times \rm log  F_{\rm 0.5-2keV} - 8.8
.\end{equation}
We can confirm the validity of such a selection by overlaying the confirmed AGNs in the SPIDERS sample, which all clearly lie above the cut-off line. We then use the AllWISE W1-W2, W2-W3, and W3-W4 relative magnitudes to isolate AGNs from blazars and starbust and normal galaxies, as developed in \cite{assef_mid-infrared_2013}. In the W1-W2 versus W2-W3 digram (middle panel of Fig. \ref{fig:IR_color_color}), the stars and elliptical galaxies exhibit colors near zero and are located in the lower left quadrant, while the spiral galaxies are red in W2–W3, but not in W1-W2, and ultraluminous infrared galaxies (ULIRGS) are red in both, lying in the upper right quadrant of the diagram \citep{wright_wide-field_2010}. We selected the sources for which $ 1.5 < W2-W3 < 4.5$ and $ 0.2 < W1-W2 < 1.75$ (black square in middle panel of Fig. \ref{fig:IR_color_color}). Once again, we used the SPIDERS AGN sample to justify the selection criteria being made in the W1-W2 versus W2-W3 color-color space. Similarly, a visual cut was made on the W1-W2 vs W3-W4 plane, following the SPIDERS-AGN locus (black line of bottom panel in Fig. \ref{fig:IR_color_color}) \citep{abbasi_search_2022}.

\begin{figure}
\centering
\includegraphics[width=0.45\textwidth]{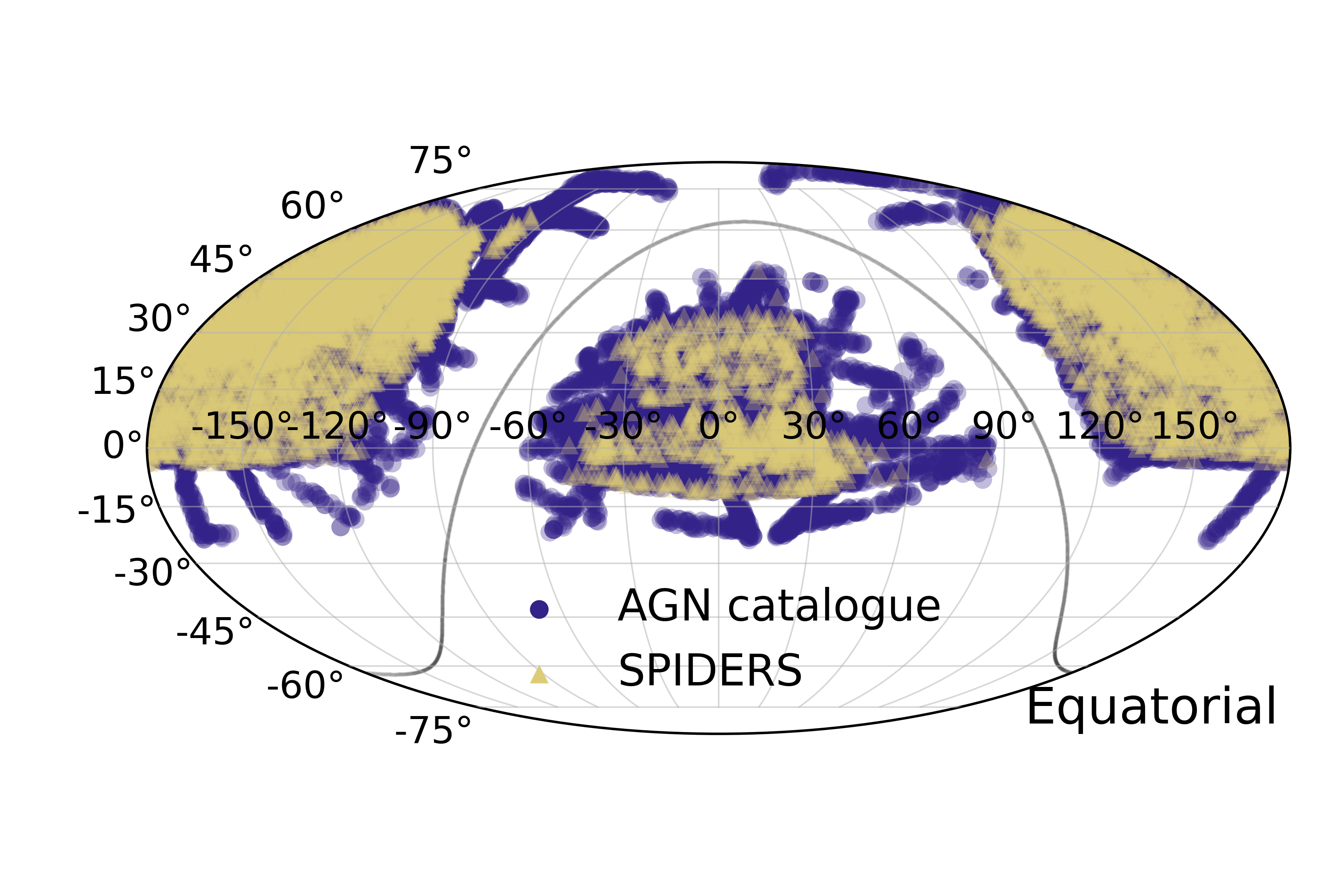}
\caption{Spatial distribution of sources in equatorial Mollweide projection for the for the selected AGN sample (in blue) and the SPIDERS AGN sample (yellow). The requirement for all sources to have been observed by SDSS constrain their distribution to the Northern Sky footprint. The galactic plane is shown as a gray line.}
\label{fig:map}
\end{figure}

After these selections, \num{60952} AGN are identified, from an original sample of \num{123436} X-ray-AllWISE sources. 
The spatial distribution of the final catalogue is shown in Fig. \ref{fig:map}: the sources follow the SDSS footprint, as the requirement to have been observed photometrically by SDSS marks the most stringent cut on the data.

\section{Measurement uncertainties and pseudo-sets} \label{sec:pseudo-sets}

\begin{figure}
\centering
\includegraphics[width=0.35\textwidth]{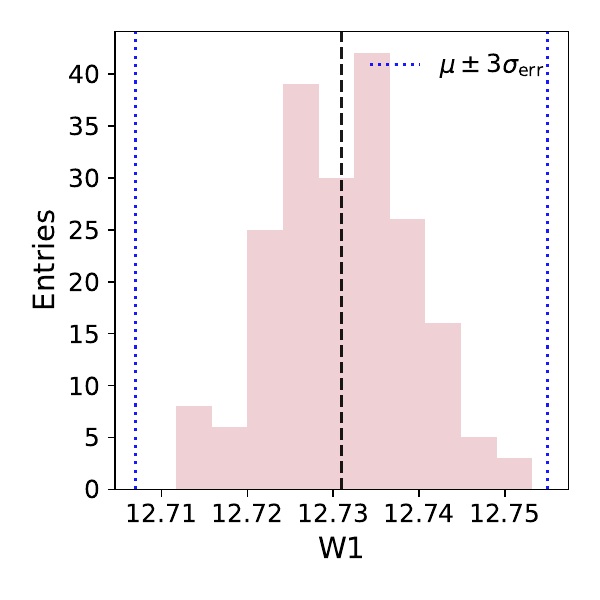}
\caption{Distribution of W1 input smeared by the measurement uncertainty for a single source. Each point is drawn from a normal distribution centered at the given catalogue input feature $\mu_{\rm value}$ (black dashed line) and extending to $\pm 3 \sigma_{\rm err}$ (blue dotted lines), from the given photometric measurement error.}
\label{fig:smearing_with_err}
\end{figure}

Each photometric observation used as an input (presented in Table \ref{tab:inputs}) comes with a measurement uncertainty of non-constant variance  (also called ``heteroscedastic'' error). Properly taking them into account is an active area of study in astrostatistics \citep{feigelson_twenty-first-century_2021}. Considering the stochastic nature of both input features and ML models, we adopted the approach outlined in \cite{shy_incorporating_2022}: all measurement errors, $\sigma_{\rm err}$, are assumed to be Gaussian, so that any photometric input for a single source is represented as a normal distribution, centered around the given value $\mu_{\rm value}$, extending to $\pm 3 \sigma_{\rm err}$. Figure \ref{fig:smearing_with_err} shows such an example of the W1 for a single training source with the measurement given by $\mu_{\rm value}$ (black dashed line) and $\mu_{\rm value} \pm 3 \sigma_{\rm err}$  (blue dotted lines). \newline
Drawing randomly from each independent ``smeared'' input distributions, we can thus create $N$ pseudo-sets for each AGN source, where the photometric inputs differ within $\pm \sigma_{\rm err}$ between each realization. We create $N$=200 pseudo-sets of both the training (for the regression) and the reconstructed datasets. The given photometric errors are not added as explicit features of the ML training, but are included through the scatter present in the photometric features across the $N$ pseudo-sets. Sections \ref{sec:classification} and \ref{sec:training} further develop the way this simulation-based treatment of measurement errors helps characterize both the performance and reconstruction of unlabeled or unknown data in the context of ML classification and regression tasks.

\section{Machine learning classifier for type 2 AGN identification} \label{sec:classification}

Once the AGNs have been identified, a further step is needed, namely: since our training sample exclusively contains type 1 AGNs (broad emission line, unobscured), we must distinguish type 1 from type 2 (narrow emission line, obscured) in our reconstructed sample to study any potential biases in the spectroscopic predictions. Obscured AGNs (also called type 2) are systems where the emission from the accretion disk gets absorbed and scattered by dust or gas surrounding it, masking some of the characteristic signature of the AGN: the AGN unification model \citep{antonucci_unified_1993, urry_unified_1995} states that the obscuration effect is merely a by-product of the observer's orientation, while others \citep{laor_nature_2003,elitzur_agn-obscuring_2006,ricci_reflection_2011} claim there to be a structural difference in the narrow- and broadline regions of AGNs.

The impact of such a suppression is wavelength-dependent, and a reliable and complete identification method of this population remains challenging and important. Distinctions between type 1 and type 2 AGNs can be made across multiple photometric and spectroscopic observation types. For a review on obscured AGNs, we refer to \cite{hickox_obscured_2018}. The most classical way to identify an AGN class is through UV-NIR spectroscopy \citep{comparat_final_2020, koss_bass_2022}. Type 1 AGNs have broad emission lines showing velocity dispersion $>$1000 km s$^{-1}$, while type 2 AGNs have narrow emission lines only, with a velocity dispersion $<$ 1000 km s$^{-1}$ \citep{padovani_active_2017}. However, since the purpose of this study is to characterize AGNs that have not been spectroscopically observed, we must circumvent the absence of such of information. \\

Our sources were originally selected based on their soft X-ray flux \citep{salvato_dissecting_2011}: this already skews the sample towards a majority of type 1 sources, as soft X-rays get absorbed by the high hydrogen column density, $N_{\rm H}$, around the accretion disk \citep{hasinger_absorption_2008}, while harder X-ray are less suppressed in obscured AGNs \citep{ananna_bass_2022}. These are also known to have stronger emission in the MIR than they do in other bands, as larger dust column reprocesses radiation from other bands. We thus expect weak UV/optical/NIR emission compared to that of the MIR \citep{hickox_obscured_2018}.
As described in Sect. \ref{subsubsec:spectroscopy_and_redshift}, we already know the AGN class for a sub-sample of sources that have a classification $\tt{CLASS\_BEST}$, so that the correlations between photometric observations and the object type can be studied. One could try to identify a single feature that best allows the distinction between obscured and unobscured AGNs, such as the ratio of W2/W1 IR emission. We developed this ``classical'' method  following the method used in \cite{abbasi_search_2022} and detailed in in Appendix \ref{appendix:feature selection}. However, a more judicious use of the multiwavelength information collected would be to train a classification machine learning model. Taking the \num{13415} sources for which SDSS classification is known as a training sample, we add a new feature called ``obscuration'': its value is 0 for type 1 AGNs and 1 for type 2 AGNs. We followed the classification into type 1 or 2 employed in the SDSS pipeline, whereby spectral templates to distinguish between the classes \citep{bolton_spectral_2012}.
We then sought to characterize whether the remaining \num{15533} are obscured or not.

\subsection{Imbalanced classification} \label{subsec:imbalanced_class}
Of the \num{13415} labeled sources, \num{12629} are unobscured  (including SPIDERS sources which are all type 1), while \num{786} are obscured AGNs, a ratio of 16:1. This classification task is thus an imbalanced one; this is a frequent situation, where a classifier must learn to identify a minority case, although it is trained on a dataset over-represented by a majority case. This ultimately leads to bias in the reconstructed sample.

Many strategies exist to mitigate this issue \citep{barandela_strategies_2003,batista_study_2004, wang_review_2021, kim_empirical_2022}. One approach is to randomly resample the training dataset: we can perform either a random undersampling (RUS) of the majority cased by selecting a sub-sample of type 1 AGNs or a random oversampling (ROS), where points in the minority class are duplicated. In both cases this results in a changed $n_{\rm 1}$/$n_{\rm 2}$ ratio, brought closer to parity. Both methods present some disadvantages: in the case of RUS, the greater part of the majority class is discarded and lost to the model training, while in ROS, the naive duplication of minority samples can lead to overfitting. One can also perform a more sophisticated type of oversampling, via SMOTE (Synthetic Minority Oversampling Technique), where \textit{k}-nearest neighbors are found to create synthetic minority class points \citep{chawla_smote_2002}.\newline
We tested baseline (no mitigation strategy), undersampling, oversampling, and SMOTE methods by training a random forest classifier (RFC) \citep{breiman_random_2001} with 18 smeared photometric measurement as input features (see Sect. \ref{sec:pseudo-sets}). We used a stratified five-fold cross-validation method \citep{stone_cross-validatory_1974} to separate the training and testing sets,  ensuring that all points were used for training and validation at least once. The resampling of the data was applied to the training datasets only, while the testing set retained the original imbalance of our AGN catalog.\newline

We call a true positive (TP) a type 2 AGNs classified as type 2, a false positive (FP) a type 1 classified as type 2, a true negative (TN) a type 1 classified as type 1, and a false negative (FN) a type 2 classified as type 1. The precision of the classification is then defined as: 
\begin{equation} \
P = \dfrac{TP}{TP + FP},
 \end{equation}
 that is, the ability of the classifier not to label as positive a sample that is negative. The recall, also called sensitivity or true positive rate (TPR), is calculated with:
 \begin{equation}
 R = \dfrac{TP}{TP + FN},
 \end{equation}
  that is, the ability of the classifier to find all the positive samples. For completion, we provide the definition of the false positive rate (FPR), used in the receiving operating characteristic (ROC) curve,\footnote{In instances of classification task on imbalanced datasets, the precision-recall curve (PRC) is more informative than the ROC \citep{saito_precision-recall_2015}.}
   \begin{equation}
 FPR = \dfrac{FP}{TN + FP},
 \end{equation}
so that it is a measure of finding all the negative samples. 

\begin{figure}
\centering
\includegraphics[width=0.3\textwidth]{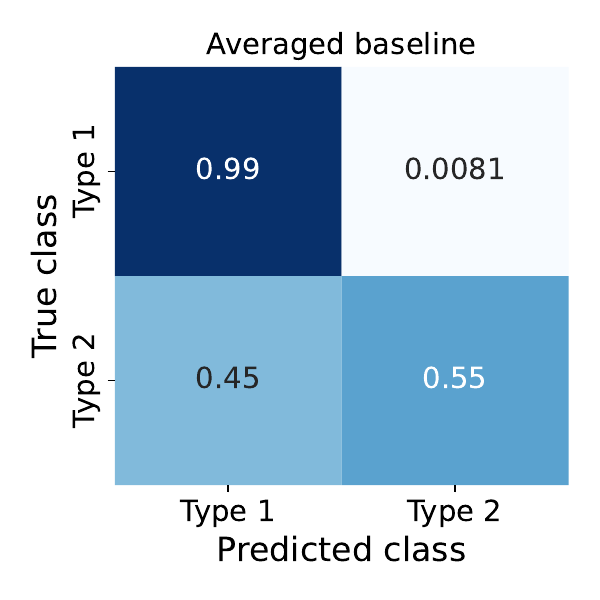}
\includegraphics[width=0.3\textwidth]{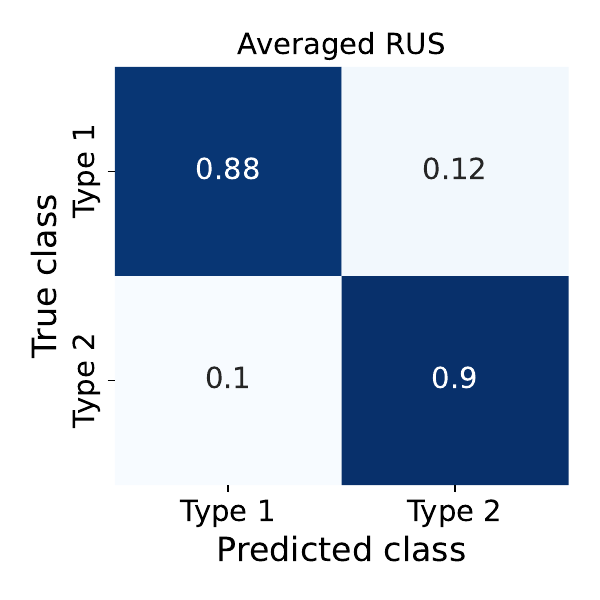}
\includegraphics[width=0.3\textwidth]{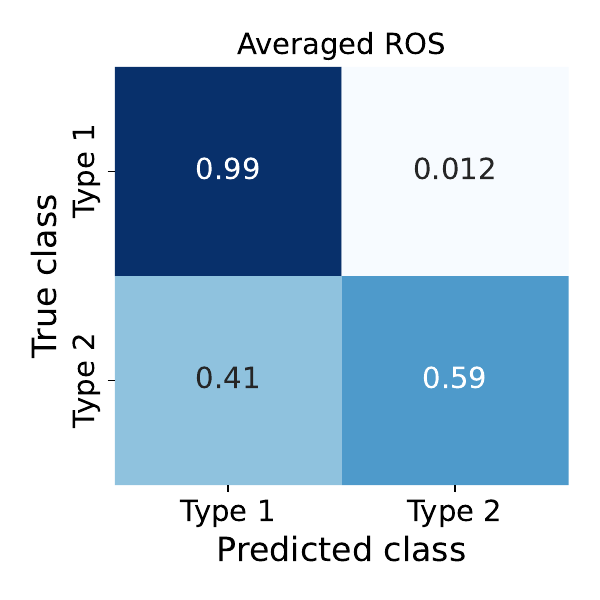}
\includegraphics[width=0.3\textwidth]{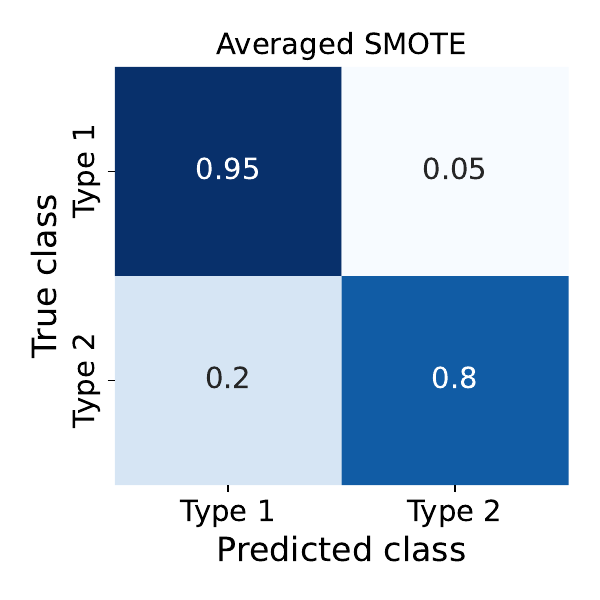}
\caption{Confusion matrices for different imbalanced classification mitigation techniques.. The naturally imbalanced set (baseline), while accurately selecting Type 1 AGN 99\% of the time, performs poorly in finding the rarer Type 2 AGN. Both RUS and SMOTE techniques show great classification improvements.}
\label{fig:classification_obscureness}
\end{figure}

Figure \ref{fig:classification_obscureness} shows the confusion matrices for all tested resampling method  presented in Fig. \ref{fig:classification_obscureness}. As expected, the baseline model, while it is able to identify type 1 AGNs with ease (99\% TPR), it is not able to reliably identify type 2 AGNs . The same can be said of the ROS, where the naive duplication of minority samples seems to improve the classification only slightly. Clear improvements start to show for the RUS and SMOTE trained samples. The RUS method shows a larger type 2 AGNs recall than the SMOTE technique (90\% vs 80\%), but a decreased precision, with a larger fraction of type 1 AGNs falsely identified as type 2 AGNs (12\% for RUS, 5\% for SMOTE). Using SMOTE to train the RF classifier thus appears to be a good compromise between precision and recall, with a larger fraction of properly classified type 1 AGNs (the greater dataset), along with an acceptable identification of type 2 AGNs.

\subsection{Classification accuracy on the labeled set} \label{subsec:accuracy_classifier}

\begin{figure}
\centering
\includegraphics[width=0.35\textwidth]{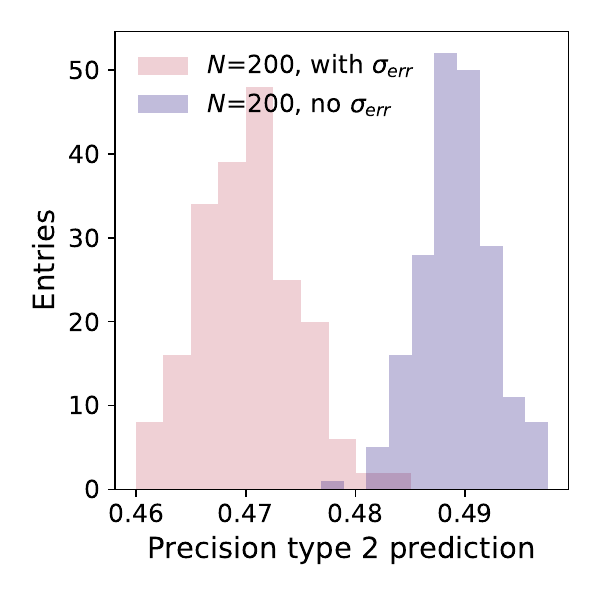}
\includegraphics[width=0.35\textwidth]{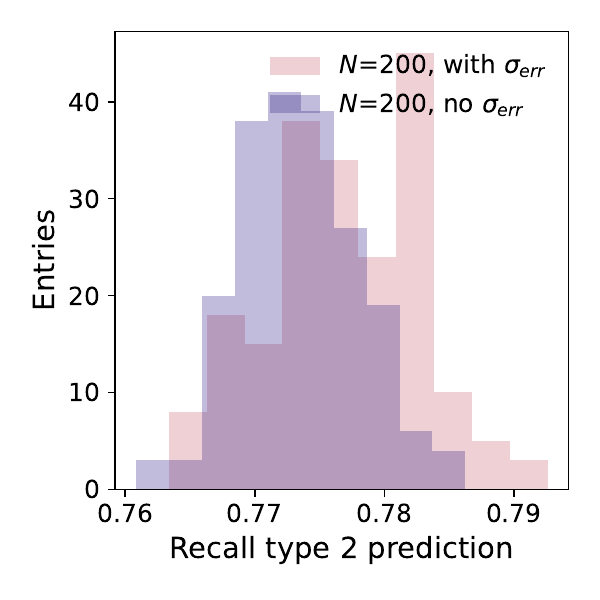}
\caption{Posterior distributions for the Type 2 precision (\textit{left}) and recall predictions  (\textit{right}) from fitting the labeled dataset $N$=200 times. The blue distribution indicate the value obtained from $N$ RF reconstructions without inclusion of measurement errors, while the red distribution correspond to the reconstructions of the measurement uncertainty propagated pseudo-sets. In certain cases, the performance of the classifier is overestimated when photometric uncertainties are not taken into account.}
\label{fig:posterior_dist_precision}
\end{figure}

After the study on a single pseudo-set, we settled on training a random forest classifier using the SMOTE method. To propagate the photometric measurement uncertainties in the input features, we made use of the $N$=200 pseudo-sets previously generated (see Sect. \ref{sec:pseudo-sets}) to label the unknown AGNs. We follow the steps outlined in \cite{shy_incorporating_2022}: for each simulation set, a classifier is fit to a realization of the labeled data, then used to reconstruct a realization of the unlabeled data. This way, all sources are reconstructed $N$ times, whether they belong to the unlabeled or the validation datasets. For all performance metrics defined in Sect. \ref{subsec:imbalanced_class}, we thus obtained a posterior predictive distribution comprising of the results of each set's classification (Fig. \ref{fig:posterior_dist_precision}). The variation across multiple fits reflects the propagated uncertainty through all steps of the procedure. The RF classifier prediction is inherently stochastic, as the blue distributions indicate: these result from running the classifier $N$ times on the exact same dataset, that is,  not using the error-propagated pseudo-sets. For some performance metrics, such as the precision of type 2 predictions (left panel of in Fig. \ref{fig:posterior_dist_precision}) and the type 1 recall, ignoring measurement uncertainties leads to an overestimation of the predictive power of a classifier, in agreement with the findings of \cite{shy_incorporating_2022}. For other metrics (type 1 precision and type 2 recall, shown in the right panel of Fig. \ref{fig:posterior_dist_precision}), while the mean prediction of the two methods does not significantly differ, the pseudo-set generated distributions display a greater variance. The final scores with uncertainties on the AGN classifier can be found in Table \ref{tab:scores_RF}. \newline

\begin{table}
\centering
\resizebox{0.45\textwidth}{!}
{\begin{tabular}{|c|c|c|}
\hline
AGN Type & Precision & Recall  \\   
\hline
Type 1 &  0.986 $\pm$ 0.000 & 0.945 $\pm$ 0.001 \\
Type 2 &  0.470 $\pm$ 0.004 & 0.777 $\pm$ 0.006 \\
\hline
\end{tabular}}
\caption{Precision and recall scores and uncertainties for Type 1 and Type 2 prediction using $N$=200 fits to pseudo-sets with measurement uncertainties.}
\label{tab:scores_RF}
\end{table}

\subsection{Further softening a soft classifier} \label{subsec:softening_classifier}

\begin{figure}
\centering
\includegraphics[width=0.35\textwidth]{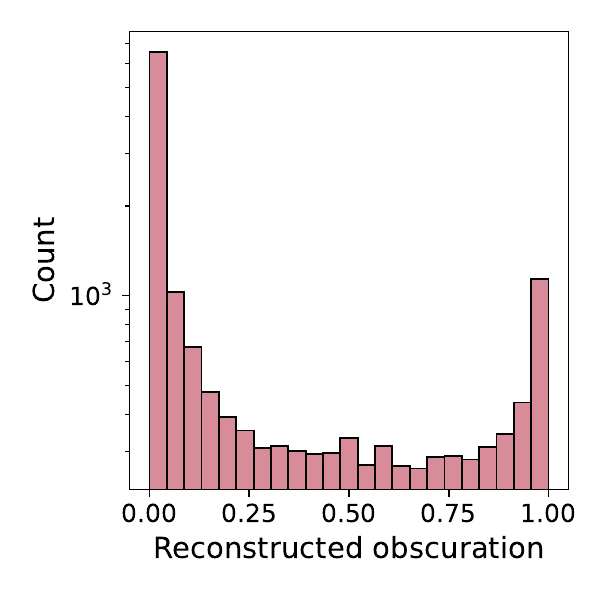}
\caption{Histogram of the averaged reconstructed obscuration values for all unlabeled data. While the majority of sources have an obscuration value equal to 0 or 1, a non-negligible number of them lie in the region between the two.}
\label{fig:reconstructed_obscuration}
\end{figure}

In addition to giving a more accurate view of the classifier's performance thanks to the validation set, this simulation-based method introduces further nuances into the reconstruction of the unlabeled obscuration level. With each source now being reconstructed $N$=200 times, the reconstructed obscuration is then the mean of the relative probability of a source to be in each class; that is, the RF is known as a ``soft'' classifier, as it provides for each object a continuous, relative confidence value of belonging to a class, between 0 and 1 \footnote{Whereas a hard classifier would only give  0 or 1 values based on class threshold, usually 0.5.}. For the unlabeled data, we recorded the relative class probability for each $N$ reconstruction. The final obscuration level of each source, $\mu_{\rm obscuration}$, is thus the   arithmetic mean of the $N$ classification results, with its associated standard deviation value $\sigma_{\rm obscuration}$: we have further softened a soft classifier. Fig. \ref {fig:reconstructed_obscuration} shows the reconstructed obscuration level for the unlabeled data, which lies in a continuous spectrum between 0 and 1. We established a custom decision threshold $t$ on $\mu_{\rm obscuration}$ and $\sigma_{\rm obscuration}$ for an AGN source to be considered as either type 1 or type 2. This threshold can be set to be more or less stringent. Choosing to enhance the purity of the classification, we set $t$=0.7, so that an AGN source is considered of type 2 if $\mu_{\rm obscuration}>0.7$ and of type 1 if $\mu_{\rm obscuration}<0.3$. \newline

\begin{figure}
\centering
\includegraphics[width=0.45\textwidth]{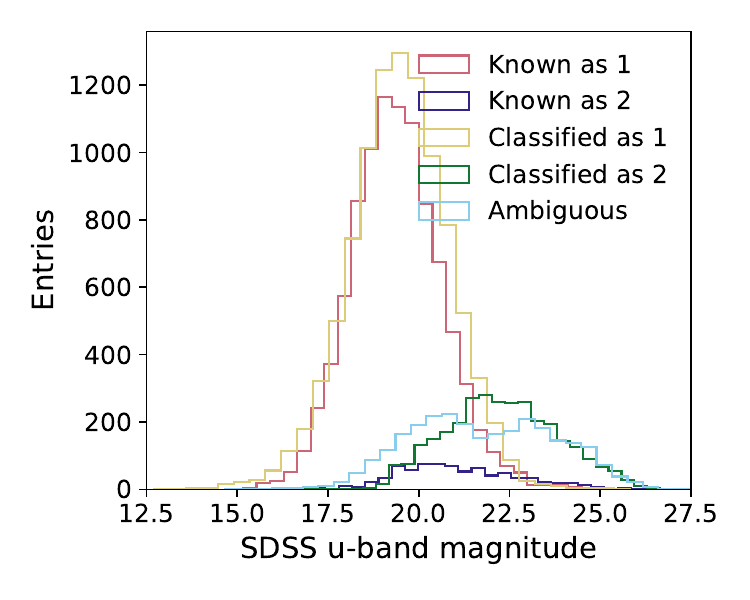}
\caption{SDSS \textit{u}-band magnitude for all labeled and reconstructed datasets. The unlabeled sources classified as Type 2 AGN are fainter than the labeled Type 2 sources, which have made the photometric limit criteria for SDSS spectroscopic observations. In general, obscured AGN have a fainter optical spectra than unobscured ones.}
\label{fig:reconstructed_obscuration_distribution}
\end{figure}
Doing so, we find that \num{9747} are marked as type 1, \num{3062} as type 2, and \num{2719} as ``ambiguous.'' This corresponds to a $n_{\rm 1}$/$n_{\rm 2}$ ratio of $\sim$ 3.2:1, which is markedly smaller than the 16:1 ratio from the labeled dataset. The reason for such a stark discrepancy can be found in Fig. \ref{fig:reconstructed_obscuration_distribution}, which shows how the labeled dataset (``known as 1'', ``known as 2'') is biased towards optically brighter (\textit{u}-band mag < 24). This follows from the target requirements established by the various SDSS surveys prior to the spectroscopic observations of the AGN targets \citep{alam_eleventh_2015}. The AGN classified as T=type 2 (green histogram) constitute the fainter end of our catalog, too faint to have been spectroscopically followed-up. Because the RF classifier infers that fainter sources are more likely to be of type 2, the reconstructed unlabeled catalog naturally results in a more balanced AGN ratio.

\section{Machine learning for AGN property estimations} \label{sec:training}

The following section presents the detail of selecting a suitable machine learning model to predict the parameters of Table \ref{tab:outputs}, using  the features presented in Table \ref{tab:inputs} as inputs. Since redshift measurements are available for almost half of the \num{21050} AGN sources (but not for the other), we trained and tested two separate models, which we call $ML_{\rm w/z}$, where $z$ is added as an input and  $ML_{\rm wo/z}$, where $z$ is one of the outputs of the regressor. Just as is was done in Sect. \ref{sec:classification} for ML classification, we develop how measurement errors are taken into account in ML regression using the pseudo-sets generated, thereby offering a more complete picture of the performance and quality of the reconstruction. For the training, we transformed our target parameters: $z$, $L_{\rm X}$, $L_{\rm Bol}$, $M_{\rm BH}$, and $\lambda_{\rm Edd}$ are often expressed in log scale to representatively describe the span of values across several decades.
\subsection{Exploratory data analysis} \label{subsec:eda}

\begin{figure}
\centering
\includegraphics[width=0.45\textwidth]{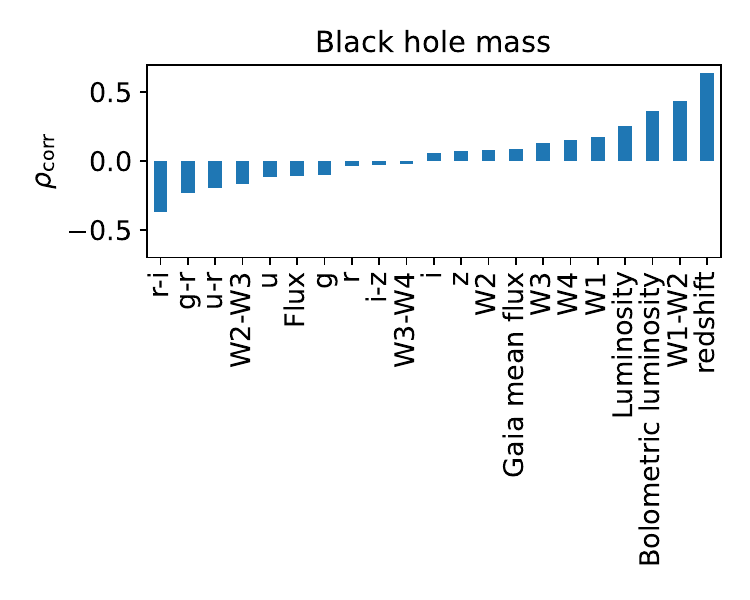}
\caption{Bar chart showing the Pearson correlation score of input variables and the black hole mass, from the training sample data. The redshift, luminosity, and bolometric luminosity correlations are included in this chart, since $z$ (and thus $L_{\rm X}$) are known for almost half of the sources, and the outputs will be predicted before the black hole mass, underlining the logic behind the chain regression.}
\label{fig:barplot_corr_plot}
\end{figure}

\begin{figure}
\centering
\includegraphics[width=0.35\textwidth]{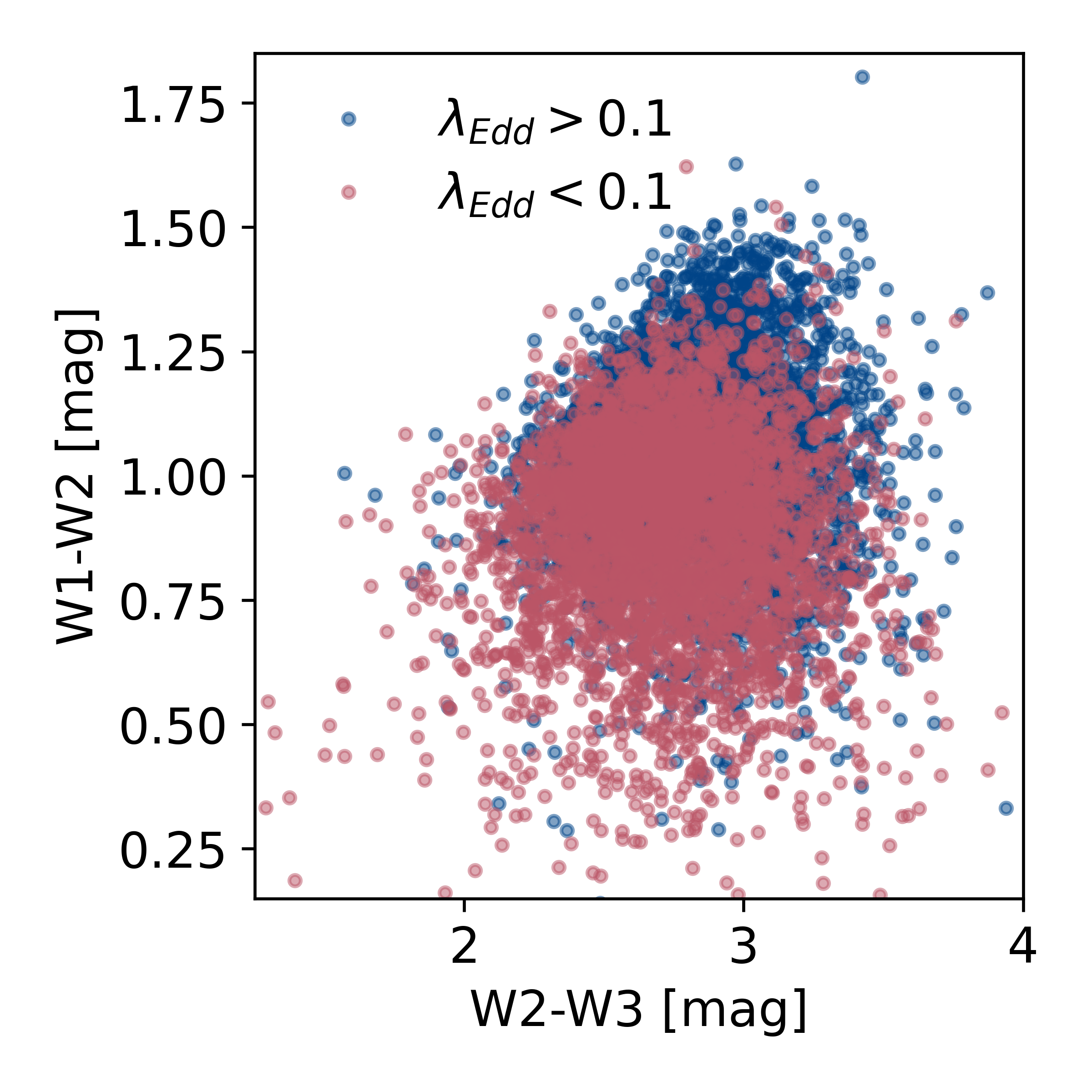}
\caption{W1-W2 magnitudes as a function of W2-W3 magnitudes for sources in the training sample with low (red dots) and high (blue dots) $\lambda_{\rm Edd}$. The low and high samples are separated by the median value of the $\lambda_{\rm Edd}$ distribution, 0.1.}
\label{fig:corr_plot}
\end{figure}

\indent Before choosing and training a machine-learning algorithm on the SPIDERS sample, we explored the relationship between the input variables and the target parameters. Figure \ref{fig:barplot_corr_plot} shows the sorted Pearson's correlation coefficients for all inputs and one of the outputs, the black hole mass of the AGN. The relative IR and optical color magnitudes demonstrate the highest level correlation. This is even more visually evident when one looks, once more, at the IR color-color plot in Fig. \ref{fig:corr_plot}. The scatter plot presents the W1/W2 versus W2/W3 AllWISE colors for AGNs with $\lambda_{\rm Edd} < 0.1$ and $\lambda_{\rm Edd} > 0.1$, the median value of $\lambda_{\rm Edd}$ in the training sample. We observe that strong accretion disks (higher $\lambda_{\rm Edd}$) are redder in both W1/W2 and W2/W3 than lower $\lambda_{\rm Edd}$ values. These clear connections between IR photometry and spectroscopic observables, already accessible with a naive and straightforward data analysis, are encouraging indications that our goal --- the estimation of AGN physical properties --- is suited for a machine-learning task.

\subsection{Machine learning model parameters} \label{subsec:ML_parameters}

\begin{table*}
\centering
\resizebox{1.\linewidth}{!}{
\begin{tabular}{|cccc|}
\hline
Model features & Type & Properties & Notes \\   
\hline
Model type  & Random Forest Regression & \makecell{Max. depth = 25,\\ Max. features=0.9,\\Number of estimators=100} & \makecell{Training a chain regressor\\connects the non-independent target\\parameter to one another} \\
\hline

Data scaling & Max-min normalization & Inputs and outputs are scaled &  Aids the model to learn the problem \\
\hline
Validation method & K-fold cross-validation & \makecell{$k$=10\\with data shuffle}  & \makecell{\\Ensures the model gets trained\\on every single data point}\\
\hline
\end{tabular}
}
\caption {Properties of the final RF regressor ML-model properties chosen to be trained on.}
\label{tab:ML_properties}
\end{table*}
Ours is essentially a multi-dimensional linear regression task, with 18 multi-wavelength inputs, listed in Table \ref{tab:inputs}, and 5 or 6 target parameters, depending on whether $z$ is known for a source (see Table \ref{tab:outputs}). Many ML applications are readily available to use for such a supervised learning task notably through the $\tt{scikit-learn}$ python library \citep{pedregosa_scikit-learn_2011}.
We used a single-output, multi-step chain regression \citep{demirel_ensemble_2019} so that the ML model can learn the correlations between target parameters, as previously done in \cite{cunha_photometric_2022}. In the first pass of the chain regressor, the initial 18 inputs are used to predict the first output, namely, the redshift $z$. In the next pass, the model takes 18+1 inputs, the extra-one being the predicted $z$, and outputs the next parameter, $L_{\rm X}$, and so on.
 
\subsubsection{Selection of the ML model} \label{subsubsec:estimator}
We detail in this section our non-exhaustive search for the most suitable estimator. All supervised learning ML models essentially learn a mapping of inputs to outputs given an example of such a map. However,  there is a plethora of model types available to choose from. For instance, linear models expect the output to be a linear combination of the features: certain regressors simplify the model by introducing penalty coefficients that will minimize (in the case of ridge regression \citep{hilt_ridge_1977}) or reduce (for Lasso regression \citep{tibshirani_regression_1996}) the input parameters, if some are found to contribute less to the learning. This procedure, called regularization, aims to reduce the overall error in the validation dataset. On the other hand, support vector regression (SVR), an application of the kernel-based support vector machines \citep{cortes_support-vector_1995}, allows us to tune the tolerance, $\epsilon,$ to such errors, while introducing non-linearity parametrization through hyperplane fits to the data. Non-linearity is also a feature of neural networks, for instance, in a multi-layer perceptron (MLP) \citep{murtagh_multilayer_1991}, via an activation function connecting the neural layers. Finally, decision tree predictors, such as random forests \citep{breiman_random_2001}, also present the advantages of handling non-linear relationships between the input and output parameters. We compare these different ML models in the next section.

\subsubsection{Model evaluation} \label{subsubec:model_evaluation}

\begin{figure}
\centering
\includegraphics[width=0.35\textwidth]{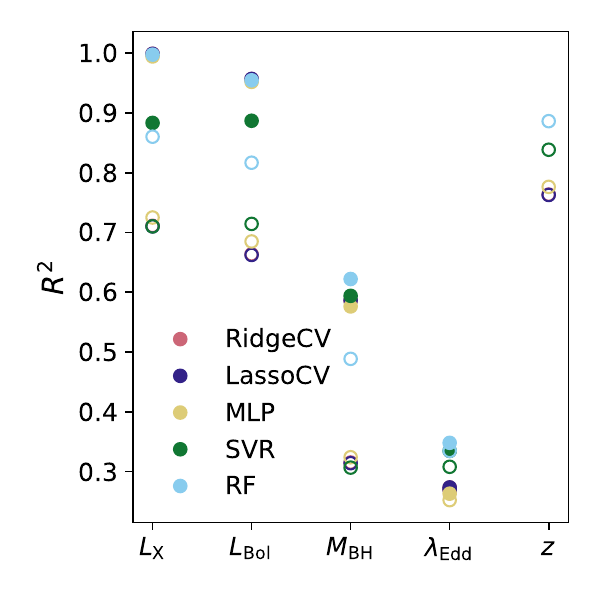}
\caption{Comparison of $R^{2}$ for ML-models tested on all target parameters. Full circles represent $ML_{\rm w/z}$ and open circles $ML_{\rm wo/z}$, the learning done for sources with unknown redshift. The RF algorithm performs best on crucial variables ($M_{\rm BH}$ and $\lambda_{\rm Edd}$).}
\label{fig:comparison_models}
\end{figure}

To determine the best model, performance metrics are defined based on the validation dataset, where $y_{\rm true}$ and $y_{\rm reco}$ are known. We again used a \textit{K}-fold cross validation method to insure a non-biased evaluation of the model, as each sample (in our case, each AGN source), will be used as a validation point once, and as a training set $k-1$ times: the following metrics are thus calculated for a sample size $N$ = \num{7616}.
As is common for regression problems, we use the $R^{2}$ score, also called coefficient of determination, for each parameter to assess the performance of each model. This coefficient is calculated as: 
\begin{equation}
\centering
R^{2} = 1 - \frac{\sum_{i} (y_{i_{\rm true}} - y_{i_{\rm reco}})^{2}}{\sum_{i} (y_{i_{\rm true}} - \overline{y_{\rm true}})^{2}},
\end{equation}
with the numerator being the residual sum of squares, and the denominator the total sum of squares. For a perfect regressor, we have $R^{2} = 1 $.\
Figure \ref{fig:comparison_models} presents the target-by-target comparison between the two linear models (ridge and Lasso regression), a support vector regression model, a multi-layered perceptron (MLP) deep neural network model\footnote{As default parameters, the MLP has one hidden layer with 100 neurons, use the ``relu'' activation function and the ``adam'' optimizer. More details can be found \url{https://scikit-learn.org/stable/modules/generated/sklearn.neural_network.MLPRegressor.html}}, and a RF regressor. We stress that for this test, none of the model parameters have been tuned. The RF model is better at predicting all target outputs, in particular $M_{\rm BH}$, $L_{\rm Edd}$, and $\lambda_{\rm Edd}$, with the SVR model coming in closely. This trend is also confirmed in the more difficult case of unknown $z$, represented by open circles in Fig. \ref{fig:comparison_models}. From a pragmatic aspect, the runtime speed and low number of tuning parameters of the RF regressor were clear advantages compared to the vast phase space of MLP neural networks, for instance.  

Once the performance of the RF model has been assessed, we completed a grid search over its hyperparameters to find their optimal values: the maximum depth, that is, the number of splits that each decision tree is allowed to make, the maximum number of features considered on a per-split level and the number of estimators, and the number of decisions trees in the forest. This step is important as the final result can quite vary between default and optimized parameters. We summarize the final parameters of the machine learning model that is selected for training in Table \ref{tab:ML_properties}. 

\subsubsection{Regression metrics on N pseudo-sets}

As was done for the classification task, we then used the $N$=200 pseudo-sets to propagate both the uncertainties in the photometric measurements in the training and reconstructed datasets, as well as fluctuations of the regressor's reconstruction. Here, we adopt an iterative method once more, where, for each $i$-th training sample, $T_{i}$, that  the RF model is fitted with, values of $T_{i}$ are predicted (for performance evaluation studies). This $i$-th fitted RF is then used to reconstruct the target parameters in $C_{i}$, which is the $i$-th pseudo-set of the unreconstructed catalog. The process was repeated $N$ times.

%\begin{algorithm}
%\caption{Iterative fitting and reconstruction on $N$ pseudo-sets}\label{algo_regressor}
%\begin{algorithmic}[1]
%\Procedure{ML regressor}{$T,C$}
%\For{$i=1....N$}
%\State Fit RF to $T_{i}$ for $ML_{i}_{\rm w/z}$ \Comment{Using $z$ an an input}
%\State Fit RFF to $T_{i}$ for $ML_{i}_{\rm wo/z}$ \Comment{$z$ is the first output}
%\State Predict $T_{i}$ using $ML_{i}_{\rm w/z}$ and $ML_{i}_{\rm wo/z}$  
%\State Reconstruct \num{9944} $C_{i}^$ sources with $ML_{i}_{\rm w/z}$ 
%\State Reconstruct \num{11363} $C_{i}^$ sources with $ML_{i}_{\rm wo/z}$ 
%\EndFor
%\EndProcedure
%\end{algorithmic}
%\end{algorithm}

\begin{figure}
\centering
\includegraphics[width=0.45\textwidth]{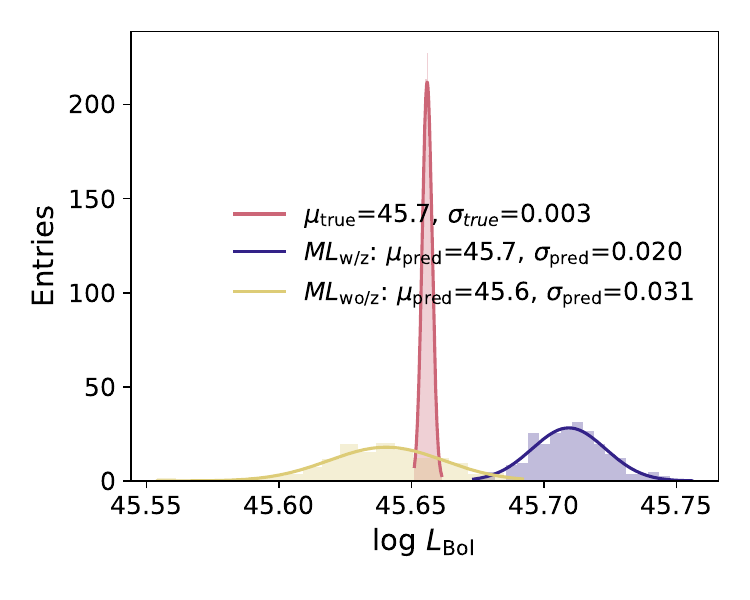}
\includegraphics[width=0.45\textwidth]{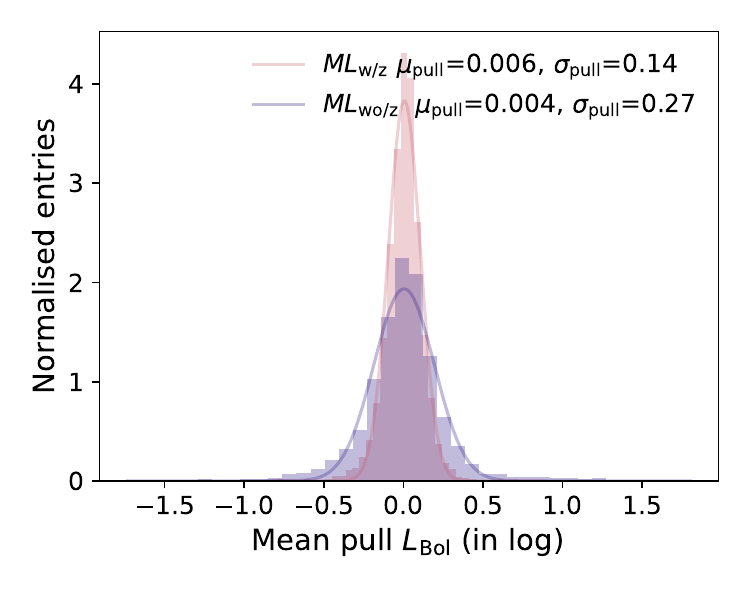}
\caption{\textit{Top}: True (red), and predicted distributions reconstructed $N$ times with $ML_{\rm w/z}$ (purple) and $ML_{\rm wo/z}$ (yellow) of the bolometric luminosity for a training source. The true value is represented by a normal distribution by taking into account the measurement error $\sigma_{\rm true}$ and assuming it to be gaussian. \textit{Bottom}: Mean pull distribution for the $L_{\rm Bol}$ for all training sources, taking all $\mu_{\rm true}-\mu_{\rm pred}$ values for $ML_{\rm w/z}$ (red) and $ML_{\rm wo/z}$ (purple).}
\label{fig:single_source_lbol_reco_}
\end{figure}

The top panel of Fig. \ref{fig:single_source_lbol_reco_} displays the true and predicted distributions for the bolometric luminosity of a single source in the training sample. Here, $\mu_{\rm true}$ and $\sigma_{\rm true}$ are given by the SPIDERS-AGN catalog for all target parameters and all sources, while $\mu_{\rm pred}$ and $\sigma_{\rm pred}$ were obtained by fitting a Gaussian function to the $N$=200 reconstructed values for each source. From these, we constructed the ``pseudo-pull" distribution, $\mu_{\rm true}-\mu_{\rm pred}$. We note how, in this single source, the $ML_{\rm w/z}$ (purple) and $ML_{\rm wo/z}$ (yellow) recontructed values of $L_{\rm Bol}$ are multiple $\sigma_{\rm pred}$ apart. This is also clear from the bottom panel of Fig. \ref{fig:single_source_lbol_reco_}, which shows the distribution of all pulls for the reconstructed $L_{\rm Bol}$. Here, the difference in the quality of reconstruction between $ML_{\rm w/z}$ and $ML_{\rm wo/z}$ is apparent in the greater smearing of the pull distribution.

The precision of the reconstruction is also derived from the normalized median absolute deviation $\sigma_{\rm NMAD} = 1.48 \times$ median($\mid \Delta \mu$ - median$(\Delta \mu) \mid / \mu_{\rm true}$, expressed in \%, with $\Delta \mu = \mu_{\rm true}-\mu_{\rm pred}$. It is a robust measure of the deviation, one that is insensitive to outliers. 

\begin{figure}
\centering
\includegraphics[width=0.45\textwidth]{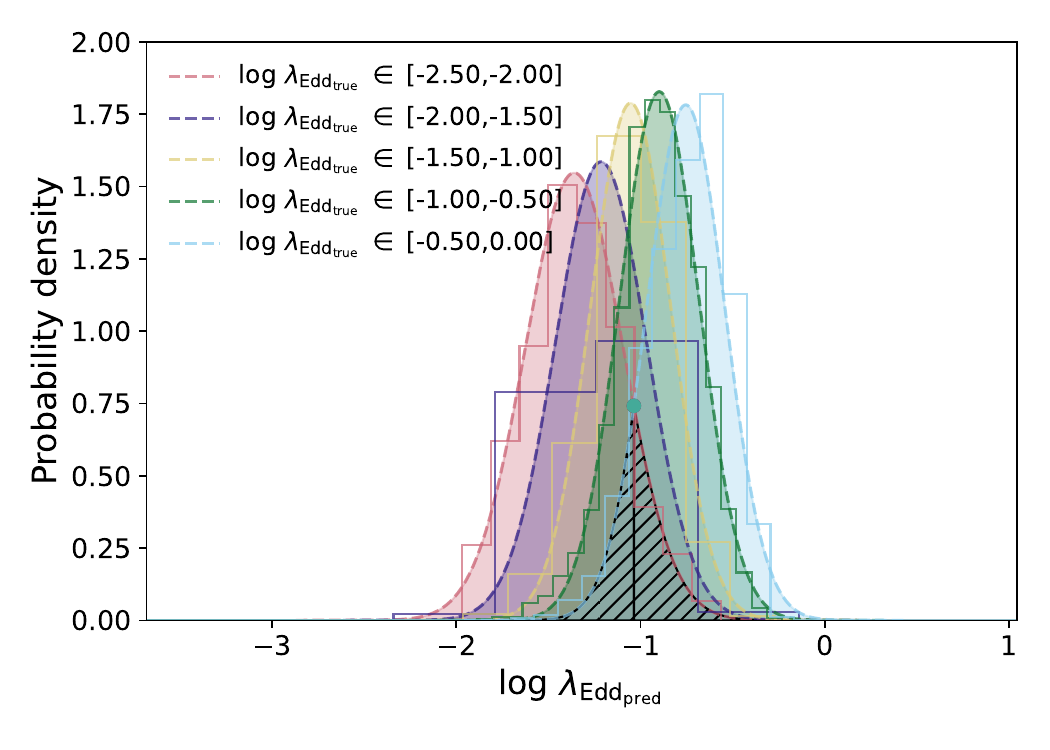}
\caption{Distributions of $\lambda_{\rm Edd_{pred}}$ for various bins in $\lambda_{\rm Edd_{true}}$, in log space. Gaussian PDFs are fitted and shown over their respective histograms. The hatched area corresponds to the contamination between the lowest and highest range PDFs: the values quoted in Table \ref{tab:ml_results} were calculated by dividing the overlapping (hatched area) with the integral of the highest range PDF.}
\label{fig:contamination_edd_ratio}
\end{figure}

Finally, the last metric we establish is the contamination level of the reconstruction: for each bin in $\mu_{\rm true}$, we look at the $\mu_{\rm pred}$ distribution and fit a Gaussian probability density function (PDF), as seen in Fig. \ref{fig:contamination_edd_ratio}. We then defined the contamination to be the overlapping area between the lowest and highest true intervals reconstructed PDF, represented by the hashed area in Fig. \ref{fig:contamination_edd_ratio}. In the case of the $\lambda_{\rm Edd}$ parameter, it is the measure of how often our ML-model mistakes a low accretion-rate AGN with a high accretion-rate AGN and vice versa. This is a useful measurement in the event that the regressor lacks the precision to carry on single source studies, but still provides enough information to consider the features of a larger population, as it does for $\lambda_{\rm Edd}$. Thus, even though the mean of the binned reconstructed values do not correspond to the center of the true bins, the scaling relation between lower and higher $\lambda_{\rm Edd}$ is preserved. 

All performance metrics described in this section are presented in Table \ref{tab:ml_results} for the $ML_{\rm w/z}$ and $ML_{\rm wo/z}$ cases. In the following section, we discuss in greater details the ability of our ML model to predict the physical parameters of AGN cores.

\subsection{Prediction performance} \label{subsec:performance}

\begin{figure*}

\includegraphics[width=0.5\hsize]{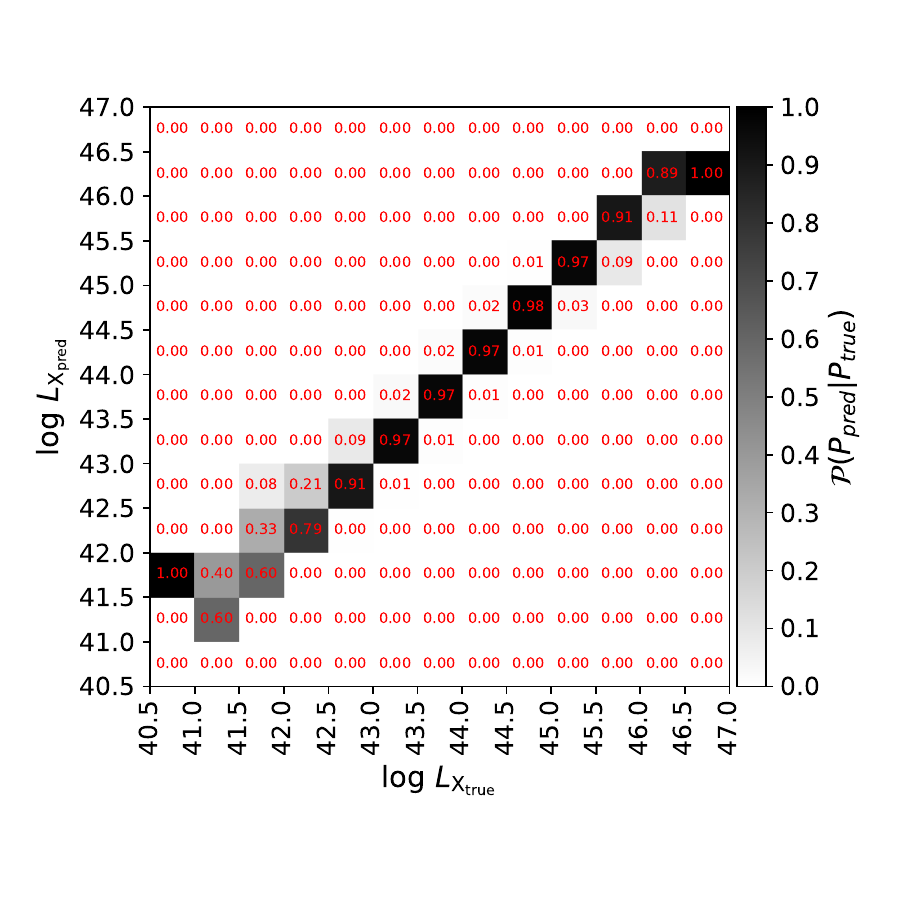}
\includegraphics[width=0.5\hsize]{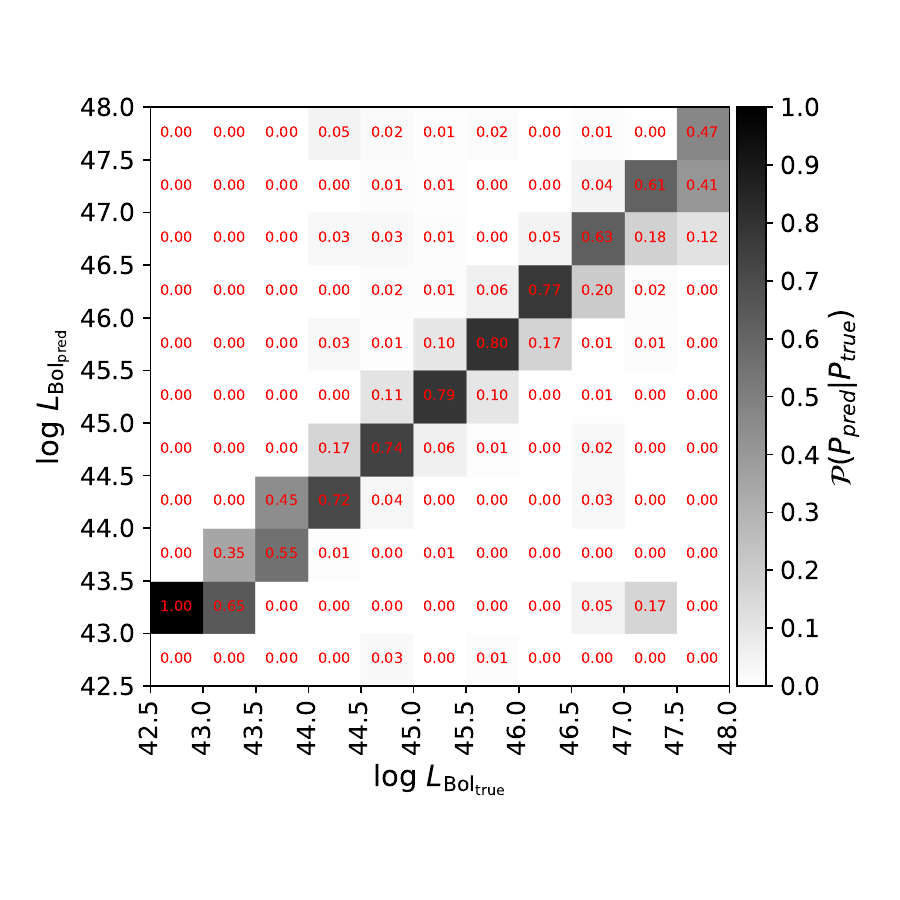}
\vspace{-8.00mm}
\includegraphics[width=0.5\hsize]{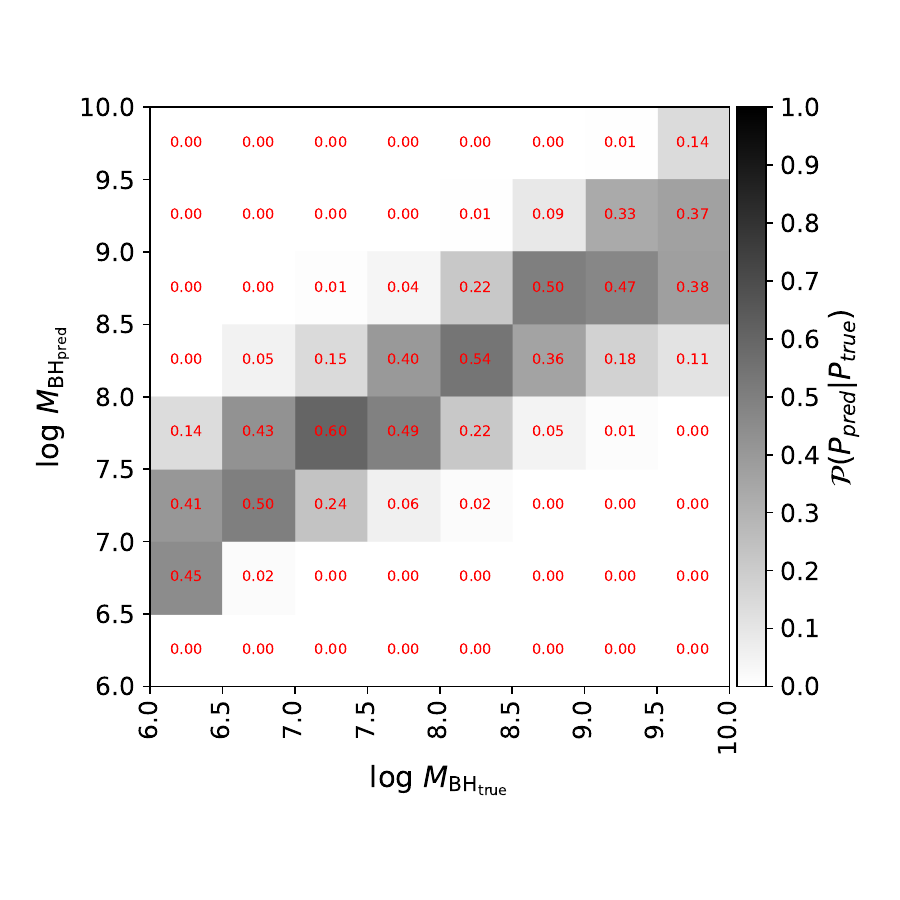}
\includegraphics[width=0.5\hsize]{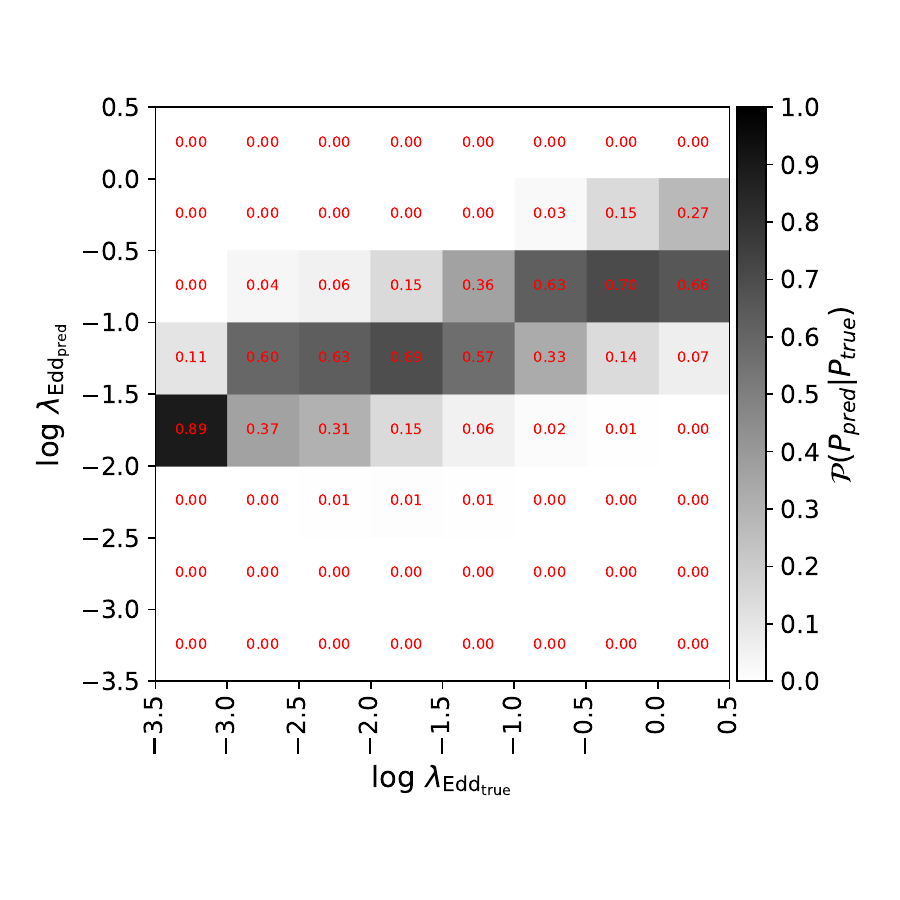}
\caption{Normalized performance matrices for ML-estimator with known redshift as an input ($ML_{\rm w/z}$). The true and reconstructed parameters are plotted on the x and y axis, respectively. The error on the reconstruction is used as a weight to the histogram.}
\label{fig:response_matrices_with_z}
\end{figure*}

\begin{figure*}

	\includegraphics[width=0.5\hsize]{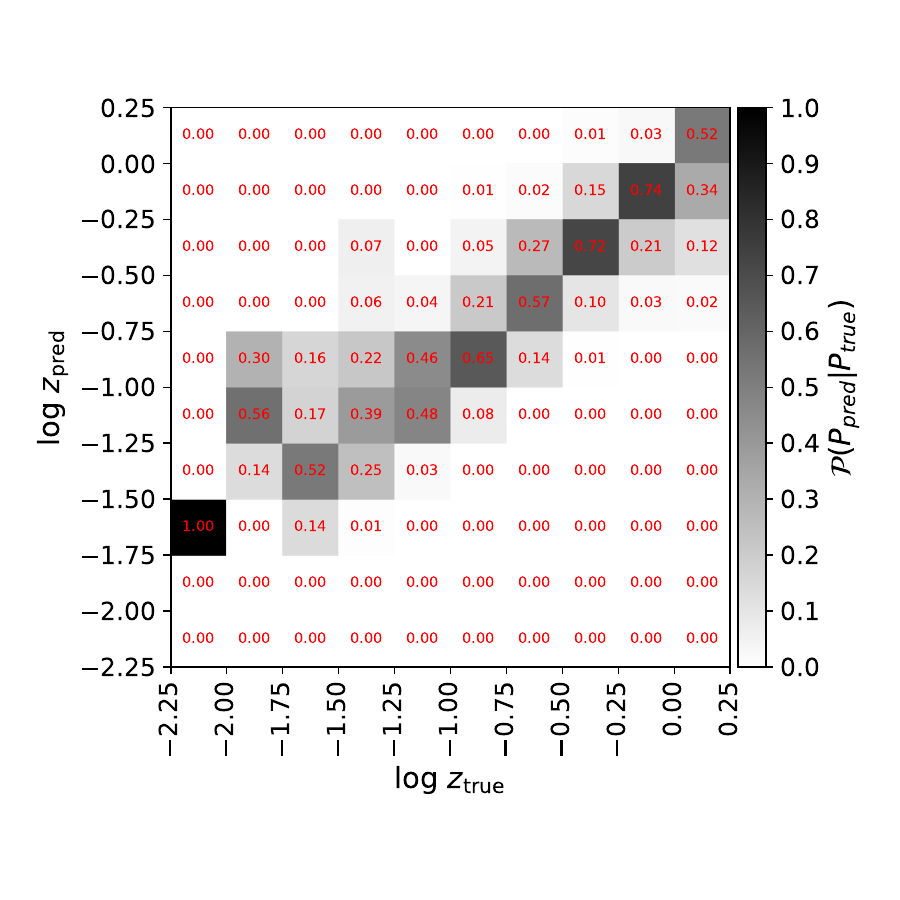}
	\includegraphics[width=0.5\hsize]{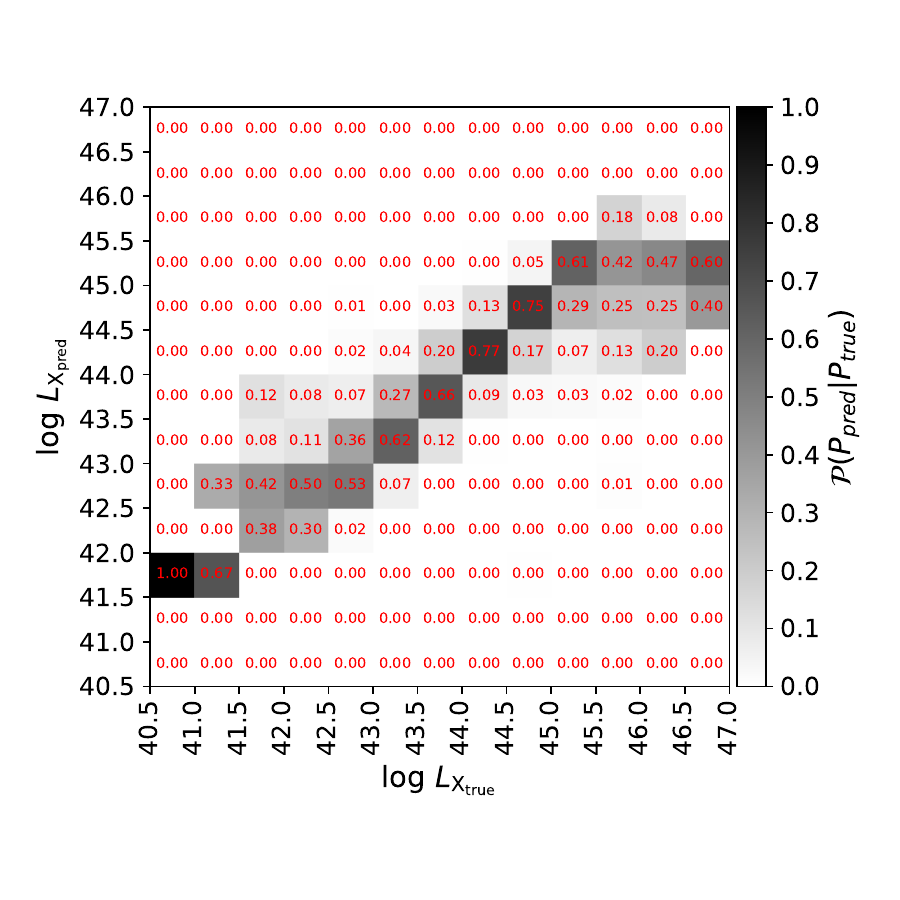}
	\vspace{-10.00mm}
	\includegraphics[width=0.5\hsize]{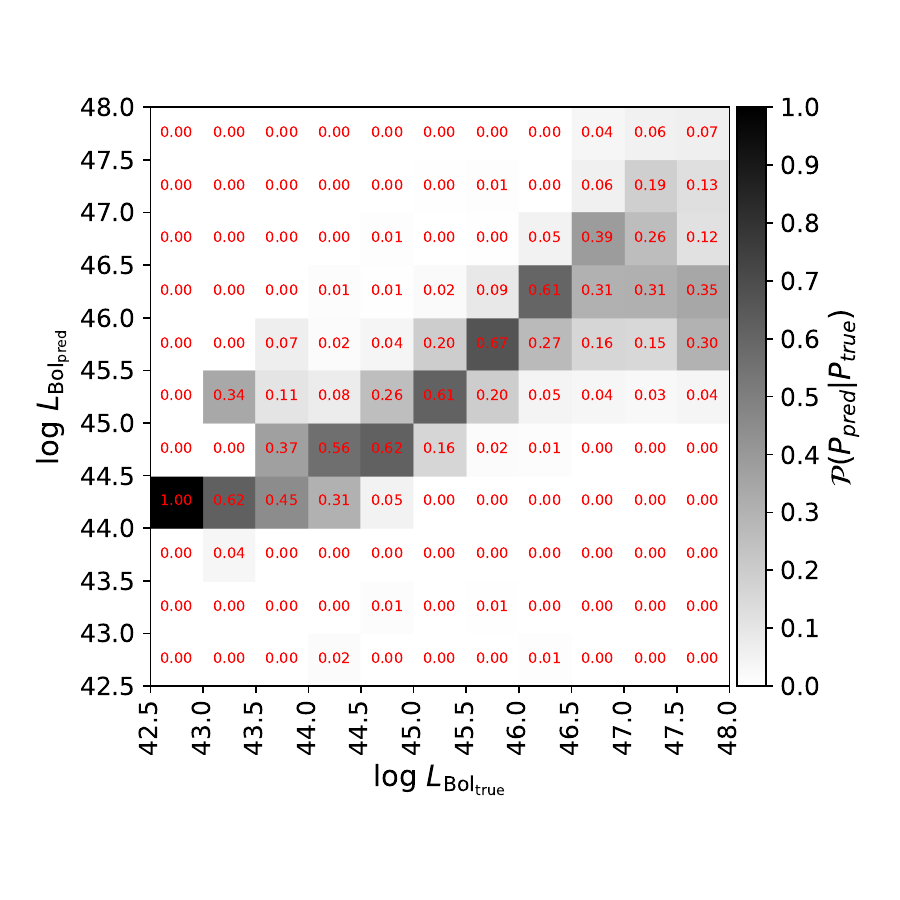}
	\includegraphics[width=0.5\hsize]{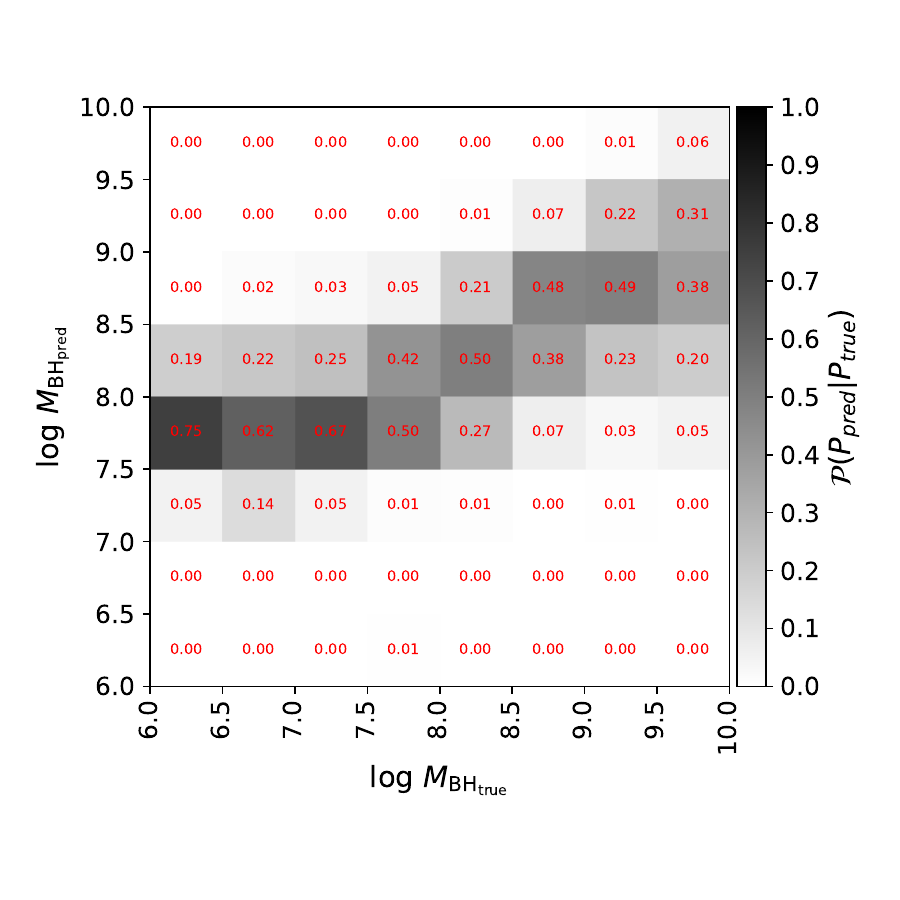}
	\vspace{-10.00mm}
	\includegraphics[width=0.5\hsize]{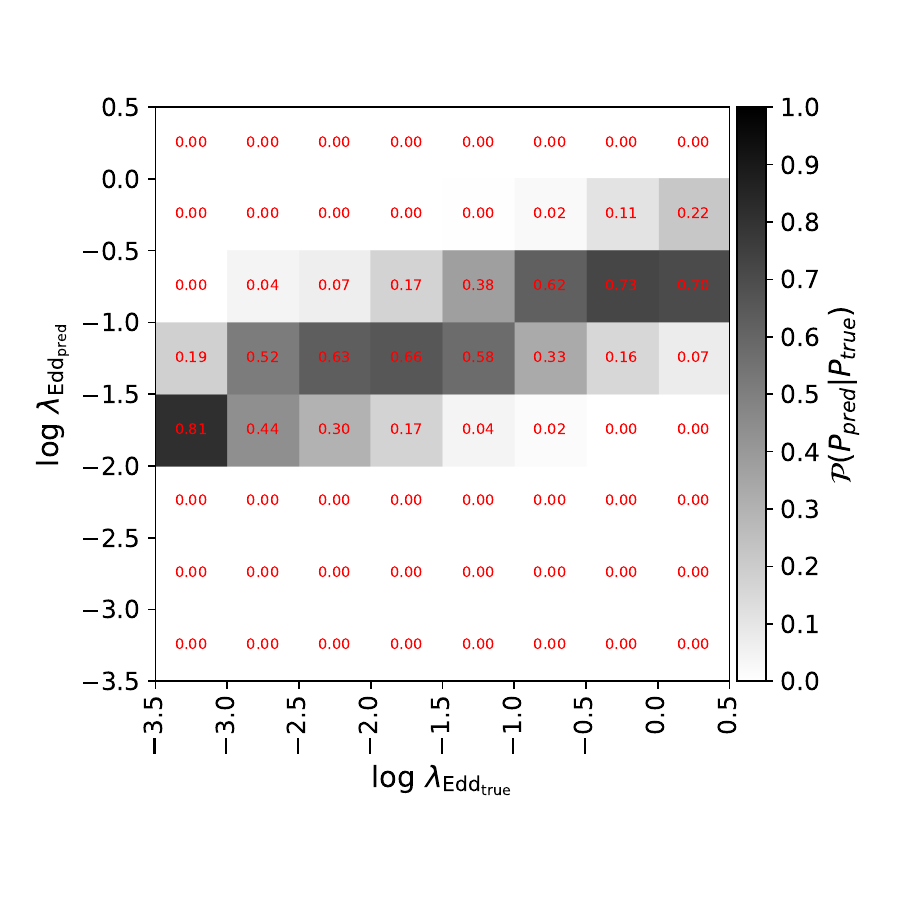}
	\caption{Normalized performance matrices for the ML-estimator without a known redshift as an input ($ML_{\rm wo/z}$). The matrix for the reconstructed $z$ is added to the variables already presented in Fig. \ref{fig:response_matrices_with_z}.}
	\label{fig:response_matrices_no_z}
\end{figure*}

Figures \ref{fig:response_matrices_with_z} and Fig. \ref{fig:response_matrices_no_z} summarizes the performance of the RF for the case with  ($ML_{\rm w/z}$) and without redshift ($ML_{\rm wo/z}$), respectively. Each bin of the response matrix is normalized to the true bins (by column). The matrix elements represent the probability for an AGN with target parameter $P_{\rm \mathrm{true}}$ to be reconstructed with a value $P_{\rm \mathrm{pred}}$. The error on the reconstructed value $\sigma_{\rm pred}$ for each source is used to weight to the histogram.

\subsubsection{Prediction of $z$} \label{z_prediction}
\begin{figure}
\centering
\includegraphics[width=0.35\textwidth]{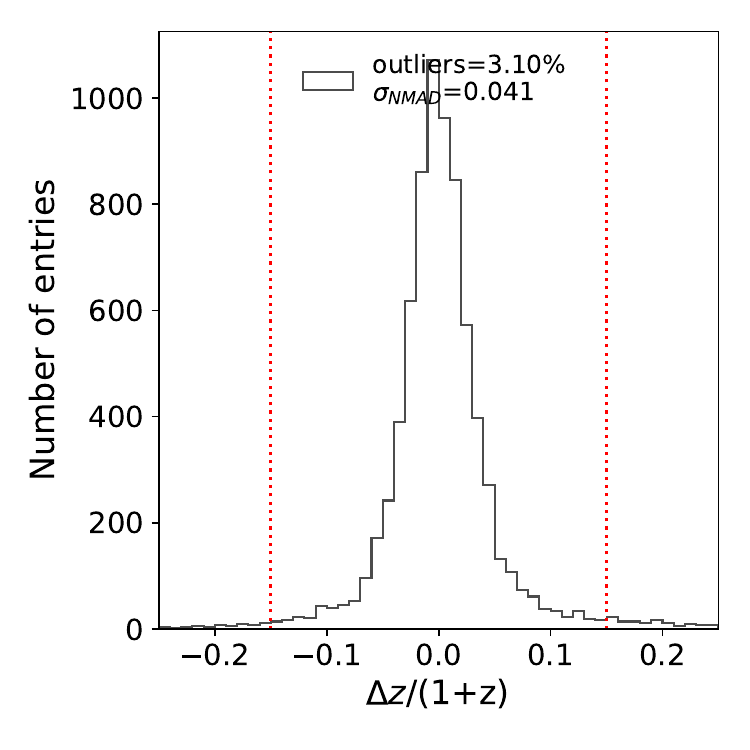}
\caption{Distribution of the predicted redshift, $z_{\rm pred}$, accuracy derived with the true spectroscopically measured redshift, $z_{\rm true}$. The dashed red lines represent the limit beyond which a prediction is counted as an outlier.}
\label{fig:redshift_accuracy}
\end{figure}

\indent Determining redshifts through spectroscopic or photometric means has always been a primary goal of large AGN surveys. The overall performance of the ML-model in predicting the redshift up to z $\sim$ 2.5 can be seen on the top left matrix of Fig. \ref{fig:response_matrices_no_z}. To better evaluate the accuracy of $\mid\Delta z\mid / (1+z_{\rm true})$ ($\Delta z=z_{\rm pred} - z_{\rm true}$), we modified our formula of $\sigma_{\rm NMAD}$ slightly to match the same estimator found in the literature \citep{brammer_eazy_2008,luo_identifications_2010}, such that $\sigma_{\rm NMAD}$=$1.48 \times$ median($\mid \Delta z$- median($\Delta z) \mid / (1 + z_{\rm true})$. An outlier is defined as having  $\mid\Delta z\mid / (1+z_{\rm true}) > 0.15$ (see Fig. \ref{fig:redshift_accuracy}).For redshifts reconstructed with the best parameter RF, we find the rate of outlier to be 1.17\% and $\sigma_{\rm NMAD}$=0.041 (accuracy of 4.1\%), performing just as well as the estimation of photometric redshifts using template-fitting methods \citep{pforr_photometric_2019,salvato_dissecting_2011,luo_identifications_2010,ilbert_accurate_2006, bolzonella_photometric_2000,baum_photoelectric_1962}.

The regressor's reconstruction slightly worsens as we move into higher $z$: this is simply because the SPIDERS dataset provides few sample for the supervised learning to train on, as the distribution of $z$ decreases sharply (see Fig. \ref{fig:redshifts}). The ML reconstruction of redshifts proves to be a very reliable estimator for a crucial parameter for AGN studies, and one that could be adapted to different depths given an appropriate training dataset. 

\subsubsection{$L_{\rm X}$ and $L_{\rm Bol}$}  \label{L_prediction}
Naturally, once the model has an estimation for $z$, it can easily find the appropriate regression for $L_{\rm X}$, given that the X-ray flux is one of the model's inputs. As anticipated, the X-ray luminosity is predicted with high accuracy and precision: when $z$ is known, the error on the estimate is log($L_{\rm X}$/erg s$^{-1}$) $\sim$ 0.037 and rises to $\sim$ 0.193 when $z$ is reconstructed. 
The bolometric luminosity $L_{\rm Bol}$, being the convolution of multiple wavelength observations presents the first moderate challenge for the model to predict: however, it gives reliable reconstructed values, with $\sigma_{\rm err}$ = 0.142 (0.268) for $ML_{\rm w/z}$ ($ML_{\rm wo/z}$).  In general, the performance worsens slightly as we move into the tails of the bin edges and the data in sample we have trained on become scarce: the reconstructed parameters tend to be overestimated for low values of the true parameter, while they are underestimated for high values.

\subsubsection{Prediction of $M_{\rm BH}$ and $\lambda_{\rm Edd}$}  \label{mbh_eddratio_prediciton}

When it comes to estimations of $M_{\rm BH}$, the $R^{2}$ score is 0.62 for $ML_{\rm w/z}$ and 0.48 for $ML_{\rm wo/z}$, and the width of the pull $\sigma_{\rm pull}$ goes from 0.56 to 0.65 in units of log($M_{\rm \odot}$). This uncertainty in the truth reconstruction is slightly higher than the systematic uncertainties in $M_{\rm BH}$ from velocity dispersion measurements \citep{tremaine_slope_2002,hu_black_2008,koss_bass_2022, ricci_bass_2022}.

\begin{figure*}
\centering
\includegraphics[width=1.0\textwidth]{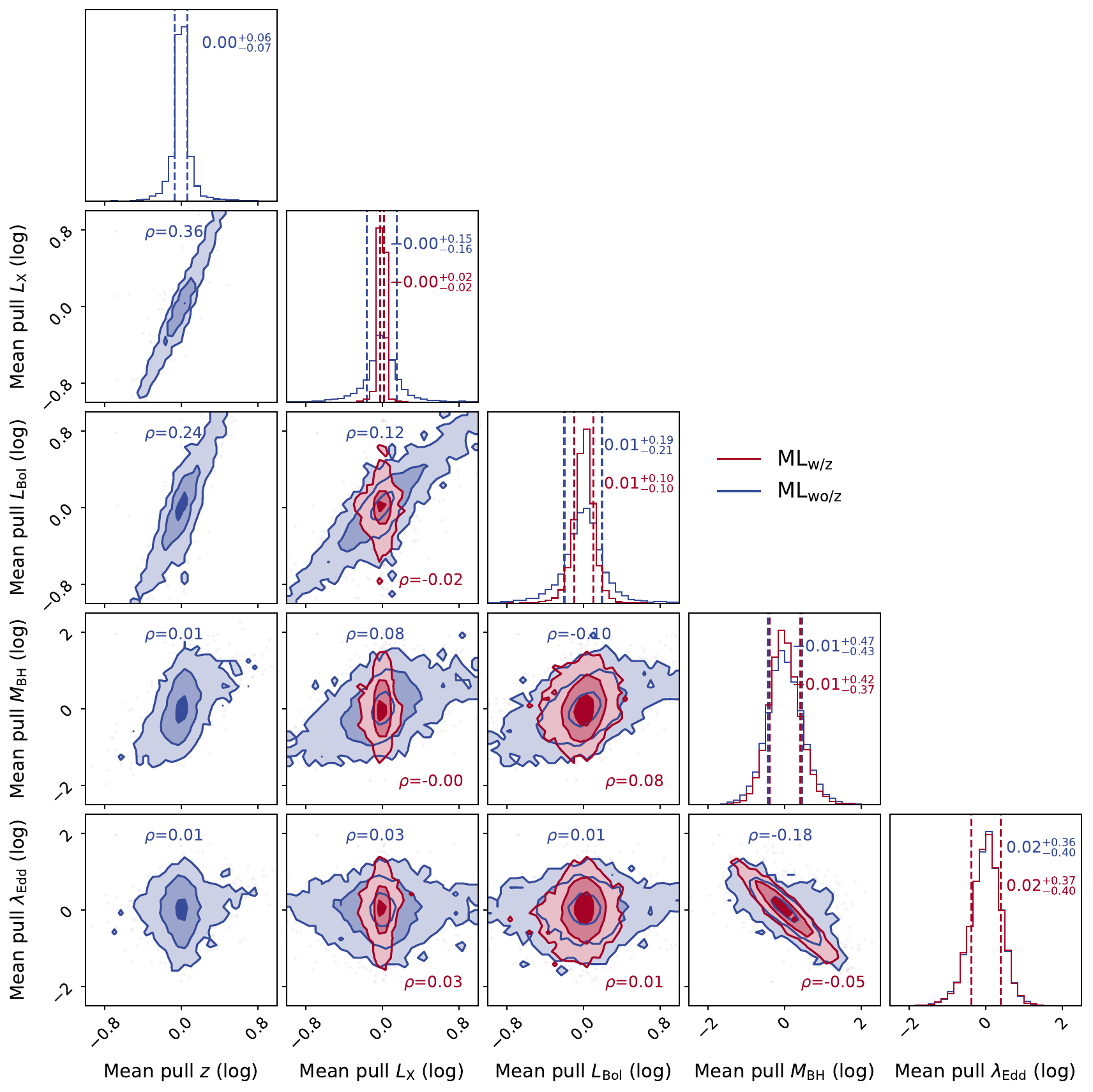}
\caption{Correlation between pull values for all parameters for $ML_{\rm w/z}$ (red) and $ML_{\rm wo/z}$ (blue). The quality of reconstruction, represented by the mean pull value, is more correlated between the variables in $ML_{\rm wo/z}$ than it is for $ML_{\rm w/z}$. The vertical dashed lines in the histograms indicate the 0.16 and 0.84 quantiles of the distributions, and the numbers show the respective medians and 0.16 and 0.84 quantiles.}
\label{fig:corner_plot} 
\end{figure*}

The Eddington ratio to be the hardest parameter to predict, since the errors from the previous predictions get propagated and compounded. However, the RF regressor manages to reconstruct it with a satisfying accuracy: the $R^{2}$ score is 0.35 for both $ML_{\rm w/z}$ for $ML_{\rm wo/z}$, and the width of the pull $\sigma_{\rm pull}$ is about 0.54 in units of log($\lambda_{\rm Edd}$).
It is also the characteristic for which the prior knowledge of $z$ has the least impact, although  the reconstruction of both $L_{\rm Bol}$ and $M_{\rm BH}$ is markedly better in the  $ML_{\rm w/z}$.
To understand why that is, we can study the correlations between the predicted parameters in the form of the mean pull values $\mu_{\rm pull}$. 
Figure \ref{fig:corner_plot} presents the correlations of pulls between all parameters, for $ML_{\rm w/z}$ (red) and $ML_{\rm wo/z}$ (blue). In the case where the redshift, $z,$ is the first predicted parameter in the chain regression ($ML_{\rm wo/z}$), all subsequent parameters remain more or less strongly correlated to one another, as the Pearson's correlation score can attest to. Specifically, if a prior parameter is poorly reconstructed, the subsequent one will be as well. On the other hand, no such correlations between the pulls is found in the case of $ML_{\rm w/z}$, where $L_{\rm X}$ is the first estimated parameter. 

\begin{table*}
\centering
\resizebox{0.75\linewidth}{!}
{\begin{tabular}{|ccccccc|}
\hline
Predicted parameter & Unit & $z$ input? & $R^{2}$ & $\mu_{\rm pull} \pm \sigma_{\rm pull}$ & $\sigma_{\rm NMAD}$  & Contamination   \\   
\hline
$z$ & log &$ML_{\rm wo/z}$ & 0.88 & 0.001 $\pm$ 0.086  & 1.52\% &  0.00\%  \\
\hline
$L_{\rm X}$ & erg.s$^{1}$(log) & $ML_{\rm w/z}$ & 0.99 & -0.003 $\pm$ 0.027  & 0.64\% & 0.0\% \\
\cline{3-6}
& & $ML_{\rm wo/z}$ & 0.85 &   -0.002 $\pm$ 0.193  & 3.83\% & 0.32\%  \\
\hline
$L_{\rm Bol}$ & erg.s$^{1}$(log) & $ML_{\rm w/z}$ & 0.95 & 0.006 $\pm$ 0.142 & 3.76\% &   0.0\% \\
\cline{3-6}
& & $ML_{\rm wo/z}$ & 0.81 &  0.004 $\pm$ 0.268 & 7.04\% &  0.0\%  \\
\hline
$M_{\rm BH}$ & log($M_{\rm \odot}$) & $ML_{\rm w/z}$ & 0.62 & -0.002 $\pm$ 0.563 & 18.8\%  & 1.70\%\\
\cline{3-6}
& & $ML_{\rm wo/z}$ & 0.48 & 0.001 $\pm$ 0.647  & 22.6\%  &  7.53\%  \\
\hline
$\lambda_{\rm Edd}$ & log & $ML_{\rm w/z}$ & 0.35 &  0.008 $\pm$ 0.544 & 18.7\% & 11.13\% \\
\cline{3-6}
& & $ML_{\rm wo/z}$ & 0.35 &  0.017 $\pm$ 0.530 & 21.9\%   & 21.56\%  \\
\hline
\end{tabular}}
\caption {Results of the best-tuned RF estimators, $ML_{\rm w/z}$ and $ML_{\rm wo/z}$, for the performance metrics presented in Sect. \ref{subsubec:model_evaluation} for all target parameters.}
\label{tab:ml_results}
\end{table*}

\section{Reconstruction of the full catalog} \label{sec:results} 

\begin{figure}
\centering
\includegraphics[width=0.35\textwidth]{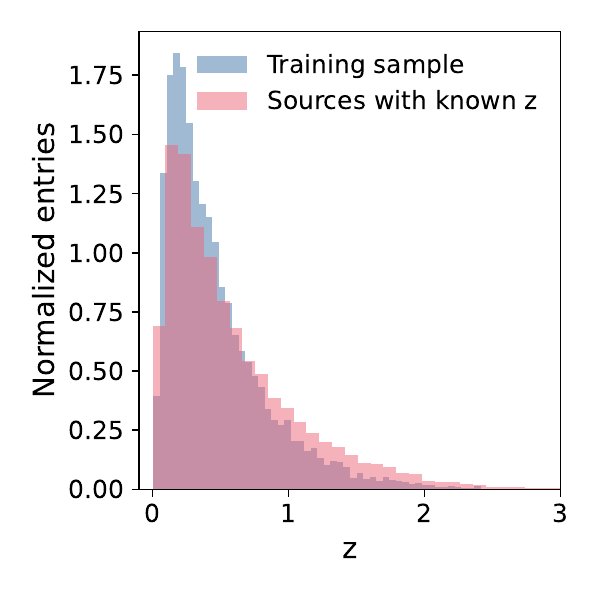}
\caption{Distribution of $z$ for the training sample coming from the SPIDERS AGN catalogue (blue) and the subsample of AGN sources in the reconstruction sample for which $z$ is known (red).}
\label{fig:redshifts}
\end{figure}

In the following section, we take a closer look at the estimated AGN physical parameters for the $\sim$\num{22000} sources without spectroscopic information. As was done for the training sample, all AGN were reconstructed $N$=200 times with the method outlined in Sect. \ref{sec:pseudo-sets}. The mean $\mu_{\rm reco}$ and standard deviation $\sigma_{\rm reco}$ from a Gaussian fit to the posterior probability distribution are recorded for each source \footnote{In the training set, these were called $\mu_{\rm pred}$ and $\sigma_{\rm pred}$, see top panel of Fig. \ref{fig:single_source_lbol_reco_}}. Pointers  are encoded in $\sigma_{\rm reco}$ to the regressor's ability to reconstruct AGNs that are further away from the input range, revealing differences between population type. 
The distribution of $z$ for the \num{9944} sources in the full catalog is overlaid on top of that of the training sample in Fig \ref{fig:redshifts}: the median is $\overline{z}=0.52$ and we can observe that the range of $z$ follows a similar trend, with slightly more AGN sources in $z>1$ range. \newline

\subsection{Reconstruction quality for type 1 and type 2 AGN}  \label{subsec:type1_type2_reco} 

\begin{figure}
\centering
\includegraphics[width=0.35\textwidth]{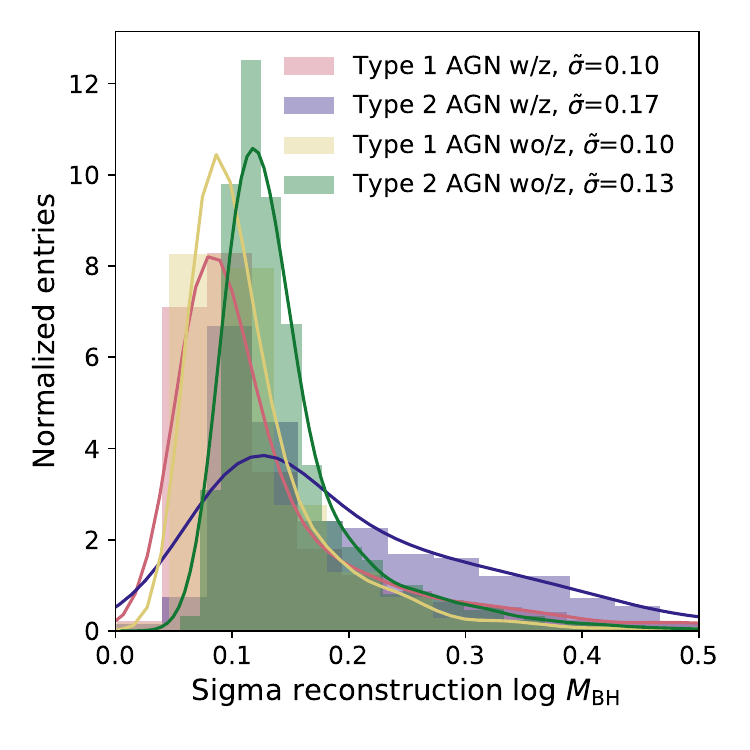}
\caption{Uncertainty $\sigma_{\rm reco}$ on the log of the reconstructed black hole mass $M_{\rm BH}$ for Type 1 and 2 AGN, with and without $z$ information (estimated with $ML_{\rm w/z}$ and $ML_{\rm wo/z}$ respectively). The $\tilde{\sigma}$ values correspond to the median of their distributions. All AGN types can be reconstructed, as proven by their characteristic PDF structures, although the overall reconstructed uncertainties worsen as a function of the AGN class.}
\label{fig:mbh_2types_pred}
\end{figure}

Although the ML model was trained on a sample of type 1 AGNs only, we reconstructed the \num{3062} sources in our catalog identified as type 2 AGNs, using the classifier and criteria presented in Sect. \ref{sec:classification}. Figure \ref{fig:mbh_2types_pred} presents the distributions of the reconstruction uncertainty $\sigma_{\rm reco}$ on the $M_{\rm BH}$ parameter for type 1 and 2 AGNs, with and without known $z$. Sources reconstructed with $ML_{\rm w/z}$ have smaller uncertainties, following the results from the training sample (see Sect. \ref{sec:training}), and the RF has a harder time estimating parameters for type 2 AGNs. The shape of the distribution informs that the regressor is able to reconstruct the $M_{\rm BH}$ type 2 AGNs with both known and unknown $z$ (purple and green distributions).

Considering what was already shown in Fig. \ref{fig:reconstructed_obscuration_distribution}, we know that many AGNs classified as type 2 fall into the faint end of the optical magnitude distribution (e.g., SDSS \textit{u}-band mag > 22) \citep{antonucci_unified_1993m hickox_obscured_2018}. The fainter the source, the more outside of the bounds of \textit{u} magnitude the ML model has trained on, the greater the uncertainty on the reconstructed parameter. This is yet another piece of information gained by propagating the input uncertainties and reconstructing each source iteratively. In particular, the difficulty the regressor encounters when estimating points outside of its known range is translated in the spread of the posterior distribution for all output parameters. 

\subsection{Removal of outliers} \label{outliers}

As a last step, we removed the sources in the final sample for which the reconstructed values lie beyond the phase space of the training sample target parameters. This requirement for $z$ removes \num{297} sources alone.

In addition, we cross-matched our sources with the Fermi 4FGL catalog \citep{ajello_fourth_2020}, to eliminate any AGNs already identified as a blazar. Requiring a wide matching radius of 10 arcseconds, \num{24} sources were removed, bringing the final number of outliers to \num{319}.
In total, \num{21050} AGNs with newly reconstructed physical parameters are added to the catalog, on top of the \num{7613} remaining SPIDERS AGN sources.

\subsection{Type 1 AGN studies}
 
\begin{figure}
\centering
\includegraphics[width=0.35\textwidth]{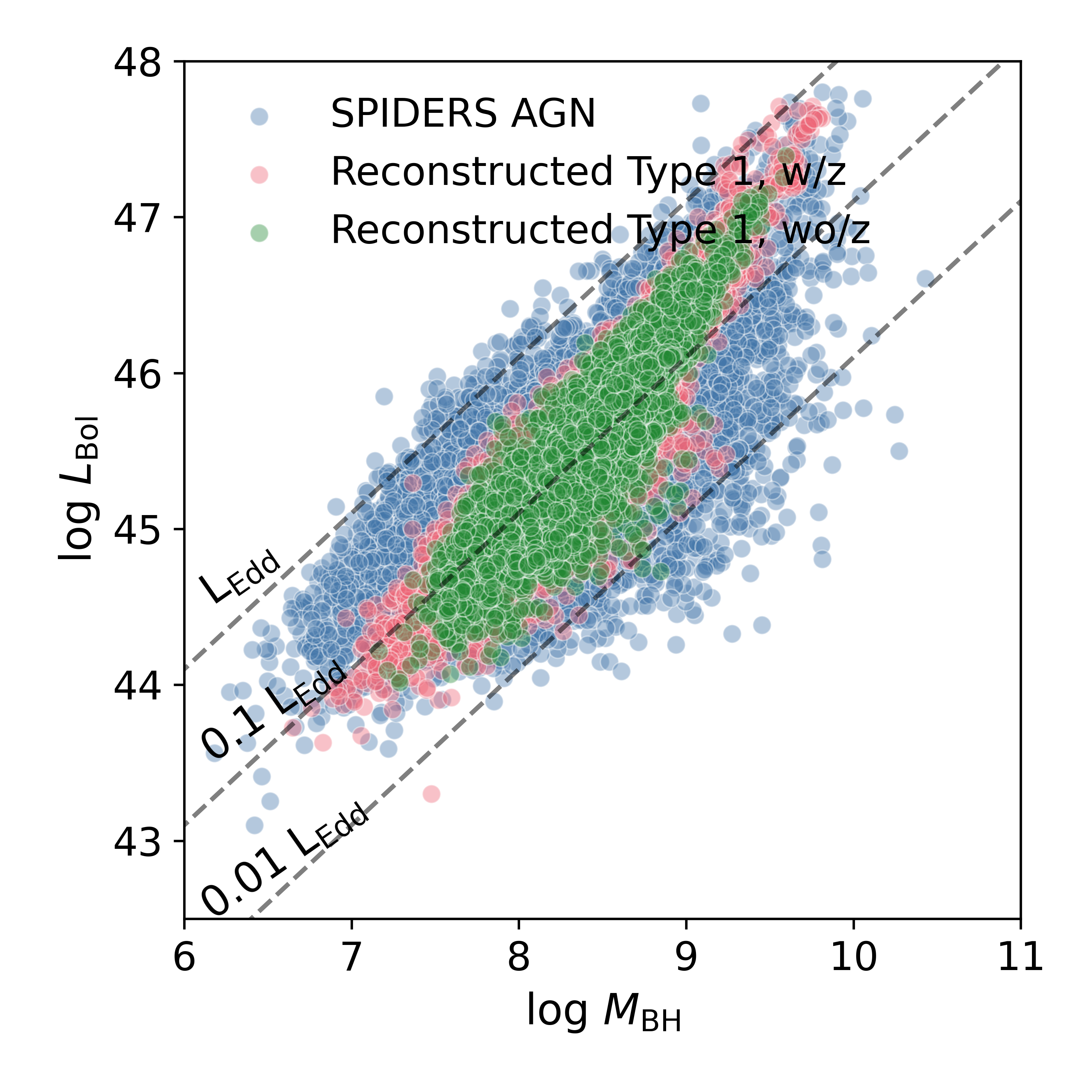}
\includegraphics[width=0.35\textwidth]{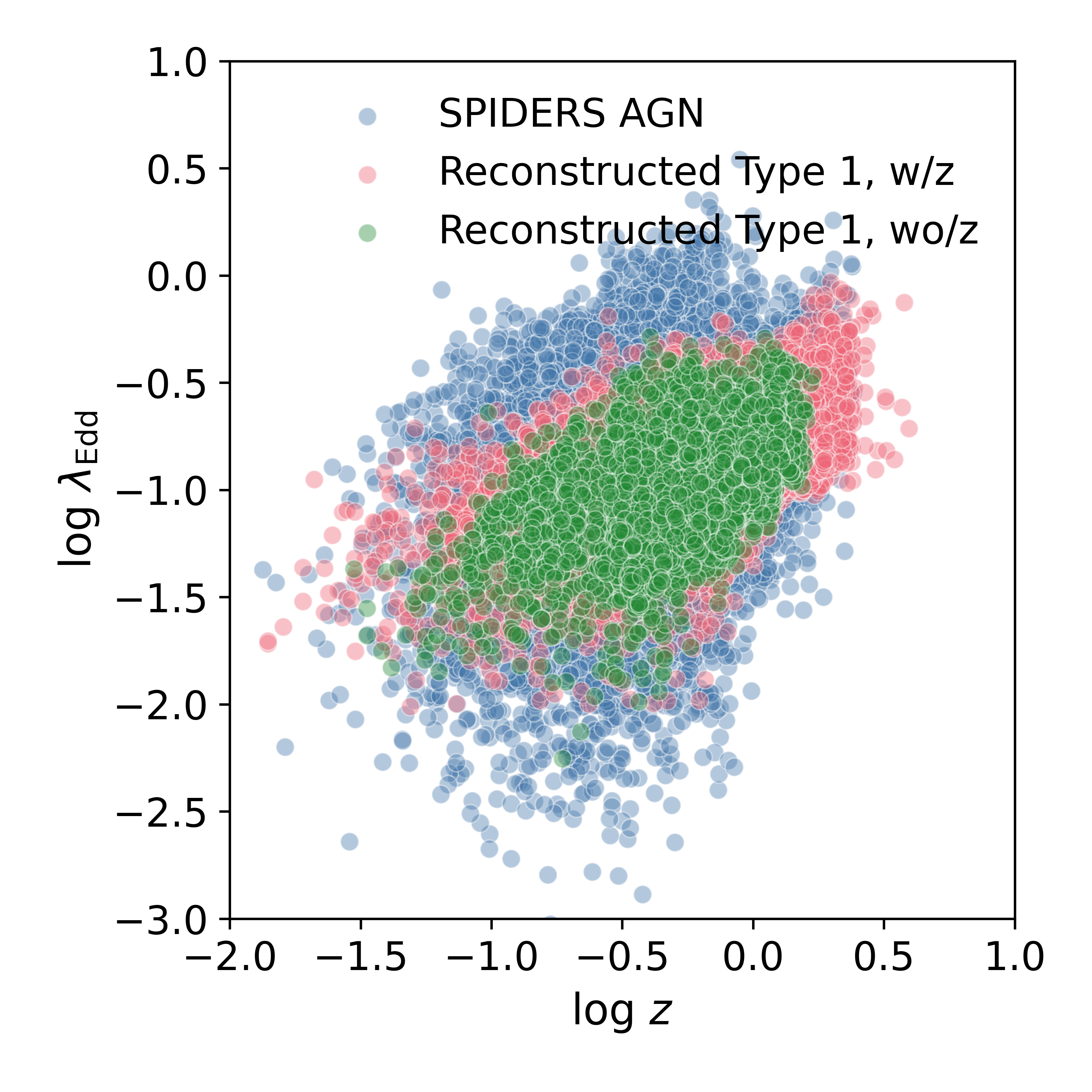}
\caption{(\textit{Top}) Scatter plot of $L_{\rm Bol}$ as a function of $M_{\rm BH}$ for the SPIDERS AGN (blue dots), and reconstructed AGN classified as Type 1. AGN with known $z$ measurement are shown in red dots, and those with reconstructed $z$ are represented in green. (\textit{Bottom}) Same three samples for the $\lambda_{\rm Edd}$ vs $z$ distribution.}
\label{fig:typeI_prediction}
\end{figure}

We now focus on the AGNs classified as type 1, as that population follows the training dataset more closely. Figure \ref{fig:typeI_prediction} presents the bolometric luminosity versus BH mass (top panel) for the type 1 AGNs presented in this work and the SPIDERS AGNs: the reconstructed sample is well bounded by the Eddington limit. The bottom panel of Fig. \ref{fig:typeI_prediction} shows the Eddington ratio as a function of the redshift for the same subsamples. As the response matrices in Figs. \ref{fig:response_matrices_with_z} and \ref{fig:response_matrices_no_z} have shown, the ML model is not very apt with respect to reconstructing the extreme cases in the lower and upper tails of the target parameter distribution. That is to say, it will overestimate low values and underestimate higher ones: this is indeed visible in the bottom panel of Fig. \ref{fig:typeI_prediction}, where the reconstructed samples occupy a smaller region of the log $\lambda_{\rm Edd}$ space than the SPIDERS AGN do.

\begin{figure}
\centering
\includegraphics[width=0.35\textwidth]{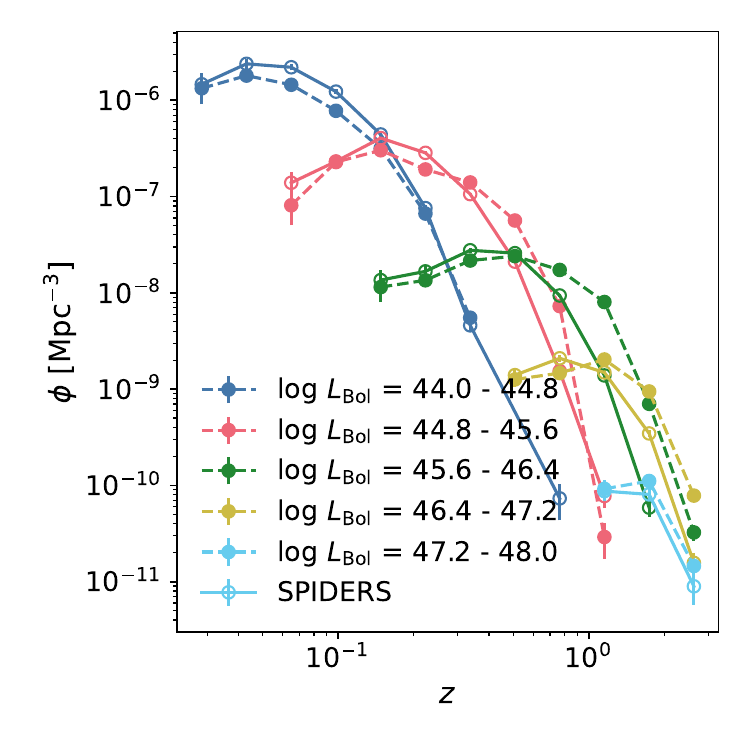}
\includegraphics[width=0.35\textwidth]{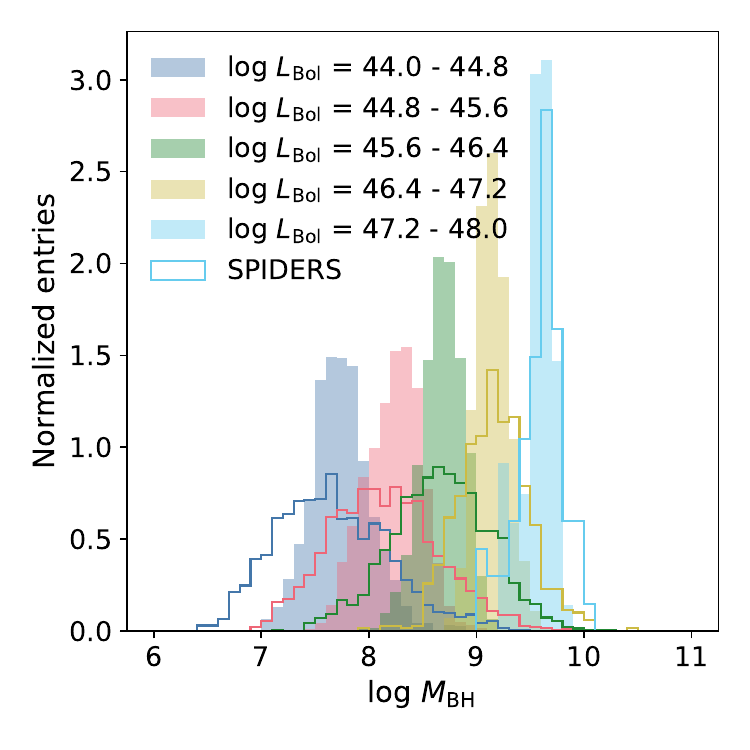}
\caption{(\textit{Top}) AGN downsizing: comoving number density vs. redshift for Type 1 AGN from this work's catalogue (full circles) and the SPIDERS AGN catalogue (open circles) for different bins of $L_{\rm Bol}$  in units of log(erg s$^{-1}$). (\textit{Bottom}) Distribution of $M_{\rm BH}$ for the same bins of bolometric luminosities, for the reconstructed AGN (colored bars) and SPIDERS AGN (colored steps). A flat $\Lambda$CDM cosmology with $H_{0}$ = 70 km s$^{-1}$ Mpc$^{-1}$, $\Omega_{\rm M}$ = 0.3, and $\Omega_{\Lambda}$ = 0.7 is assumed to calculate the comoving number density.}
\label{fig:downsizing}
\end{figure}

The top panel of Fig. \ref{fig:downsizing} presents the AGN number source density over a wide range of redshifts for several bins in bolometric luminosity. Reconstructed type 1 AGNs  are shown in full circles and SPIDERS AGNs are represented in open circles for the same luminosity bins. The same trends are observed in the spectroscopically observed and reconstructed samples: the number density of lower-luminosity AGN peaks later in cosmic time than that of more luminous ones: this effect is known as AGN downsizing \citep{ueda_cosmological_2003,ueda_toward_2014, miyaji_detailed_2015, brandt_cosmic_2015}.
The bottom panel of Fig. \ref{fig:downsizing} shows the black hole masses of these sources, using the same binning in $L_{\rm Bol}$. Although the tails of the distributions are not well represented in the reconstructed sample when matched to their SPIDERS AGN counterparts, it is important to note that not only does the scaling trend of increasing $M_{\rm BH}$ with $L_{\rm Bol}$ remain, but the peaks of the distribution are also coincident between the spectroscopically observed and ML reconstructed type 1 AGN samples. Although deriving a luminosity function from these AGNs would require the non-trivial correction of the number density for detection and selection efficiencies and biases \citep{schulze_cosmic_2015,weigel_agns_2017,ananna_bass_2022}, the two panels of Fig. \ref{fig:downsizing} show the accurate description of AGN behavior in a multi-dimensional phase space ($z$, $L_{\rm Bol}$, and $M_{\rm BH}$) when compared with the spectroscopically measured SPIDERS AGN sample.

\subsection{Validation with BASS DR2 catalog}
\begin{figure*}
\centering
\includegraphics[width=0.45\textwidth]{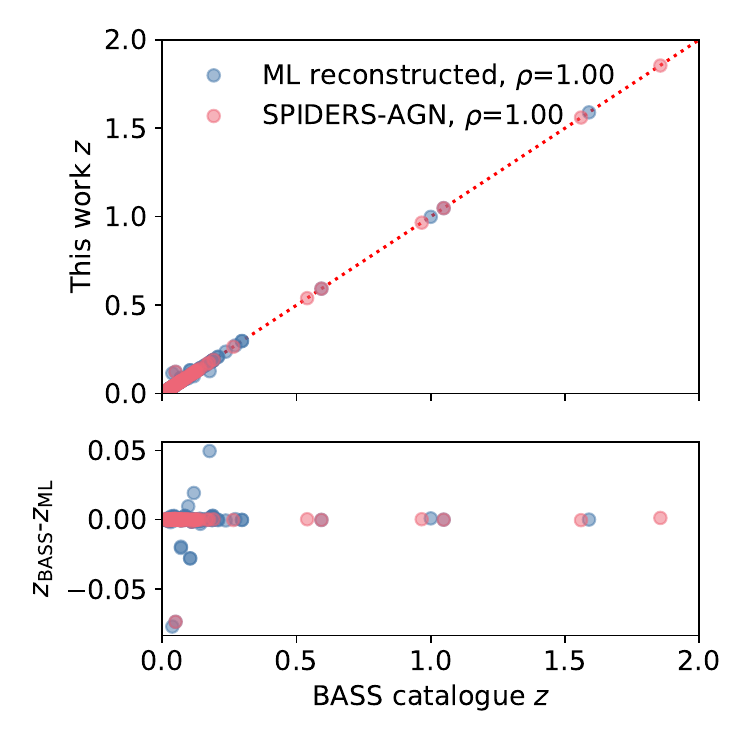}
\includegraphics[width=0.45\textwidth]{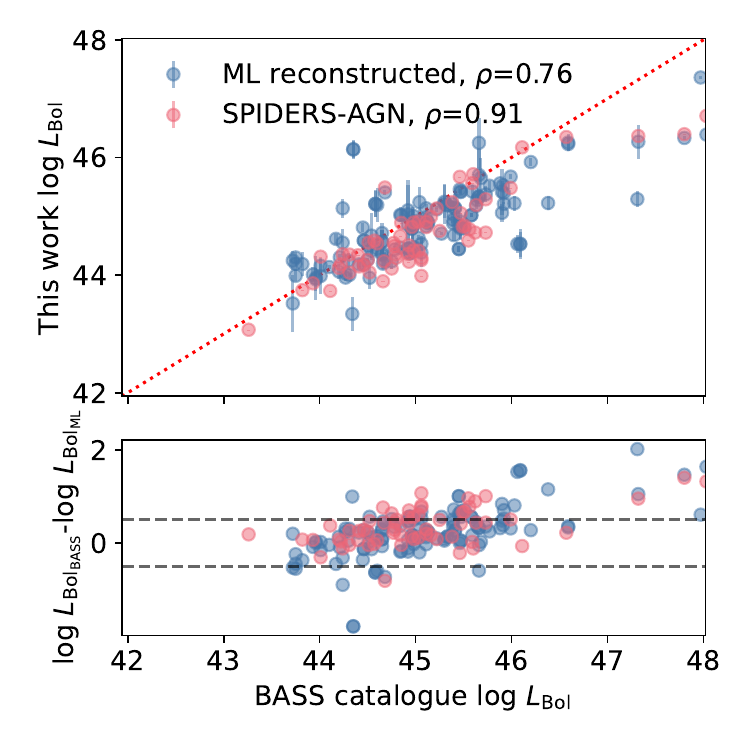}
\includegraphics[width=0.45\textwidth]{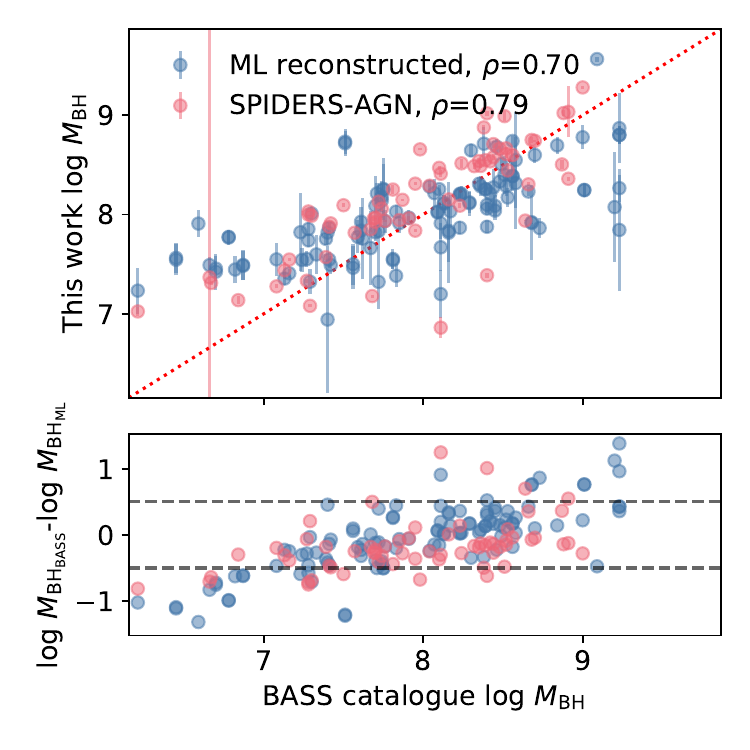}
\includegraphics[width=0.45\textwidth]{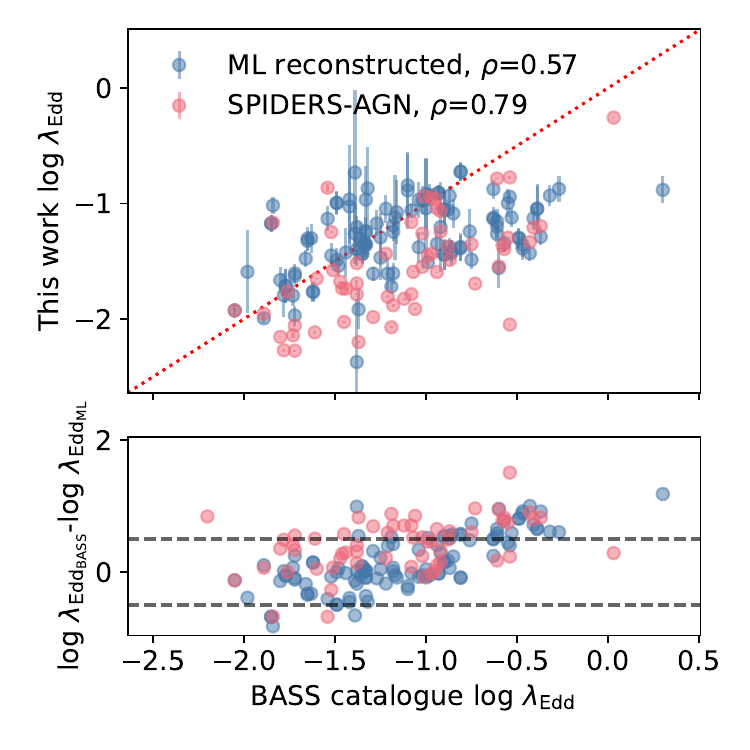}
\caption{Comparison of AGN physical paramaters from the BASS, SPIDERS-AGN and ML-reconstructed catalogues for \num{196} cross-matched sources: the redshift $z$ (\textit{top left}), $L_{\rm Bol}$ (\textit{top right}), $M_{\rm BH}$ (\textit{bottom left}) and $\lambda_{\rm Edd}$ (\textit{bottom right}). The bottom panel of each subfigure represents the difference of the BASS survey's and this work's values. The horizontal grey dashed line show the $\pm$ 0.5 dex level. }
\label{fig:BASS_study}
\end{figure*}

The BAT AGN spectroscopic survey (BASS) is a hard X-ray spectroscopic survey with the SWIFT instrument that characterizes the physical parameters ($z$, $L_{\rm Bol}$, $M_{\rm BH}$ and $\lambda_{\rm Edd}$) of $\sim$ 1000 AGNs selected in the hard (14-195 keV) X-ray \citep{koss_bass_2022}. The second data release (DR2) offers itself as an ideal independent validation set for our ML study. 
We cross-matched our catalog with the BASS DR2, demanding a strict 1 arcsec maximum radius, and found \num{196} counterparts; of these, \num{64} come from the SPIDERS AGN training sample \citep{coffey_sdss-ivspiders_2019} and \num{115} were ML-reconstructed. 
We compared the key AGN parameters from this work and the BASS catalog in Fig. \ref{fig:BASS_study}. The red points represent the SPIDERS-AGN values, allowing for a direct comparison from two optical spectroscopy measurements, while the blue points represent ML-reconstructed values. We observe that for $z,$ the three datasets are in excellent agreement, while 67\% (65\%) of ML-reconstructed (SPIDERS-AGN) $L_{\rm Bol}$ are within 0.5 dex of their BASS counterpart measurement.

For black hole masses, while the correlation scores are lower overall, the reconstructed values are well validated by the BASS survey values: 64\% (77\%) of ML-reconstructed (SPIDERS-AGN) $M_{\rm BH}$ values are within 0.5 dex of their BASS counterpart. The Eddington ratio, however, shows the greatest level of discrepancy, with a visible systematic shift between spectroscopic measurement. While they are closer to one another than ML-reconstructed $\lambda_{\rm Edd}$ values,  the differences might be explained by the two instrument's observations of changing-look AGN during different phases of accretion activity, with the emergence or disappearance of broad emission lines \citep{ulrich_variability_1997,soldi_long-term_2014, mereghetti_time_2021,ricci_changing-look_2022}.

\section{Summary and conclusions} \label{sec:summary}

We have released the first photometry based estimates of bolometric luminosity $L_{\rm Bol}$, black hole mass, $M_{\rm BH}$, and Eddington ratio, $\lambda_{\rm Edd}$, for \num{21050} sources ranging over 6 dex in luminosity and up to $z$=2.4 in redshift. For \num{11363} of these sources, the redshift was previously determined spectroscopically and was used in the estimation of the remaining parameters, as well as for the verification of the redshift estimate. For \num{11363} sources without a previously known redshift, the reconstructed $z$ is provided, with excellent accuracy. An uncertainty is given for all estimated parameters, thanks to a simulation based technique which incorporates measurement errors in the fit and reconstruction of the ML regressor. In addition, we have demonstrated how ML classification tools can help identify obscured AGN, a crucial challenge in the field (see review by \cite{hickox_obscured_2018}. Finally, we used the existing BASS survey catalog \citep{koss_bass_2022} to benchmark this work's ML-reconstructed values with an independent spectroscopic measurement of AGN physical parameters. \\

While the addition of $\sim$\num{21000} AGN sources from this catalog might not dramatically improve our knowledge of the luminosity function, considering that the \num{8000} SPIDERS AGN sources were measured with greater accuracy in the same phase space, the release of this new dataset is of particular use for multimessenger astronomy studies, where one needs to know these physical parameters for a large sample of sources, while maximizing the sky coverage.
AGN have been favored to be strong cosmic ray emitters \citep{halzen_neutrino_1997, murase_neutrinos_2022,murase_high-energy_2022}, with the recent discovery showing the nearby obscured AGN NGC 1068 to be a steady source of neutrinos \citep{icecube_collaboration_evidence_2022}. Searches for a cumulative signal from different AGN populations, such as \cite{abbasi_search_2022}, can help characterize which sources contribute most to the flux of neutrinos, based on their accretion parameters.
By strategically targeting a subset of sources observed spectroscopically, we would be able to train similar ML-algorithms and reconstruct a larger sample of photometrically measured AGNs. The method can easily be expanded to other cosmic demographics (e.g., higher $z$) granted a corresponding dataset is provided to train a ML algorithm. 
In this work, we were limited by demanding that sources had been observed with SDSS photometry: this constrained the coverage to a quarter of the full sky. A natural next step would be to expand the optical sky coverage by cross-matching sources with the Pan-STARRS 3$\pi$ survey \cite{flewelling_pan-starrs1_2020}, and recover most AGNs identified with IR and X-ray telescopes. Finally, eROSITA has been scanning the full-sky with unprecedented sensitivity in the soft (0.2–2.3 keV) and hard (2.3–8 keV) bands \citep{merloni_erosita_2012,predehl_erosita_2021}, having recently published its first data release \citep{merloni_srgerosita_2024}. Incorporating this dataset will also offer new understanding of obscured AGNs, as harder X-ray photons are transparent to obscuring dust \citep{salvato_erosita_2022,waddell_erosita_2023, waddell_srgerosita_2024}.

%Large new datasets are poised to be made available with the launch of great new missions: eROSITA has been scanning full-sky in the soft (0.2–2.3 keV) and hard (2.3–8 keV) X-ray sky, a welcome addition, as harder X-ray photons are transparent to obscuring dust in Type 2 AGN. The number density of eROSITA targets, in their majority Type 1 AGN, will peak at $z \sim$ 1 \citep{merloni_erosita_2012,predehl_erosita_2021}, an ideal range for this model trained on the SPIDERS dataset \citep{coffey_sdss-ivspiders_2019}. The MIR band is necessary to identify AGN from the large backdrop of X-ray bright sources: only the W1 and W2 bands are still functional on NEOWISE \citep{mainzer_preliminary_2011}, the repurposed mission of WISE \citep{wright_wide-field_2010}. Thankfully, the Mid-Infrared Instrument (MIRI) \citep{rieke_mid-infrared_2015} on board of the James Webb Space Telescope (\textit{JSWT}) covers a range from 5-28.5 $\micron$, allowing for color-color diagnostics and identify dust-obscured AGN at $z$ = 1-3 \citep{kirkpatrick_agnstar_2017}. The future of optical survey is also secure with the imminent first light of the Vera Rubin Observatory: it is expected to observe 20 billion galaxies over ten years, with six filters (\textit{ugrizy}) covering the wavelength range 320–1050 nm \citep{ivezic_lsst_2019}. \\

\begin{acknowledgements} 
The author wishes to thank Iftach Sadeh, Anna Franckowiak, Sjoert Van Velzen, Jannis Necker, Simone Garrapa and Jonas Sinapius for their fruitful comments and support. Timo Karg is to be thanked for helping with various technicalities. 
\end{acknowledgements}

% style aa.bst
\bibliographystyle{aa} 
\bibpunct{(}{)}{;}{a}{}{,} % to follow the A&A style
\bibliography{AGN_Paper}

\begin{thebibliography}{116}
\expandafter\ifx\csname natexlab\endcsname\relax\def\natexlab#1{#1}\fi

\bibitem[{Aartsen {et~al.}(2017)Aartsen, Abraham, Ackermann, Adams, Aguilar,
  Ahlers, Ahrens, Altmann, Andeen, Anderson, Ansseau, Anton, Archinger,
  Arguelles, Arlen, Auffenberg, Axani, Bai, Barwick, Baum, Bay, Beatty, Tjus,
  Becker, BenZvi, Berghaus, Berley, Bernardini, Bernhard, Besson, Binder,
  Bindig, Bissok, Blaufuss, Blot, Boersma, Bohm, Börner, Bos, Bose, Böser,
  Botner, Braun, Brayeur, Bretz, Burgman, Casey, Casier, Cheung, Chirkin,
  Christov, Clark, Classen, Coenders, Collin, Conrad, Cowen, Silva, Daughhetee,
  Davis, Day, André, Clercq, Rosendo, Dembinski, Ridder, Desiati, Vries,
  Wasseige, With, DeYoung, Díaz-Vélez, Lorenzo, Dujmovic, Dumm, Dunkman,
  Eberhardt, Ehrhardt, Eichmann, Euler, Evenson, Fahey, Fazely, Feintzeig,
  Felde, Filimonov, Finley, Flis, Fösig, Franckowiak, Fuchs, Gaisser, Gaior,
  Gallagher, Gerhardt, Ghorbani, Giang, Gladstone, Glagla, Glüsenkamp,
  Goldschmidt, Golup, Gonzalez, Góra, Grant, Griffith, Haack, Ismail,
  Hallgren, Halzen, Hansen, Hansmann, Hansmann, Hanson, Hebecker, Heereman,
  Helbing, Hellauer, Hickford, Hignight, Hill, Hoffman, Hoffmann, Holzapfel,
  Homeier, Hoshina, Huang, Huber, Huelsnitz, Hultqvist, In, Ishihara, Jacobi,
  Japaridze, Jeong, Jero, Jones, Jurkovic, Kappes, Karg, Karle, Katz, Kauer,
  Keivani, Kelley, Kemp, Kheirandish, Kim, Kintscher, Kiryluk, Kittler, Klein,
  Kohnen, Koirala, Kolanoski, Konietz, Köpke, Kopper, Kopper, Koskinen,
  Kowalski, Krings, Kroll, Krückl, Krüger, Kunnen, Kunwar, Kurahashi,
  Kuwabara, Labare, Lanfranchi, Larson, Lennarz, Lesiak-Bzdak, Leuermann,
  Leuner, Lu, Lünemann, Madsen, Maggi, Mahn, Mancina, Mandelartz, Maruyama,
  Mase, Maunu, McNally, Meagher, Medici, Meier, Meli, Menne, Merino, Meures,
  Miarecki, Middell, Mohrmann, Montaruli, Moulai, Nahnhauer, Naumann, Neer,
  Niederhausen, Nowicki, Nygren, Pollmann, Olivas, Omairat, O'Murchadha,
  Palczewski, Pandya, Pankova, Penek, Pepper, Heros, Pfendner, Pieloth, Pinat,
  Posselt, Price, Przybylski, Quinnan, Raab, Rädel, Rameez, Rawlins, Reimann,
  Relich, Resconi, Rhode, Richman, Riedel, Robertson, Rongen, Rott, Ruhe,
  Ryckbosch, Rysewyk, Sabbatini, Herrera, Sandrock, Sandroos, Sarkar,
  Satalecka, Schimp, Schlunder, Schmidt, Schoenen, Schöneberg, Schönwald,
  Schumacher, Seckel, Seunarine, Soldin, Song, Spiczak, Spiering, Stahlberg,
  Stamatikos, Stanev, Stasik, Steuer, Stezelberger, Stokstad, Stößl, Ström,
  Strotjohann, Sullivan, Sutherland, Taavola, Taboada, Tatar, Ter-Antonyan,
  Terliuk, Tešić, Tilav, Toale, Tobin, Toscano, Tosi, Tselengidou, Turcati,
  Unger, Usner, Vallecorsa, Vandenbroucke, Eijndhoven, Vanheule, Rossem,
  Santen, Veenkamp, Vehring, Voge, Vraeghe, Walck, Wallace, Wallraff,
  Wandkowsky, Weaver, Wendt, Westerhoff, Whelan, Wickmann, Wiebe, Wiebusch,
  Wille, Williams, Wills, Wissing, Wolf, Wood, Woolsey, Woschnagg, Xu, Xu, Xu,
  Yanez, Yodh, Yoshida, Zoll, \& Collaboration)}]{aartsen_contribution_2017}
Aartsen, M.~G., Abraham, K., Ackermann, M., {et~al.} 2017, The Astrophysical
  Journal, 835, 45, publisher: The American Astronomical Society

\bibitem[{Aartsen {et~al.}(2020)Aartsen, Ackermann, Adams, Aguilar, Ahlers,
  Ahrens, Alispach, Andeen, Anderson, Ansseau, Anton, Argüelles, Auffenberg,
  Axani, Bagherpour, Bai, V, Barbano, Barwick, Bastian, Baum, Baur, Bay,
  Beatty, Becker, Tjus, BenZvi, Berley, Bernardini, Besson, Binder, Bindig,
  Blaufuss, Blot, Bohm, Böser, Botner, Böttcher, Bourbeau, Bourbeau,
  Bradascio, Braun, Bron, Brostean-Kaiser, Burgman, Buscher, Busse, Carver,
  Chen, Cheung, Chirkin, Choi, Clark, Clark, Classen, Coleman, Collin, Conrad,
  Coppin, Correa, Cowen, Cross, Dave, Clercq, DeLaunay, Dembinski, Deoskar,
  Ridder, Desiati, Vries, Wasseige, With, DeYoung, Diaz, Díaz-Vélez,
  Dujmovic, Dunkman, Dvorak, Eberhardt, Ehrhardt, Eller, Engel, Evenson, Fahey,
  Fazely, Felde, Filimonov, Finley, Fox, Franckowiak, Friedman, Fritz, Gaisser,
  Gallagher, Ganster, Garrappa, Gerhardt, Ghorbani, Glauch, Glüsenkamp,
  Goldschmidt, Gonzalez, Grant, Grégoire, Griffith, Griswold, Günder,
  Gündüz, Haack, Hallgren, Halliday, Halve, Halzen, Hanson, Haungs, Hebecker,
  Heereman, Heix, Helbing, Hellauer, Henningsen, Hickford, Hignight, Hill,
  Hoffman, Hoffmann, Hoinka, Hokanson-Fasig, Hoshina, Huang, Huber, Huber,
  Hultqvist, Hünnefeld, Hussain, In, Iovine, Ishihara, Jansson, Japaridze,
  Jeong, Jero, Jones, Jonske, Joppe, Kang, Kang, Kappes, Kappesser, Karg, Karl,
  Karle, Katz, Kauer, Kellermann, Kelley, Kheirandish, Kim, Kintscher, Kiryluk,
  Kittler, Klein, Koirala, Kolanoski, Köpke, Kopper, Kopper, Koskinen,
  Kowalski, Krings, Krückl, Kulacz, Kurahashi, Kyriacou, Lanfranchi, Larson,
  Lauber, Lazar, Leonard, Leszczyńska, Liu, Lohfink, Mariscal, Lu, Lucarelli,
  Ludwig, Lünemann, Luszczak, Lyu, Ma, Madsen, Maggi, Mahn, Makino, Mallik,
  Mallot, Mancina, Mariş, Maruyama, Mase, Maunu, McNally, Meagher, Medici,
  Medina, Meier, Meighen-Berger, Merino, Meures, Micallef, Mockler, Momenté,
  Montaruli, Moore, Morse, Moulai, Muth, Nagai, Naumann, Neer, Nguyên,
  Niederhausen, Nisa, Nowicki, Nygren, Pollmann, Oehler, Olivas, O'Murchadha,
  O'Sullivan, Palczewski, Pandya, Pankova, Park, Peiffer, Heros, Philippen,
  Pieloth, Pieper, Pinat, Pizzuto, Plum, Porcelli, Price, Przybylski, Raab,
  Raissi, Rameez, Rauch, Rawlins, Rea, Rehman, Reimann, Relethford, Renschler,
  Renzi, Resconi, Rhode, Richman, Robertson, Rongen, Rott, Ruhe, Ryckbosch,
  Cantu, Safa, Herrera, Sandrock, Sandroos, Santander, Sarkar, Sarkar,
  Satalecka, Schaufel, Schieler, Schlunder, Schmidt, Schneider, Schneider,
  Schröder, Schumacher, Sclafani, Seckel, Seunarine, Shefali, Silva, Snihur,
  Soedingrekso, Soldin, Song, Spiczak, Spiering, Stachurska, Stamatikos,
  Stanev, Stein, Stettner, Steuer, Stezelberger, Stokstad, Stößl,
  Strotjohann, Stürwald, Stuttard, Sullivan, Taboada, Tenholt, Ter-Antonyan,
  Terliuk, Tilav, Tollefson, Tomankova, Tönnis, Toscano, Tosi, Trettin,
  Tselengidou, Tung, Turcati, Turcotte, Turley, Ty, Unger, Elorrieta, Usner,
  Vandenbroucke, Driessche, Eijk, Eijndhoven, Santen, Verpoest, Vraeghe, Walck,
  Wallace, Wallraff, Wandkowsky, Watson, Weaver, Weindl, Weiss, Weldert, Wendt,
  Werthebach, Whelan, Whitehorn, Wiebe, Wiebusch, Wille, Williams, Wills, Wolf,
  Wood, Wood, Woschnagg, Wrede, Xu, Xu, Xu, Yanez, Yodh, Yoshida, Yuan,
  Zöcklein, \& Collaboration}]{aartsen_icecube_2020}
Aartsen, M.~G., Ackermann, M., Adams, J., {et~al.} 2020, The Astrophysical
  Journal, 898, 117, publisher: The American Astronomical Society

\bibitem[{Abbasi {et~al.}(2023)Abbasi, Ackermann, Adams, Agarwalla, Aguilar,
  Ahlers, Alameddine, Amin, Andeen, Anton, Argüelles, Ashida, Athanasiadou,
  Axani, Bai, V, Baricevic, Barwick, Basu, Bay, Beatty, Becker, Tjus, Beise,
  Bellenghi, BenZvi, Berley, Bernardini, Besson, Binder, Bindig, Blaufuss,
  Blot, Bontempo, Book, Meneguolo, Böser, Botner, Böttcher, Bourbeau, Braun,
  Brinson, Brostean-Kaiser, Burley, Busse, Butterfield, Campana, Carloni,
  Carnie-Bronca, Chattopadhyay, Chen, Chen, Chirkin, Choi, Clark, Classen,
  Coleman, Collin, Connolly, Conrad, Coppin, Correa, Countryman, Cowen, Dave,
  Clercq, DeLaunay, López, Dembinski, Deoskar, Desai, Desiati, Vries,
  Wasseige, DeYoung, Diaz, Díaz-Vélez, Dittmer, Domi, Dujmovic, DuVernois,
  Ehrhardt, Eller, Engel, Erpenbeck, Evans, Evenson, Fan, Fang, Fazely,
  Fedynitch, Feigl, Fiedlschuster, Finley, Fischer, Fox, Franckowiak, Friedman,
  Fritz, Fürst, Gaisser, Gallagher, Ganster, Garcia, Garrappa, Gerhardt,
  Ghadimi, Glaser, Glauch, Glüsenkamp, Goehlke, Gonzalez, Goswami, Grant,
  Gray, Griffin, Griswold, Günther, Gutjahr, Haack, Hallgren, Halliday, Halve,
  Halzen, Hamdaoui, Minh, Hanson, Hardin, Harnisch, Hatch, Haungs, Helbing,
  Hellrung, Henningsen, Heuermann, Hickford, Hidvegi, Hill, Hill, Hoffman,
  Hoshina, Hou, Huber, Hultqvist, Hünnefeld, Hussain, Hymon, In, Iovine,
  Ishihara, Jacquart, Jansson, Japaridze, Jayakumar, Jeong, Jin, Jones, Kang,
  Kang, Kang, Kappes, Kappesser, Kardum, Karg, Karl, Karle, Katz, Kauer,
  Kelley, Zathul, Kheirandish, Kin, Kiryluk, Klein, Kochocki, Koirala,
  Kolanoski, Kontrimas, Köpke, Kopper, Koskinen, Koundal, Kovacevich,
  Kowalski, Kozynets, Kruiswijk, Krupczak, Kumar, Kun, Kurahashi, Lad, Gualda,
  Lamoureux, Larson, Lauber, Lazar, Lee, DeHolton, Leszczyńska, Lincetto, Liu,
  Liubarska, Lohfink, Love, Mariscal, Lu, Lucarelli, Ludwig, Luszczak, Lyu, Ma,
  Madsen, Mahn, Makino, Mancina, Sainte, Mariş, Marka, Marka, Marsee,
  Martinez-Soler, Maruyama, Mayhew, McElroy, McNally, Mead, Meagher, Mechbal,
  Medina, Meier, Meighen-Berger, Merckx, Merten, Micallef, Mockler, Montaruli,
  Moore, Morii, Morse, Moulai, Mukherjee, Naab, Nagai, Nakos, Naumann, Necker,
  Neumann, Niederhausen, Nisa, Noell, Nowicki, Pollmann, O'Dell, Oehler,
  Oeyen, Olivas, Orsoe, Osborn, O'Sullivan, Pandya, Park, Parker, Paudel,
  Paul, Heros, Peterson, Philippen, Pieper, Pizzuto, Plum, Popovych, Rodriguez,
  Pries, Procter-Murphy, Przybylski, Raab, Rack-Helleis, Rawlins, Rechav,
  Rehman, Reichherzer, Renzi, Resconi, Reusch, Rhode, Richman, Riedel, Roberts,
  Robertson, Rodan, Roellinghoff, Rongen, Rott, Ruhe, Ruohan, Ryckbosch, Safa,
  Saffer, Salazar-Gallegos, Sampathkumar, Herrera, Sandrock, Santander, Sarkar,
  Sarkar, Savelberg, Savina, Schaufel, Schieler, Schindler, Schlüter, Schmidt,
  Schneider, Schröder, Schumacher, Schwefer, Sclafani, Seckel, Seunarine,
  Sharma, Shefali, Shimizu, Silva, Skrzypek, Smithers, Snihur, Soedingrekso,
  Søgaard, Soldin, Sommani, Spannfellner, Spiczak, Spiering, Stamatikos,
  Stanev, Stasik, Stein, Stezelberger, Stürwald, Stuttard, Sullivan, Taboada,
  Ter-Antonyan, Thompson, Thwaites, Tilav, Tollefson, Tönnis, Toscano, Tosi,
  Trettin, Tung, Turcotte, Twagirayezu, Ty, Elorrieta, Upadhyay, Upshaw,
  Valtonen-Mattila, Vandenbroucke, Eijndhoven, Vannerom, Santen, Vara,
  Veitch-Michaelis, Venugopal, Verpoest, Veske, Walck, Watson, Weaver, Weigel,
  Weindl, Weldert, Wendt, Werthebach, Weyrauch, Whitehorn, Wiebusch, Willey,
  Williams, Wolf, Wrede, Wulff, Xu, Yanez, Yildizci, Yoshida, Yu, Yu, Yuan,
  Zhang, Zhelnin, \& Collaboration}]{abbasi_constraining_2023}
Abbasi, R., Ackermann, M., Adams, J., {et~al.} 2023, The Astrophysical Journal
  Letters, 949, L12, publisher: The American Astronomical Society

\bibitem[{Abbasi {et~al.}(2022)Abbasi, Ackermann, Adams, Aguilar, Ahlers,
  Ahrens, Alameddine, Alispach, Alves, Amin, Andeen, Anderson, Anton,
  Argüelles, Ashida, Axani, Bai, Balagopal~V., Barbano, Barwick, Bastian,
  Basu, Baur, Bay, Beatty, Becker, Becker~Tjus, Bellenghi, BenZvi, Berley,
  Bernardini, Besson, Binder, Bindig, Blaufuss, Blot, Boddenberg, Bontempo,
  Borowka, Böser, Botner, Böttcher, Bourbeau, Bradascio, Braun, Brinson,
  Bron, Brostean-Kaiser, Browne, Burgman, Burley, Busse, Campana,
  Carnie-Bronca, Chen, Chen, Chirkin, Choi, Clark, Clark, Classen, Coleman,
  Collin, Conrad, Coppin, Correa, Cowen, Cross, Dappen, Dave, De~Clercq,
  DeLaunay, Delgado~López, Dembinski, Deoskar, Desai, Desiati, de~Vries,
  de~Wasseige, de~With, DeYoung, Diaz, Díaz-Vélez, Dittmer, Dujmovic,
  Dunkman, DuVernois, Dvorak, Ehrhardt, Eller, Engel, Erpenbeck, Evans,
  Evenson, Fan, Fazely, Fedynitch, Feigl, Fiedlschuster, Fienberg, Filimonov,
  Finley, Fischer, Fox, Franckowiak, Friedman, Fritz, Fürst, Gaisser,
  Gallagher, Ganster, Garcia, Garrappa, Gerhardt, Ghadimi, Glaser, Glauch,
  Glüsenkamp, Gonzalez, Goswami, Grant, Grégoire, Griswold, Günther,
  Gutjahr, Haack, Hallgren, Halliday, Halve, Halzen, Ha~Minh, Hanson, Hardin,
  Harnisch, Haungs, Hebecker, Helbing, Henningsen, Hettinger, Hickford,
  Hignight, Hill, Hill, Hoffman, Hoffmann, Hokanson-Fasig, Hoshina, Huang,
  Huber, Huber, Hultqvist, Hünnefeld, Hussain, Hymon, In, Iovine, Ishihara,
  Jansson, Japaridze, Jeong, Jin, Jones, Kang, Kang, Kang, Kappes, Kappesser,
  Kardum, Karg, Karl, Karle, Katz, Kauer, Kellermann, Kelley, Kheirandish, Kin,
  Kintscher, Kiryluk, Klein, Koirala, Kolanoski, Kontrimas, Köpke, Kopper,
  Kopper, Koskinen, Koundal, Kovacevich, Kowalski, Kozynets, Kun, Kurahashi,
  Lad, Lagunas~Gualda, Lanfranchi, Larson, Lauber, Lazar, Lee, Leonard,
  Leszczyńska, Li, Lincetto, Liu, Liubarska, Lohfink, Lozano~Mariscal, Lu,
  Lucarelli, Ludwig, Luszczak, Lyu, Ma, Madsen, Mahn, Makino, Mancina, Mariş,
  Martinez-Soler, Maruyama, Mase, McElroy, McNally, Mead, Meagher, Mechbal,
  Medina, Meier, Meighen-Berger, Micallef, Mockler, Montaruli, Moore, Morse,
  Moulai, Naab, Nagai, Naumann, Necker, Nguyễn, Niederhausen, Nisa, Nowicki,
  Obertacke~Pollmann, Oehler, Oeyen, Olivas, O'Sullivan, Pandya, Pankova,
  Park, Parker, Paudel, Paul, Pérez de~los Heros, Peters, Peterson, Philippen,
  Pieper, Pittermann, Pizzuto, Plum, Popovych, Porcelli, Prado~Rodriguez,
  Price, Pries, Przybylski, Raab, Raissi, Rameez, Rawlins, Rea, Rehman,
  Reichherzer, Reimann, Renzi, Resconi, Reusch, Rhode, Richman, Riedel,
  Roberts, Robertson, Roellinghoff, Rongen, Rott, Ruhe, Ryckbosch,
  Rysewyk~Cantu, Safa, Saffer, Sanchez~Herrera, Sandrock, Sandroos, Santander,
  Sarkar, Sarkar, Satalecka, Schaufel, Schieler, Schindler, Schmidt, Schneider,
  Schneider, Schröder, Schumacher, Schwefer, Sclafani, Seckel, Seunarine,
  Sharma, Shefali, Silva, Skrzypek, Smithers, Snihur, Soedingrekso, Soldin,
  Spannfellner, Spiczak, Spiering, Stachurska, Stamatikos, Stanev, Stein,
  Stettner, Steuer, Stezelberger, Stürwald, Stuttard, Sullivan, Taboada,
  Ter-Antonyan, Tilav, Tischbein, Tollefson, Tönnis, Toscano, Tosi, Trettin,
  Tselengidou, Tung, Turcati, Turcotte, Turley, Twagirayezu, Ty,
  Unland~Elorrieta, Valtonen-Mattila, Vandenbroucke, van Eijndhoven, Vannerom,
  van Santen, Verpoest, Walck, Watson, Weaver, Weigel, Weindl, Weiss, Weldert,
  Wendt, Werthebach, Weyrauch, Whitehorn, Wiebusch, Williams, Wolf, Woschnagg,
  Wrede, Wulff, Xu, Yanez, Yoshida, Yu, Yuan, Zhang, Zhelnin, \& {IceCube
  Collaboration}}]{abbasi_search_2022}
Abbasi, R., Ackermann, M., Adams, J., {et~al.} 2022, Physical Review D, 106,
  022005

\bibitem[{Achterberg {et~al.}(2006)Achterberg, Ackermann, Adams, Ahrens, Atlee,
  Bahcall, Bai, Baret, Bartelt, Barwick, Bay, Beattie, Becka, Becker, Becker,
  Berghaus, Berley, Bernardini, Bertrand, Besson, Blaufuss, Boersma, Bohm,
  Böser, Botner, Bouchta, Braun, Burgess, Burgess, Castermans, Chirkin, Clem,
  Collin, Conrad, Cooley, Cowen, D'Agostino, Davour, Day, De~Clercq, Desiati,
  DeYoung, Dreyer, Duvoort, Edwards, Ehrlich, Ellsworth, Evenson, Fazely,
  Feser, Filimonov, Gaisser, Gallagher, Ganugapati, Geenen, Gerhardt,
  Goldschmidt, Goodman, Greene, Grullon, Groß, Gunasingha, Hallgren, Halzen,
  Han, Hanson, Hardtke, Hardtke, Harenberg, Hart, Hauschildt, Hays, Heise,
  Helbing, Hellwig, Herquet, Hill, Hodges, Hoffman, Hoshina, Hubert, Hughey,
  Hulth, Hultqvist, Hundertmark, Ishihara, Jacobsen, Japaridze, Jones, Joseph,
  Kampert, Karle, Kawai, Kelley, Kestel, Kitamura, Klein, Klepser, Kohnen,
  Kolanoski, Köpke, Krasberg, Kuehn, Landsman, Lang, Leich, Leuthold,
  Liubarsky, Lundberg, Madsen, Mase, Matis, McCauley, McParland, Meli,
  Messarius, Mészáros, Minor, Miočinović, Miyamoto, Mokhtarani, Montaruli,
  Morey, Morse, Movit, Münich, Nahnhauer, Nam, Nießen, Nygren, Ögelman,
  Olbrechts, Olivas, Patton, Peña-Garay, Pérez de~los Heros, Pieloth, Pohl,
  Porrata, Pretz, Price, Przybylski, Rawlins, Razzaque, Refflinghaus, Resconi,
  Rhode, Ribordy, Richter, Rizzo, Robbins, Rott, Rutledge, Sander, Schlenstedt,
  Schneider, Seckel, Seo, Seunarine, Silvestri, Smith, Solarz, Song, Sopher,
  Spiczak, Spiering, Stamatikos, Stanev, Steffen, Steele, Stezelberger,
  Stokstad, Stoufer, Stoyanov, Sulanke, Sullivan, Sumner, Taboada, Tarasova,
  Tepe, Thollander, Tilav, Toale, Turčan, van Eijndhoven, Vandenbroucke,
  Voigt, Wagner, Walck, Waldmann, Walter, Wang, Wendt, Wiebusch, Wikström,
  Williams, Wischnewski, Wissing, Woschnagg, Xu, Yodh, Yoshida, Zornoza, \&
  Biermann}]{achterberg_selection_2006}
Achterberg, A., Ackermann, M., Adams, J., {et~al.} 2006, Astroparticle Physics,
  26, 282

\bibitem[{Ahumada(2020)}]{ahumada_16th_2020}
Ahumada, R. 2020, The Astrophysical Journal Supplement Series, 21

\bibitem[{Ajello {et~al.}(2020)Ajello, Angioni, Axelsson, Ballet, Barbiellini,
  Bastieri, Becerra~Gonzalez, Bellazzini, Bissaldi, Bloom, Bonino, Bottacini,
  Bruel, Buson, Cafardo, Cameron, Cavazzuti, Chen, Cheung, Ciprini, Costantin,
  Cutini, D'Ammando, de~la Torre~Luque, de~Menezes, de~Palma, Desai,
  Di~Lalla, Di~Venere, Domínguez, Dirirsa, Ferrara, Finke, Franckowiak,
  Fukazawa, Funk, Fusco, Gargano, Garrappa, Gasparrini, Giglietto, Giordano,
  Giroletti, Green, Grenier, Guiriec, Harita, Hays, Horan, Itoh, Jóhannesson,
  Kovac'evic', Krauss, Kreter, Kuss, Larsson, Leto, Li, Liodakis, Longo,
  Loparco, Lott, Lovellette, Lubrano, Madejski, Maldera, Manfreda,
  Martí-Devesa, Massaro, Mazziotta, Mereu, Meyer, Migliori, Mirabal, Mizuno,
  Monzani, Morselli, Moskalenko, Negro, Nemmen, Nuss, Ojha, Ojha, Omodei,
  Orienti, Orlando, Ormes, Paliya, Pei, Peña-Herazo, Persic, Pesce-Rollins,
  Petrov, Piron, Poon, Principe, Rainò, Rando, Rani, Razzano, Razzaque,
  Reimer, Reimer, Schinzel, Serini, Sgrò, Siskind, Spandre, Spinelli, Suson,
  Tachibana, Thompson, Torres, Torresi, Troja, Valverde, van Zyl, \&
  Yassine}]{ajello_fourth_2020}
Ajello, M., Angioni, R., Axelsson, M., {et~al.} 2020, The Astrophysical
  Journal, 892, 105

\bibitem[{Alam {et~al.}(2015)Alam, Albareti, Prieto, Anders, Anderson,
  Anderton, Andrews, Armengaud, Aubourg, Bailey, Basu, Bautista, Beaton, Beers,
  Bender, Berlind, Beutler, Bhardwaj, Bird, Bizyaev, Blake, Blanton, Blomqvist,
  Bochanski, Bolton, Bovy, Bradley, Brandt, Brauer, Brinkmann, Brown,
  Brownstein, Burden, Burtin, Busca, Cai, Capozzi, Rosell, Carr, Carrera,
  Chambers, Chaplin, Chen, Chiappini, Chojnowski, Chuang, Clerc, Comparat,
  Covey, Croft, Cuesta, Cunha, Costa, Rio, Davenport, Dawson, Lee, Delubac,
  Deshpande, Dhital, Dutra-Ferreira, Dwelly, Ealet, Ebelke, Edmondson,
  Eisenstein, Ellsworth, Elsworth, Epstein, Eracleous, Escoffier, Esposito,
  Evans, Fan, Fernández-Alvar, Feuillet, Ak, Finley, Finoguenov, Flaherty,
  Fleming, Font-Ribera, Foster, Frinchaboy, Galbraith-Frew, García,
  García-Hernández, Pérez, Gaulme, Ge, Génova-Santos, Georgakakis, Ghezzi,
  Gillespie, Girardi, Goddard, Gontcho, Hernández, Grebel, Green, Grieb,
  Grieves, Gunn, Guo, Harding, Hasselquist, Hawley, Hayden, Hearty, Hekker, Ho,
  Hogg, Holley-Bockelmann, Holtzman, Honscheid, Huber, Huehnerhoff, Ivans,
  Jiang, Johnson, Kinemuchi, Kirkby, Kitaura, Klaene, Knapp, Kneib, Koenig,
  Lam, Lan, Lang, Laurent, Goff, Leauthaud, Lee, Lee, Licquia, Liu, Long,
  López-Corredoira, Lorenzo-Oliveira, Lucatello, Lundgren, Lupton, Iii,
  Mahadevan, Maia, Majewski, Malanushenko, Malanushenko, Manchado, Manera, Mao,
  Maraston, Marchwinski, Margala, Martell, Martig, Masters, Mathur, McBride,
  McGehee, McGreer, McMahon, Ménard, Menzel, Merloni, Mészáros, Miller,
  Miralda-Escudé, Miyatake, Montero-Dorta, More, Morganson, Morice-Atkinson,
  Morrison, Mosser, Muna, Myers, Nandra, Newman, Neyrinck, Nguyen, Nichol,
  Nidever, Noterdaeme, Nuza, O'Connell, O'Connell, O'Connell, Ogando,
  Olmstead, Oravetz, Oravetz, Osumi, Owen, Padgett, Padmanabhan, Paegert,
  Palanque-Delabrouille, Pan, Parejko, Pâris, Park, Pattarakijwanich,
  Pellejero-Ibanez, Pepper, Percival, Pérez-Fournon, Pe´rez-Ra`fols,
  Petitjean, Pieri, Pinsonneault, Mello, Prada, Prakash, Price-Whelan,
  Protopapas, Raddick, Rahman, Reid, Rich, Rix, Robin, Rockosi, Rodrigues,
  Rodríguez-Torres, Roe, Ross, Ross, Rossi, Ruan, Rubiño-Martín, Rykoff,
  Salazar-Albornoz, Salvato, Samushia, Sánchez, Santiago, Sayres, Schiavon,
  Schlegel, Schmidt, Schneider, Schultheis, Schwope, Scóccola, Scott,
  Sellgren, Seo, Serenelli, Shane, Shen, Shetrone, Shu, Aguirre, Sivarani,
  Skrutskie, Slosar, Smith, Sobreira, Souto, Stassun, Steinmetz, Stello,
  Strauss, Streblyanska, Suzuki, Swanson, Tan, Tayar, Terrien, Thakar, Thomas,
  Thomas, Thompson, Tinker, Tojeiro, Troup, Vargas-Magaña, Vazquez, Verde,
  Viel, Vogt, Wake, Wang, Weaver, Weinberg, Weiner, White, Wilson, Wisniewski,
  Wood-Vasey, Ye`che, York, Zakamska, Zamora, Zasowski, Zehavi, Zhao, Zheng,
  Zhou~(周旭), Zhou~(周志民), Zou~(邹虎), \& Zhu}]{alam_eleventh_2015}
Alam, S., Albareti, F.~D., Prieto, C.~A., {et~al.} 2015, The Astrophysical
  Journal Supplement Series, 219, 12

\bibitem[{Ananna {et~al.}(2022)Ananna, Weigel, Trakhtenbrot, Koss, Urry, Ricci,
  Hickox, Treister, Bauer, Ueda, Mushotzky, Ricci, Oh, Mejía-Restrepo, Brok,
  Stern, Powell, Caglar, Ichikawa, Wong, Harrison, \&
  Schawinski}]{ananna_bass_2022}
Ananna, T.~T., Weigel, A.~K., Trakhtenbrot, B., {et~al.} 2022, The
  Astrophysical Journal Supplement Series, 261, 9

\bibitem[{Antonucci(1993)}]{antonucci_unified_1993}
Antonucci, R. 1993, Annual Review of Astronomy and Astrophysics, 31, 473

\bibitem[{Arenou {et~al.}(2017)Arenou, Luri, Babusiaux, Fabricius, Helmi,
  Robin, Vallenari, Blanco-Cuaresma, Cantat-Gaudin, Findeisen, Reylé,
  Ruiz-Dern, Sordo, Turon, Walton, Shih, Antiche, Barache, Barros, Breddels,
  Carrasco, Costigan, Diakité, Eyer, Figueras, Galluccio, Heu, Jordi,
  Krone-Martins, Lallement, Lambert, Leclerc, Marrese, Moitinho, Mor,
  Romero-Gómez, Sartoretti, Soria, Soubiran, Souchay, Veljanoski, Ziaeepour,
  Giuffrida, Pancino, \& Bragaglia}]{arenou_gaia_2017}
Arenou, F., Luri, X., Babusiaux, C., {et~al.} 2017, Astronomy \& Astrophysics,
  599, A50

\bibitem[{Assef {et~al.}(2013)Assef, Stern, Kochanek, Blain, Brodwin, Brown,
  Donoso, Eisenhardt, Jannuzi, Jarrett, Stanford, Tsai, Wu, \&
  Yan}]{assef_mid-infrared_2013}
Assef, R.~J., Stern, D., Kochanek, C.~S., {et~al.} 2013, The Astrophysical
  Journal, 772, 26

\bibitem[{Barandela {et~al.}(2003)Barandela, Sánchez, Garcı́a, \&
  Rangel}]{barandela_strategies_2003}
Barandela, R., Sánchez, J.~S., Garcı́a, V., \& Rangel, E. 2003, Pattern
  Recognition, 36, 849

\bibitem[{Batista {et~al.}(2004)Batista, Prati, \& Monard}]{batista_study_2004}
Batista, G. E. A. P.~A., Prati, R.~C., \& Monard, M.~C. 2004, ACM SIGKDD
  Explorations Newsletter, 6, 20

\bibitem[{Baum(1962)}]{baum_photoelectric_1962}
Baum, W.~A. 1962, 15, 390, conference Name: Problems of Extra-Galactic Research
  ADS Bibcode: 1962IAUS...15..390B

\bibitem[{Blanton(2017)}]{blanton_sloan_2017}
Blanton, M.~R. 2017, The Astronomical Journal, 35

\bibitem[{Boller {et~al.}(2016)Boller, Freyberg, Trümper, Haberl, Voges, \&
  Nandra}]{boller_second_2016}
Boller, T., Freyberg, M.~J., Trümper, J., {et~al.} 2016, Astronomy \&
  Astrophysics, 588, A103

\bibitem[{Bolton {et~al.}(2012)Bolton, Schlegel, Aubourg, Bailey, Bhardwaj,
  Brownstein, Burles, Chen, Dawson, Eisenstein, Gunn, Knapp, Loomis, Lupton,
  Maraston, Muna, Myers, Olmstead, Padmanabhan, Pâris, Percival, Petitjean,
  Rockosi, Ross, Schneider, Shu, Strauss, Thomas, Tremonti, Wake, Weaver, \&
  Wood-Vasey}]{bolton_spectral_2012}
Bolton, A.~S., Schlegel, D.~J., Aubourg, Ã., {et~al.} 2012, The Astronomical
  Journal, 144, 144, aDS Bibcode: 2012AJ....144..144B

\bibitem[{Bolzonella {et~al.}(2000)Bolzonella, Miralles, \&
  Pelló}]{bolzonella_photometric_2000}
Bolzonella, M., Miralles, J.~M., \& Pelló, R. 2000, Astronomy and
  Astrophysics, 363, 476, aDS Bibcode: 2000A\&A...363..476B

\bibitem[{Bonjean {et~al.}(2019)Bonjean, Aghanim, Salomé, Beelen, Douspis, \&
  Soubrié}]{bonjean_star_2019}
Bonjean, V., Aghanim, N., Salomé, P., {et~al.} 2019, Astronomy \&
  Astrophysics, 622, A137

\bibitem[{Brammer {et~al.}(2008)Brammer, van Dokkum, \&
  Coppi}]{brammer_eazy_2008}
Brammer, G.~B., van Dokkum, P.~G., \& Coppi, P. 2008, The Astrophysical
  Journal, 686, 1503

\bibitem[{Brandt \& Hasinger(2005)}]{brandt_deep_2005}
Brandt, W. \& Hasinger, G. 2005, Annual Review of Astronomy and Astrophysics,
  43, 827

\bibitem[{Brandt \& Alexander(2015)}]{brandt_cosmic_2015}
Brandt, W.~N. \& Alexander, D.~M. 2015, The Astronomy and Astrophysics Review,
  23, 1

\bibitem[{Breiman(2001)}]{breiman_random_2001}
Breiman, L. 2001, Machine Learning, 45, 5

\bibitem[{Burbidge(1958)}]{burbidge_nuclear_1958}
Burbidge, G.~R. 1958, Publications of the Astronomical Society of the Pacific,
  70, 83

\bibitem[{Chawla {et~al.}(2002)Chawla, Bowyer, Hall, \&
  Kegelmeyer}]{chawla_smote_2002}
Chawla, N.~V., Bowyer, K.~W., Hall, L.~O., \& Kegelmeyer, W.~P. 2002, Journal
  of Artificial Intelligence Research, 16, 321

\bibitem[{Clarke {et~al.}(2020)Clarke, Scaife, Greenhalgh, \&
  Griguta}]{clarke_identifying_2020}
Clarke, A.~O., Scaife, A. M.~M., Greenhalgh, R., \& Griguta, V. 2020, Astronomy
  \& Astrophysics, 639, A84, publisher: EDP Sciences

\bibitem[{Clerc {et~al.}(2016)Clerc, Merloni, Zhang, Finoguenov, Dwelly,
  Nandra, Collins, Dawson, Kneib, Rozo, Rykoff, Sadibekova, Brownstein, Lin,
  Ridl, Salvato, Schwope, Steinmetz, Seo, \& Tinker}]{clerc_spiders_2016}
Clerc, N., Merloni, A., Zhang, Y.-Y., {et~al.} 2016, Monthly Notices of the
  Royal Astronomical Society, 463, 4490

\bibitem[{Coffey {et~al.}(2019)Coffey, Salvato, Merloni, Boller, Nandra,
  Dwelly, Comparat, Schulze, Del~Moro, \&
  Schneider}]{coffey_sdss-ivspiders_2019}
Coffey, D., Salvato, M., Merloni, A., {et~al.} 2019, Astronomy \& Astrophysics,
  625, A123

\bibitem[{Comparat {et~al.}(2020)Comparat, Merloni, Dwelly, Salvato, Schwope,
  Coffey, Wolf, Arcodia, Liu, Buchner, Nandra, Georgakakis, Clerc, Brusa,
  Brownstein, Schneider, Pan, \& Bizyaev}]{comparat_final_2020}
Comparat, J., Merloni, A., Dwelly, T., {et~al.} 2020, Astronomy \&
  Astrophysics, 636, A97

\bibitem[{Cortes \& Vapnik(1995)}]{cortes_support-vector_1995}
Cortes, C. \& Vapnik, V. 1995, Machine Learning, 20, 273

\bibitem[{Cunha \& Humphrey(2022)}]{cunha_photometric_2022}
Cunha, P. A.~C. \& Humphrey, A. 2022, Astronomy \& Astrophysics, 666, A87

\bibitem[{Cutri {et~al.}(2021)Cutri, Wright, Conrow, Fowler, Eisenhardt,
  Grillmair, Kirkpatrick, Masci, McCallon, Wheelock, Fajardo-Acosta, Yan,
  Benford, Harbut, Jarrett, Lake, Leisawitz, Ressler, Stanford, Tsai, Liu,
  Helou, Mainzer, Gettngs, Gonzalez, Hoffman, Marsh, Padgett, Skrutskie, Beck,
  Papin, \& Wittman}]{cutri_vizier_2021}
Cutri, R.~M., Wright, E.~L., Conrow, T., {et~al.} 2021, VizieR Online Data
  Catalog, II/328, aDS Bibcode: 2014yCat.2328....0C

\bibitem[{Dainotti {et~al.}(2021)Dainotti, Bogdan, Narendra, Gibson,
  Miasojedow, Liodakis, Pollo, Nelson, Wozniak, Nguyen, \&
  Larrson}]{dainotti_predicting_2021}
Dainotti, M.~G., Bogdan, M., Narendra, A., {et~al.} 2021, The Astrophysical
  Journal, 920, 118, publisher: The American Astronomical Society

\bibitem[{Demirel {et~al.}(2019)Demirel, Şahin, \&
  Albey}]{demirel_ensemble_2019}
Demirel, K., Şahin, A., \& Albey, E. 2019, in Proceedings of the 8th
  {International} {Conference} on {Data} {Science}, {Technology} and
  {Applications} (Prague, Czech Republic: SCITEPRESS - Science and Technology
  Publications), 267--274

\bibitem[{Dwelly {et~al.}(2017)Dwelly, Salvato, Merloni, Brusa, Buchner,
  Anderson, Boller, Brandt, Budavári, Clerc, Coffey, Del~Moro, Georgakakis,
  Green, Jin, Menzel, Myers, Nandra, Nichol, Ridl, Schwope, \&
  Simm}]{dwelly_spiders_2017}
Dwelly, T., Salvato, M., Merloni, A., {et~al.} 2017, Monthly Notices of the
  Royal Astronomical Society, 469, 1065

\bibitem[{Edelson {et~al.}(1996)Edelson, Alexander, Crenshaw, Kaspi, Malkan,
  Peterson, Warwick, Clavel, Filippenko, Horne, Korista, Kriss, Krolik, Maoz,
  Nandra, O'Brien, Penton, Yaqoob, Albrecht, Alloin, Ayres, Balonek, Barr,
  Barth, Bertram, Bromage, Carini, Carone, Cheng, Chuvaev, Dietrich,
  Dultzin-Hacyan, Gaskell, Glass, Goad, Hemar, Ho, Huchra, Hutchings, Johnson,
  Kazanas, Kollatschny, Koratkar, Kovo, Laor, MacAlpine, Magdziarz, Martin,
  Matheson, McCollum, Miller, Morris, Oknyanskij, Penfold, Perez, Perola, Pike,
  Pogge, Ptak, Qian, Recondo-Gonzalez, Reichert, Rodriguez-Espinoza,
  Rodriguez-Pascual, Rokaki, Roland, Sadun, Salamanca, Santos-Lleo, Shields,
  Shull, Smith, Smith, Snijders, Stirpe, Stoner, Sun, Ulrich, van Groningen,
  Wagner, Wagner, Wanders, Welsh, Weymann, Wilkes, Wu, Wurster, Xue, Zdziarski,
  Zheng, \& Zou}]{edelson_multiwavelength_1996}
Edelson, R.~A., Alexander, T., Crenshaw, D.~M., {et~al.} 1996, The
  Astrophysical Journal, 470, 364, aDS Bibcode: 1996ApJ...470..364E

\bibitem[{Elitzur \& Shlosman(2006)}]{elitzur_agn-obscuring_2006}
Elitzur, M. \& Shlosman, I. 2006, The Astrophysical Journal, 648, L101,
  publisher: IOP Publishing

\bibitem[{Elvis {et~al.}(1994)Elvis, Wilkes, McDowell, Green, Bechtold,
  Willner, Oey, Polomski, \& Cutri}]{elvis_atlas_1994}
Elvis, M., Wilkes, B.~J., McDowell, J.~C., {et~al.} 1994, The Astrophysical
  Journal Supplement Series, 95, 1

\bibitem[{Feigelson {et~al.}(2021)Feigelson, de~Souza, Ishida, \&
  Babu}]{feigelson_twenty-first-century_2021}
Feigelson, E.~D., de~Souza, R.~S., Ishida, E.~E., \& Babu, G.~J. 2021, Annual
  Review of Statistics and Its Application, 8, 493, \_eprint:
  https://doi.org/10.1146/annurev-statistics-042720-112045

\bibitem[{Ferrarese \& Merritt(2000)}]{ferrarese_fundamental_2000}
Ferrarese, L. \& Merritt, D. 2000, The Astrophysical Journal, 539, L9,
  publisher: IOP Publishing

\bibitem[{Flewelling {et~al.}(2020)Flewelling, Magnier, Chambers, Heasley,
  Holmberg, Huber, Sweeney, Waters, Calamida, Casertano, Chen, Farrow,
  Hasinger, Henderson, Long, Metcalfe, Narayan, Nieto-Santisteban, Norberg,
  Rest, Saglia, Szalay, Thakar, Tonry, Valenti, Werner, White, Denneau, Draper,
  Hodapp, Jedicke, Kaiser, Kudritzki, Price, Wainscoat, Chastel, McLean,
  Postman, \& Shiao}]{flewelling_pan-starrs1_2020}
Flewelling, H.~A., Magnier, E.~A., Chambers, K.~C., {et~al.} 2020, The
  Astrophysical Journal Supplement Series, 251, 7

\bibitem[{Fotopoulou \& Paltani(2018)}]{fotopoulou_cpz_2018}
Fotopoulou, S. \& Paltani, S. 2018, Astronomy \& Astrophysics, 619, A14

\bibitem[{{Gaia Collaboration} {et~al.}(2018){Gaia Collaboration}, Brown,
  Vallenari, Prusti, de~Bruijne, Babusiaux, Bailer-Jones, Biermann, Evans,
  Eyer, Jansen, Jordi, Klioner, Lammers, Lindegren, Luri, Mignard, Panem,
  Pourbaix, Randich, Sartoretti, Siddiqui, Soubiran, van Leeuwen, Walton,
  Arenou, Bastian, Cropper, Drimmel, Katz, Lattanzi, Bakker, Cacciari,
  Castañeda, Chaoul, Cheek, De~Angeli, Fabricius, Guerra, Holl, Masana,
  Messineo, Mowlavi, Nienartowicz, Panuzzo, Portell, Riello, Seabroke, Tanga,
  Thévenin, Gracia-Abril, Comoretto, Garcia-Reinaldos, Teyssier, Altmann,
  Andrae, Audard, Bellas-Velidis, Benson, Berthier, Blomme, Burgess, Busso,
  Carry, Cellino, Clementini, Clotet, Creevey, Davidson, De~Ridder, Delchambre,
  Dell'Oro, Ducourant, Fernández-Hernández, Fouesneau, Frémat, Galluccio,
  García-Torres, González-Núñez, González-Vidal, Gosset, Guy, Halbwachs,
  Hambly, Harrison, Hernández, Hestroffer, Hodgkin, Hutton, Jasniewicz,
  Jean-Antoine-Piccolo, Jordan, Korn, Krone-Martins, Lanzafame, Lebzelter,
  Löffler, Manteiga, Marrese, Martín-Fleitas, Moitinho, Mora, Muinonen,
  Osinde, Pancino, Pauwels, Petit, Recio-Blanco, Richards, Rimoldini, Robin,
  Sarro, Siopis, Smith, Sozzetti, Süveges, Torra, van Reeven, Abbas,
  Abreu~Aramburu, Accart, Aerts, Altavilla, Álvarez, Alvarez, Alves, Anderson,
  Andrei, Anglada~Varela, Antiche, Antoja, Arcay, Astraatmadja, Bach, Baker,
  Balaguer-Núñez, Balm, Barache, Barata, Barbato, Barblan, Barklem, Barrado,
  Barros, Barstow, Bartholomé~Muñoz, Bassilana, Becciani, Bellazzini,
  Berihuete, Bertone, Bianchi, Bienaymé, Blanco-Cuaresma, Boch, Boeche,
  Bombrun, Borrachero, Bossini, Bouquillon, Bourda, Bragaglia, Bramante,
  Breddels, Bressan, Brouillet, Brüsemeister, Brugaletta, Bucciarelli,
  Burlacu, Busonero, Butkevich, Buzzi, Caffau, Cancelliere, Cannizzaro,
  Cantat-Gaudin, Carballo, Carlucci, Carrasco, Casamiquela, Castellani,
  Castro-Ginard, Charlot, Chemin, Chiavassa, Cocozza, Costigan, Cowell, Crifo,
  Crosta, Crowley, Cuypers†, Dafonte, Damerdji, Dapergolas, David, David,
  de~Laverny, De~Luise, De~March, de~Martino, de~Souza, de~Torres, Debosscher,
  del Pozo, Delbo, Delgado, Delgado, Di~Matteo, Diakite, Diener, Distefano,
  Dolding, Drazinos, Durán, Edvardsson, Enke, Eriksson, Esquej,
  Eynard~Bontemps, Fabre, Fabrizio, Faigler, Falcão, Farràs~Casas, Federici,
  Fedorets, Fernique, Figueras, Filippi, Findeisen, Fonti, Fraile, Fraser,
  Frézouls, Gai, Galleti, Garabato, García-Sedano, Garofalo, Garralda, Gavel,
  Gavras, Gerssen, Geyer, Giacobbe, Gilmore, Girona, Giuffrida, Glass, Gomes,
  Granvik, Gueguen, Guerrier, Guiraud, Gutiérrez-Sánchez, Haigron,
  Hatzidimitriou, Hauser, Haywood, Heiter, Helmi, Heu, Hilger, Hobbs, Hofmann,
  Holland, Huckle, Hypki, Icardi, Janßen, Jevardat~de Fombelle, Jonker,
  Juhász, Julbe, Karampelas, Kewley, Klar, Kochoska, Kohley, Kolenberg,
  Kontizas, Kontizas, Koposov, Kordopatis, Kostrzewa-Rutkowska, Koubsky,
  Lambert, Lanza, Lasne, Lavigne, Le~Fustec, Le~Poncin-Lafitte, Lebreton,
  Leccia, Leclerc, Lecoeur-Taibi, Lenhardt, Leroux, Liao, Licata, Lindstrøm,
  Lister, Livanou, Lobel, López, Managau, Mann, Mantelet, Marchal, Marchant,
  Marconi, Marinoni, Marschalkó, Marshall, Martino, Marton, Mary, Massari,
  Matijevič, Mazeh, McMillan, Messina, Michalik, Millar, Molina, Molinaro,
  Molnár, Montegriffo, Mor, Morbidelli, Morel, Morris, Mulone, Muraveva,
  Musella, Nelemans, Nicastro, Noval, O'Mullane, Ordénovic,
  Ordóñez-Blanco, Osborne, Pagani, Pagano, Pailler, Palacin, Palaversa,
  Panahi, Pawlak, Piersimoni, Pineau, Plachy, Plum, Poggio, Poujoulet, Prša,
  Pulone, Racero, Ragaini, Rambaux, Ramos-Lerate, Regibo, Reylé, Riclet,
  Ripepi, Riva, Rivard, Rixon, Roegiers, Roelens, Romero-Gómez, Rowell, Royer,
  Ruiz-Dern, Sadowski, Sagristà~Sellés, Sahlmann, Salgado, Salguero, Sanna,
  Santana-Ros, Sarasso, Savietto, Schultheis, Sciacca, Segol, Segovia,
  Ségransan, Shih, Siltala, Silva, Smart, Smith, Solano, Solitro, Sordo,
  Soria~Nieto, Souchay, Spagna, Spoto, Stampa, Steele, Steidelmüller,
  Stephenson, Stoev, Suess, Surdej, Szabados, Szegedi-Elek, Tapiador, Taris,
  Tauran, Taylor, Teixeira, Terrett, Teyssandier, Thuillot, Titarenko,
  Torra~Clotet, Turon, Ulla, Utrilla, Uzzi, Vaillant, Valentini, Valette, van
  Elteren, Van~Hemelryck, van Leeuwen, Vaschetto, Vecchiato, Veljanoski, Viala,
  Vicente, Vogt, von Essen, Voss, Votruba, Voutsinas, Walmsley, Weiler, Wertz,
  Wevers, Wyrzykowski, Yoldas, Žerjal, Ziaeepour, Zorec, Zschocke, Zucker,
  Zurbach, \& Zwitter}]{gaia_collaboration_gaia_2018}
{Gaia Collaboration}, Brown, A. G.~A., Vallenari, A., {et~al.} 2018, Astronomy
  \& Astrophysics, 616, A1

\bibitem[{Gebhardt {et~al.}(2000)Gebhardt, Bender, Bower, Dressler, Faber,
  Filippenko, Green, Grillmair, Ho, Kormendy, Lauer, Magorrian, Pinkney,
  Richstone, \& Tremaine}]{gebhardt_relationship_2000}
Gebhardt, K., Bender, R., Bower, G., {et~al.} 2000, The Astrophysical Journal,
  539, L13

\bibitem[{Ginsburg {et~al.}(2019)Ginsburg, Sipőcz, Brasseur, Cowperthwaite,
  Craig, Deil, Guillochon, Guzman, Liedtke, Lim, Lockhart, Mommert, Morris,
  Norman, Parikh, Persson, Robitaille, Segovia, Singer, Tollerud, de~Val-Borro,
  Valtchanov, Woillez, \& {The Astroquery collaboration, a subset of the
  astropy collaboration}}]{ginsburg_astroquery_2019}
Ginsburg, A., Sipőcz, B.~M., Brasseur, C.~E., {et~al.} 2019, The Astronomical
  Journal, 157, 98

\bibitem[{Halzen \& Zas(1997)}]{halzen_neutrino_1997}
Halzen, F. \& Zas, E. 1997, The Astrophysical Journal, 488, 669, publisher: IOP
  Publishing

\bibitem[{Hasinger(2008)}]{hasinger_absorption_2008}
Hasinger, G. 2008, Astronomy \& Astrophysics, 490, 905

\bibitem[{Hickox \& Alexander(2018)}]{hickox_obscured_2018}
Hickox, R.~C. \& Alexander, D.~M. 2018, Annual Review of Astronomy and
  Astrophysics, 56, 625

\bibitem[{Hilt {et~al.}(1977)Hilt, Hilt, Seegrist, States, \& Northeastern
  Forest Experiment Station~(Radnor}]{hilt_ridge_1977}
Hilt, D.~E., Hilt, D.~E., Seegrist, D.~W., States, U., \& Northeastern Forest
  Experiment Station~(Radnor, P.~. 1977, Ridge, a computer program for
  calculating ridge regression estimates (Upper Darby, Pa: Dept. of
  Agriculture, Forest Service, Northeastern Forest Experiment Station)

\bibitem[{Hu(2008)}]{hu_black_2008}
Hu, J. 2008, Monthly Notices of the Royal Astronomical Society, 386, 2242

\bibitem[{{IceCube Collaboration} {et~al.}(2022){IceCube Collaboration},
  Abbasi, Ackermann, Adams, Aguilar, Ahlers, Ahrens, Alameddine, Alispach,
  Alves, Amin, Andeen, Anderson, Anton, Argüelles, Ashida, Axani, Bai,
  Balagopal~V., Barbano, Barwick, Bastian, Basu, Baur, Bay, Beatty, Becker,
  Becker~Tjus, Bellenghi, BenZvi, Berley, Bernardini, Besson, Binder, Bindig,
  Blaufuss, Blot, Boddenberg, Bontempo, Borowka, Böser, Botner, Böttcher,
  Bourbeau, Bradascio, Braun, Brinson, Bron, Brostean-Kaiser, Browne, Burgman,
  Burley, Busse, Campana, Carnie-Bronca, Chen, Chen, Chirkin, Choi, Clark,
  Clark, Classen, Coleman, Collin, Conrad, Coppin, Correa, Cowen, Cross,
  Dappen, Dave, De~Clercq, DeLaunay, Delgado~López, Dembinski, Deoskar, Desai,
  Desiati, de~Vries, de~Wasseige, de~With, DeYoung, Diaz, Díaz-Vélez,
  Dittmer, Dujmovic, Dunkman, DuVernois, Dvorak, Ehrhardt, Eller, Engel,
  Erpenbeck, Evans, Evenson, Fan, Fazely, Fedynitch, Feigl, Fiedlschuster,
  Fienberg, Filimonov, Finley, Fischer, Fox, Franckowiak, Friedman, Fritz,
  Fürst, Gaisser, Gallagher, Ganster, Garcia, Garrappa, Gerhardt, Ghadimi,
  Glaser, Glauch, Glüsenkamp, Goldschmidt, Gonzalez, Goswami, Grant,
  Grégoire, Griswold, Günther, Gutjahr, Haack, Hallgren, Halliday, Halve,
  Halzen, Ha~Minh, Hanson, Hardin, Harnisch, Haungs, Hebecker, Helbing,
  Henningsen, Hettinger, Hickford, Hignight, Hill, Hill, Hoffman, Hoffmann,
  Hokanson-Fasig, Hoshina, Huang, Huber, Huber, Hultqvist, Hünnefeld, Hussain,
  Hymon, In, Iovine, Ishihara, Jansson, Japaridze, Jeong, Jin, Jones, Kang,
  Kang, Kang, Kappes, Kappesser, Kardum, Karg, Karl, Karle, Katz, Kauer,
  Kellermann, Kelley, Kheirandish, Kin, Kintscher, Kiryluk, Klein, Koirala,
  Kolanoski, Kontrimas, Köpke, Kopper, Kopper, Koskinen, Koundal, Kovacevich,
  Kowalski, Kozynets, Kun, Kurahashi, Lad, Lagunas~Gualda, Lanfranchi, Larson,
  Lauber, Lazar, Lee, Leonard, Leszczyńska, Li, Lincetto, Liu, Liubarska,
  Lohfink, Lozano~Mariscal, Lu, Lucarelli, Ludwig, Luszczak, Lyu, Ma, Madsen,
  Mahn, Makino, Mancina, Mariş, Martinez-Soler, Maruyama, Mase, McElroy,
  McNally, Mead, Meagher, Mechbal, Medina, Meier, Meighen-Berger, Micallef,
  Mockler, Montaruli, Moore, Morse, Moulai, Naab, Nagai, Nahnhauer, Naumann,
  Necker, Nguyen, Niederhausen, Nisa, Nowicki, Nygren, Obertacke~Pollmann,
  Oehler, Oeyen, Olivas, O'Sullivan, Pandya, Pankova, Park, Parker, Paudel,
  Paul, Pérez de~los Heros, Peters, Peterson, Philippen, Pieper, Pittermann,
  Pizzuto, Plum, Popovych, Porcelli, Prado~Rodriguez, Price, Pries, Przybylski,
  Raab, Rack-Helleis, Raissi, Rameez, Rawlins, Rea, Rehman, Reichherzer,
  Reimann, Renzi, Resconi, Reusch, Rhode, Richman, Riedel, Roberts, Robertson,
  Roellinghoff, Rongen, Rott, Ruhe, Ryckbosch, Rysewyk~Cantu, Safa, Saffer,
  Sanchez~Herrera, Sandrock, Sandroos, Santander, Sarkar, Sarkar, Satalecka,
  Schaufel, Schieler, Schindler, Schmidt, Schneider, Schneider, Schröder,
  Schumacher, Schwefer, Sclafani, Seckel, Seunarine, Sharma, Shefali, Silva,
  Skrzypek, Smithers, Snihur, Soedingrekso, Soldin, Spannfellner, Spiczak,
  Spiering, Stachurska, Stamatikos, Stanev, Stein, Stettner, Steuer,
  Stezelberger, Stokstad, Stürwald, Stuttard, Sullivan, Taboada, Ter-Antonyan,
  Tilav, Tischbein, Tollefson, Tönnis, Toscano, Tosi, Trettin, Tselengidou,
  Tung, Turcati, Turcotte, Turley, Twagirayezu, Ty, Unland~Elorrieta,
  Valtonen-Mattila, Vandenbroucke, van Eijndhoven, Vannerom, van Santen,
  Verpoest, Walck, Watson, Weaver, Weigel, Weindl, Weiss, Weldert, Wendt,
  Werthebach, Weyrauch, Whitehorn, Wiebusch, Williams, Wolf, Woschnagg, Wrede,
  Wulff, Xu, Yanez, Yoshida, Yu, Yuan, Zhang, \&
  Zhelnin}]{icecube_collaboration_evidence_2022}
{IceCube Collaboration}, Abbasi, R., Ackermann, M., {et~al.} 2022, Science,
  378, 538, publisher: American Association for the Advancement of Science

\bibitem[{Ilbert {et~al.}(2006)Ilbert, Arnouts, McCracken, Bolzonella, Bertin,
  Le~Fèvre, Mellier, Zamorani, Pellò, Iovino, Tresse, Le~Brun, Bottini,
  Garilli, Maccagni, Picat, Scaramella, Scodeggio, Vettolani, Zanichelli,
  Adami, Bardelli, Cappi, Charlot, Ciliegi, Contini, Cucciati, Foucaud,
  Franzetti, Gavignaud, Guzzo, Marano, Marinoni, Mazure, Meneux, Merighi,
  Paltani, Pollo, Pozzetti, Radovich, Zucca, Bondi, Bongiorno, Busarello,
  de~La~Torre, Gregorini, Lamareille, Mathez, Merluzzi, Ripepi, Rizzo, \&
  Vergani}]{ilbert_accurate_2006}
Ilbert, O., Arnouts, S., McCracken, H.~J., {et~al.} 2006, Astronomy and
  Astrophysics, 457, 841, aDS Bibcode: 2006A\&A...457..841I

\bibitem[{Khramtsov {et~al.}(2020)Khramtsov, Akhmetov, \&
  Fedorov}]{khramtsov_northern_2020}
Khramtsov, V., Akhmetov, V., \& Fedorov, P. 2020, Astronomy \& Astrophysics,
  644, A69

\bibitem[{Kim \& Hwang(2022)}]{kim_empirical_2022}
Kim, M. \& Hwang, K.-B. 2022, PLoS ONE, 17, e0271260

\bibitem[{Kochanek {et~al.}(2012)Kochanek, Eisenstein, Cool, Caldwell, Assef,
  Jannuzi, Jones, Murray, Forman, Dey, Brown, Eisenhardt, Gonzalez, Green, \&
  Stern}]{kochanek_ages_2012}
Kochanek, C.~S., Eisenstein, D.~J., Cool, R.~J., {et~al.} 2012, The
  Astrophysical Journal Supplement Series, 200, 8

\bibitem[{Kormendy \& Ho(2013)}]{kormendy_coevolution_2013}
Kormendy, J. \& Ho, L.~C. 2013, Annual Review of Astronomy and Astrophysics,
  51, 511

\bibitem[{Koss {et~al.}(2022)Koss, Trakhtenbrot, Ricci, Bauer, Treister,
  Mushotzky, Urry, Ananna, Baloković, den Brok, Cenko, Harrison, Ichikawa,
  Lamperti, Lein, Mejía-Restrepo, Oh, Pacucci, Pfeifle, Powell, Privon, Ricci,
  Salvato, Schawinski, Shimizu, Smith, \& Stern}]{koss_bass_2022}
Koss, M.~J., Trakhtenbrot, B., Ricci, C., {et~al.} 2022, The Astrophysical
  Journal Supplement Series, 261, 1

\bibitem[{Laor(2003)}]{laor_nature_2003}
Laor, A. 2003, The Astrophysical Journal, 590, 86, aDS Bibcode:
  2003ApJ...590...86L

\bibitem[{Li {et~al.}(2021)Li, Zhang, Cui, Fan, Zhao, Wu, He, Xu, Li, Han, Tao,
  Mi, Yang, \& Yang}]{li_identification_2021}
Li, C., Zhang, Y., Cui, C., {et~al.} 2021, Monthly Notices of the Royal
  Astronomical Society, 506, 1651

\bibitem[{Luo {et~al.}(2010)Luo, Brandt, Xue, Brusa, Alexander, Bauer,
  Comastri, Koekemoer, Lehmer, Mainieri, Rafferty, Schneider, Silverman, \&
  Vignali}]{luo_identifications_2010}
Luo, B., Brandt, W.~N., Xue, Y.~Q., {et~al.} 2010, The Astrophysical Journal
  Supplement Series, 187, 560

\bibitem[{Lyke {et~al.}(2020)Lyke, Higley, McLane, Schurhammer, Myers, Ross,
  Dawson, Chabanier, Martini, Busca, Mas~des Bourboux, Salvato, Streblyanska,
  Zarrouk, Burtin, Anderson, Bautista, Bizyaev, Brandt, Brinkmann, Brownstein,
  Comparat, Green, Macorra, Gutiérrez, Hou, Newman, Palanque-Delabrouille,
  Pâris, Percival, Petitjean, Rich, Rossi, Schneider, Smith, Vivek, \&
  Weaver}]{lyke_sloan_2020}
Lyke, B.~W., Higley, A.~N., McLane, J.~N., {et~al.} 2020, The Astrophysical
  Journal Supplement Series, 250, 8

\bibitem[{Lynden-Bell(1969)}]{lynden-bell_galactic_1969}
Lynden-Bell, D. 1969, Nature, 223, 690

\bibitem[{Magorrian {et~al.}(1998)Magorrian, Tremaine, Richstone, Bender,
  Bower, Dressler, Faber, Gebhardt, Green, Grillmair, Kormendy, \&
  Lauer}]{magorrian_demography_1998}
Magorrian, J., Tremaine, S., Richstone, D., {et~al.} 1998, The Astronomical
  Journal, 115, 2285

\bibitem[{Mainzer {et~al.}(2011)Mainzer, Bauer, Grav, Masiero, Cutri, Dailey,
  Eisenhardt, McMillan, Wright, Walker, Jedicke, Spahr, Tholen, Alles, Beck,
  Brandenburg, Conrow, Evans, Fowler, Jarrett, Marsh, Masci, McCallon,
  Wheelock, Wittman, Wyatt, DeBaun, Elliott, Elsbury, Gautier, Gomillion,
  Leisawitz, Maleszewski, Micheli, \& Wilkins}]{mainzer_preliminary_2011}
Mainzer, A., Bauer, J., Grav, T., {et~al.} 2011, The Astrophysical Journal,
  731, 53

\bibitem[{Mannheim(1995)}]{mannheim_high-energy_1995}
Mannheim, K. 1995, Astroparticle Physics, 3, 295

\bibitem[{Matthews \& Sandage(1963)}]{matthews_optical_1963}
Matthews, T.~A. \& Sandage, A.~R. 1963, The Astrophysical Journal, 138, 30

\bibitem[{Menzel {et~al.}(2016)Menzel, Merloni, Georgakakis, Salvato, Aubourg,
  Brandt, Brusa, Buchner, Dwelly, Nandra, Pâris, Petitjean, \&
  Schwope}]{menzel_spectroscopic_2016}
Menzel, M.-L., Merloni, A., Georgakakis, A., {et~al.} 2016, Monthly Notices of
  the Royal Astronomical Society, 457, 110

\bibitem[{Mereghetti {et~al.}(2021)Mereghetti, Balman, Caballero-Garcia,
  Del~Santo, Doroshenko, Erkut, Hanlon, Hoeflich, Markowitz, Osborne, Pian,
  Rivera~Sandoval, Webb, Amati, Ambrosi, Beardmore, Blain, Bozzo, Burderi,
  Campana, Casella, D'Aí, D'Ammando, De~Colle, Della~Valle, De~Martino,
  Di~Salvo, Doyle, Esposito, Frontera, Gandhi, Ghisellini, Gotz, Grinberg,
  Guidorzi, Hudec, Iaria, Izzo, Jaisawal, Jonker, Kong, Krumpe, Kumar,
  Manousakis, Marino, Martin-Carrillo, Mignani, Miniutti, Mundell, Mukai,
  Nucita, O'Brien, Orlandini, Orio, Palazzi, Papitto, Pintore, Piranomonte,
  Porquet, Ricci, Riggio, Rigoselli, Rodriguez, Saha, Sanna, Santangelo,
  Saxton, Sidoli, Stiele, Tagliaferri, Tavecchio, Tiengo, Tsygankov,
  Turriziani, Wijnands, Zane, \& Zhang}]{mereghetti_time_2021}
Mereghetti, S., Balman, S., Caballero-Garcia, M., {et~al.} 2021, Experimental
  Astronomy, 52, 309

\bibitem[{Merloni {et~al.}(2024)Merloni, Lamer, Liu, Ramos-Ceja, Brunner,
  Bulbul, Dennerl, Doroshenko, Freyberg, Friedrich, Gatuzz, Georgakakis,
  Haberl, Igo, Kreykenbohm, Liu, Maitra, Malyali, Mayer, Nandra, Predehl,
  Robrade, Salvato, Sanders, Stewart, Tubín-Arenas, Weber, Wilms, Arcodia,
  Artis, Aschersleben, Avakyan, Aydar, Bahar, Balzer, Becker, Berger, Boller,
  Bornemann, Brüggen, Brusa, Buchner, Burwitz, Camilloni, Clerc, Comparat,
  Coutinho, Czesla, Dannhauer, Dauner, Dauser, Dietl, Dolag, Dwelly, Egg, Ehl,
  Freund, Friedrich, Gaida, Garrel, Ghirardini, Gokus, Grünwald, Grandis,
  Grotova, Gruen, Gueguen, Hämmerich, Hamaus, Hasinger, Haubner, Homan,
  Chitham, Joseph, Joyce, König, Kaltenbrunner, Khokhriakova, Kink, Kirsch,
  Kluge, Knies, Krippendorf, Krumpe, Kurpas, Li, Liu, Locatelli, Lorenz,
  Müller, Magaudda, Mannes, McCall, Meidinger, Michailidis, Migkas,
  Muñoz-Giraldo, Musiimenta, Nguyen-Dang, Ni, Olechowska, Ota, Pacaud, Pasini,
  Perinati, Pires, Pommranz, Ponti, Poppenhaeger, Pühlhofer, Rau, Reh,
  Reiprich, Roster, Saeedi, Santangelo, Sasaki, Schmitt, Schneider, Schrabback,
  Schuster, Schwope, Seppi, Serim, Shreeram, Sokolova-Lapa, Starck, Stelzer,
  Stierhof, Suleimanov, Tenzer, Traulsen, Trümper, Tsuge, Urrutia, Veronica,
  Waddell, Willer, Wolf, Yeung, Zainab, Zangrandi, Zhang, Zhang, \&
  Zheng}]{merloni_srgerosita_2024}
Merloni, A., Lamer, G., Liu, T., {et~al.} 2024, Astronomy \& Astrophysics, 682,
  A34, publisher: EDP Sciences

\bibitem[{Merloni {et~al.}(2012)Merloni, Predehl, Becker, Böhringer, Boller,
  Brunner, Brusa, Dennerl, Freyberg, Friedrich, Georgakakis, Haberl, Hasinger,
  Meidinger, Mohr, Nandra, Rau, Reiprich, Robrade, Salvato, Santangelo, Sasaki,
  Schwope, Wilms, \& Consortium}]{merloni_erosita_2012}
Merloni, A., Predehl, P., Becker, W., {et~al.} 2012, {eROSITA} {Science}
  {Book}: {Mapping} the {Structure} of the {Energetic} {Universe},
  arXiv:1209.3114 [astro-ph]

\bibitem[{Minkowski(1960)}]{minkowski_new_1960}
Minkowski, R. 1960, The Astrophysical Journal, 132, 908, aDS Bibcode:
  1960ApJ...132..908M

\bibitem[{Miyaji {et~al.}(2015)Miyaji, Hasinger, Salvato, Brusa, Cappelluti,
  Civano, Puccetti, Elvis, Brunner, Fotopoulou, Ueda, Griffiths, Koekemoer,
  Akiyama, Comastri, Gilli, Lanzuisi, Merloni, \&
  Vignali}]{miyaji_detailed_2015}
Miyaji, T., Hasinger, G., Salvato, M., {et~al.} 2015, The Astrophysical
  Journal, 804, 104, publisher: The American Astronomical Society

\bibitem[{Murase(2022)}]{murase_neutrinos_2022}
Murase, K. 2022, Science, publisher: American Association for the Advancement
  of Science

\bibitem[{Murtagh(1991)}]{murtagh_multilayer_1991}
Murtagh, F. 1991, Neurocomputing, 2, 183

\bibitem[{Padovani {et~al.}(2017)Padovani, Alexander, Assef, De~Marco, Giommi,
  Hickox, Richards, Smolčić, Hatziminaoglou, Mainieri, \&
  Salvato}]{padovani_active_2017}
Padovani, P., Alexander, D.~M., Assef, R.~J., {et~al.} 2017, The Astronomy and
  Astrophysics Review, 25, 2

\bibitem[{Pedregosa {et~al.}(2011)Pedregosa, Varoquaux, Gramfort, Michel,
  Thirion, Grisel, Blondel, Prettenhofer, Weiss, Dubourg, Vanderplas, Passos,
  \& Cournapeau}]{pedregosa_scikit-learn_2011}
Pedregosa, F., Varoquaux, G., Gramfort, A., {et~al.} 2011, MACHINE LEARNING IN
  PYTHON, 6

\bibitem[{Pforr {et~al.}(2019)Pforr, Vaccari, Lacy, Maraston, Nyland,
  Marchetti, \& Thomas}]{pforr_photometric_2019}
Pforr, J., Vaccari, M., Lacy, M., {et~al.} 2019, Monthly Notices of the Royal
  Astronomical Society, 483, 3168

\bibitem[{Plotkin {et~al.}(2008)Plotkin, Anderson, Hall, Margon, Voges,
  Schneider, Stinson, \& York}]{plotkin_large_2008}
Plotkin, R.~M., Anderson, S.~F., Hall, P.~B., {et~al.} 2008, The Astronomical
  Journal, 135, 2453

\bibitem[{Predehl {et~al.}(2021)Predehl, Andritschke, Arefiev, Babyshkin,
  Batanov, Becker, Böhringer, Bogomolov, Boller, Borm, Bornemann, Bräuninger,
  Brüggen, Brunner, Brusa, Bulbul, Buntov, Burwitz, Burkert, Clerc, Churazov,
  Coutinho, Dauser, Dennerl, Doroshenko, Eder, Emberger, Eraerds, Finoguenov,
  Freyberg, Friedrich, Friedrich, Fürmetz, Georgakakis, Gilfanov, Granato,
  Grossberger, Gueguen, Gureev, Haberl, Hälker, Hartner, Hasinger, Huber, Ji,
  Kienlin, Kink, Korotkov, Kreykenbohm, Lamer, Lomakin, Lapshov, Liu, Maitra,
  Meidinger, Menz, Merloni, Mernik, Mican, Mohr, Müller, Nandra, Nazarov,
  Pacaud, Pavlinsky, Perinati, Pfeffermann, Pietschner, Ramos-Ceja, Rau,
  Reiffers, Reiprich, Robrade, Salvato, Sanders, Santangelo, Sasaki, Scheuerle,
  Schmid, Schmitt, Schwope, Shirshakov, Steinmetz, Stewart, Strüder, Sunyaev,
  Tenzer, Tiedemann, Trümper, Voron, Weber, Wilms, \&
  Yaroshenko}]{predehl_erosita_2021}
Predehl, P., Andritschke, R., Arefiev, V., {et~al.} 2021, Astronomy \&
  Astrophysics, 647, A1

\bibitem[{Rees(1984)}]{rees_black_1984}
Rees, M.~J. 1984, Annual Review of Astronomy and Astrophysics, 22, 471, aDS
  Bibcode: 1984ARA\&A..22..471R

\bibitem[{Rhea {et~al.}(2021)Rhea, Rousseau-Nepton, Prunet, Prasow-Émond,
  Hlavacek-Larrondo, Asari, Grasha, \&
  Perreault-Levasseur}]{rhea_machine-learning_2021}
Rhea, C., Rousseau-Nepton, L., Prunet, S., {et~al.} 2021, The Astrophysical
  Journal, 910, 129

\bibitem[{Ricci {et~al.}(2011)Ricci, Walter, Courvoisier, \&
  Paltani}]{ricci_reflection_2011}
Ricci, C., Walter, R., Courvoisier, T. J.~L., \& Paltani, S. 2011, Astronomy
  and Astrophysics, 532, A102, aDS Bibcode: 2011A\&A...532A.102R

\bibitem[{Ricci {et~al.}(2022)Ricci, Treister, Bauer, Mejía-Restrepo, Koss,
  den Brok, Baloković, Bär, Bessiere, Caglar, Harrison, Ichikawa, Kakkad,
  Lamperti, Mushotzky, Oh, Powell, Privon, Ricci, Riffel, Rojas, Sani, Smith,
  Stern, Trakhtenbrot, Urry, \& Veilleux}]{ricci_bass_2022}
Ricci, F., Treister, E., Bauer, F.~E., {et~al.} 2022, The Astrophysical Journal
  Supplement Series, 261, 8

\bibitem[{Sadeh {et~al.}(2016)Sadeh, Abdalla, \& Lahav}]{sadeh_annz2_2016}
Sadeh, I., Abdalla, F.~B., \& Lahav, O. 2016, Publications of the Astronomical
  Society of the Pacific, 128, 104502

\bibitem[{Saito \& Rehmsmeier(2015)}]{saito_precision-recall_2015}
Saito, T. \& Rehmsmeier, M. 2015, PLOS ONE, 10, e0118432

\bibitem[{Salvato {et~al.}(2018)Salvato, Buchner, Budavári, Dwelly, Merloni,
  Brusa, Rau, Fotopoulou, \& Nandra}]{salvato_finding_2018}
Salvato, M., Buchner, J., Budavári, T., {et~al.} 2018, Monthly Notices of the
  Royal Astronomical Society, 473, 4937

\bibitem[{Salvato {et~al.}(2011)Salvato, Ilbert, Hasinger, Rau, Civano,
  Zamorani, Brusa, Elvis, Vignali, Aussel, Comastri, Fiore, Le~Floc'h,
  Mainieri, Bardelli, Bolzonella, Bongiorno, Capak, Caputi, Cappelluti,
  Carollo, Contini, Garilli, Iovino, Fotopoulou, Fruscione, Gilli, Halliday,
  Kneib, Kakazu, Kartaltepe, Koekemoer, Kovac, Ideue, Ikeda, Impey, Le~Fevre,
  Lamareille, Lanzuisi, Le~Borgne, Le~Brun, Lilly, Maier, Manohar, Masters,
  McCracken, Messias, Mignoli, Mobasher, Nagao, Pello, Puccetti, Perez-Montero,
  Renzini, Sargent, Sanders, Scodeggio, Scoville, Shopbell, Silvermann,
  Taniguchi, Tasca, Tresse, Trump, \& Zucca}]{salvato_dissecting_2011}
Salvato, M., Ilbert, O., Hasinger, G., {et~al.} 2011, The Astrophysical
  Journal, 742, 61

\bibitem[{Salvato {et~al.}(2022)Salvato, Wolf, Dwelly, Georgakakis, Brusa,
  Merloni, Liu, Toba, Nandra, Lamer, Buchner, Schneider, Freund, Rau, Schwope,
  Nishizawa, Klein, Arcodia, Comparat, Musiimenta, Nagao, Brunner, Malyali,
  Finoguenov, Anderson, Shen, Ibarra-Medel, Trump, Brandt, Urry, Rivera,
  Krumpe, Urrutia, Miyaji, Ichikawa, Schneider, Fresco, Boller, Haase,
  Brownstein, Lane, Bizyaev, \& Nitschelm}]{salvato_erosita_2022}
Salvato, M., Wolf, J., Dwelly, T., {et~al.} 2022, Astronomy \& Astrophysics,
  661, A3, publisher: EDP Sciences

\bibitem[{Saxton {et~al.}(2008)Saxton, Read, Esquej, Freyberg, Altieri, \&
  Bermejo}]{saxton_first_2008}
Saxton, R.~D., Read, A.~M., Esquej, P., {et~al.} 2008, Astronomy \&
  Astrophysics, 480, 611

\bibitem[{Schmidt(1963)}]{schmidt_3c_1963}
Schmidt, M. 1963, Nature, 197, 1040, number: 4872 Publisher: Nature Publishing
  Group

\bibitem[{Schulze {et~al.}(2015)Schulze, Bongiorno, Gavignaud, Schramm,
  Silverman, Merloni, Zamorani, Hirschmann, Mainieri, Wisotzki, Shankar, Fiore,
  Koekemoer, \& Temporin}]{schulze_cosmic_2015}
Schulze, A., Bongiorno, A., Gavignaud, I., {et~al.} 2015, Monthly Notices of
  the Royal Astronomical Society, 447, 2085

\bibitem[{Schulze \& Wisotzki(2010)}]{schulze_low_2010}
Schulze, A. \& Wisotzki, L. 2010, Astronomy and Astrophysics, 516, A87

\bibitem[{Shy {et~al.}(2022)Shy, Tak, Feigelson, Timlin, \&
  Babu}]{shy_incorporating_2022}
Shy, S., Tak, H., Feigelson, E.~D., Timlin, J.~D., \& Babu, G.~J. 2022, The
  Astronomical Journal, 164, 6

\bibitem[{Simet {et~al.}(2021)Simet, Chartab, Lu, \&
  Mobasher}]{simet_comparison_2021}
Simet, M., Chartab, N., Lu, Y., \& Mobasher, B. 2021, The Astrophysical
  Journal, 908, 47, publisher: The American Astronomical Society

\bibitem[{Soldi {et~al.}(2014)Soldi, Beckmann, Baumgartner, Ponti, Shrader,
  Lubiński, Krimm, Mattana, \& Tueller}]{soldi_long-term_2014}
Soldi, S., Beckmann, V., Baumgartner, W.~H., {et~al.} 2014, Astronomy \&
  Astrophysics, 563, A57, publisher: EDP Sciences

\bibitem[{Soƚtan(1982)}]{soltan_masses_1982}
Soƚtan, A. 1982, Monthly Notices of the Royal Astronomical Society, 200, 115

\bibitem[{Stern {et~al.}(2005)Stern, Eisenhardt, Gorjian, Kochanek, Caldwell,
  Eisenstein, Brodwin, Brown, Cool, Dey, Green, Jannuzi, Murray, Pahre, \&
  Willner}]{stern_midinfrared_2005}
Stern, D., Eisenhardt, P., Gorjian, V., {et~al.} 2005, The Astrophysical
  Journal, 631, 163

\bibitem[{Stone(1974)}]{stone_cross-validatory_1974}
Stone, M. 1974, Journal of the Royal Statistical Society: Series B
  (Methodological), 36, 111

\bibitem[{Tibshirani(1996)}]{tibshirani_regression_1996}
Tibshirani, R. 1996, Journal of the Royal Statistical Society. Series B
  (Methodological), 58, 267, publisher: [Royal Statistical Society, Wiley]

\bibitem[{Tremaine {et~al.}(2002)Tremaine, Gebhardt, Bender, Bower, Dressler,
  Faber, Filippenko, Green, Grillmair, Ho, Kormendy, Lauer, Magorrian, Pinkney,
  \& Richstone}]{tremaine_slope_2002}
Tremaine, S., Gebhardt, K., Bender, R., {et~al.} 2002, The Astrophysical
  Journal, 574, 740

\bibitem[{Trümper(1982)}]{trumper_rosat_1982}
Trümper, J. 1982, Advances in Space Research, 2, 241

\bibitem[{Ucci {et~al.}(2017)Ucci, Ferrara, Gallerani, \&
  Pallottini}]{ucci_inferring_2017}
Ucci, G., Ferrara, A., Gallerani, S., \& Pallottini, A. 2017, Monthly Notices
  of the Royal Astronomical Society, 465, 1144

\bibitem[{Ueda {et~al.}(2014)Ueda, Akiyama, Hasinger, Miyaji, \&
  Watson}]{ueda_toward_2014}
Ueda, Y., Akiyama, M., Hasinger, G., Miyaji, T., \& Watson, M.~G. 2014, The
  Astrophysical Journal, 786, 104, aDS Bibcode: 2014ApJ...786..104U

\bibitem[{Ueda {et~al.}(2003)Ueda, Akiyama, Ohta, \&
  Miyaji}]{ueda_cosmological_2003}
Ueda, Y., Akiyama, M., Ohta, K., \& Miyaji, T. 2003, The Astrophysical Journal,
  598, 886, aDS Bibcode: 2003ApJ...598..886U

\bibitem[{Ulrich {et~al.}(1997)Ulrich, Maraschi, \&
  Urry}]{ulrich_variability_1997}
Ulrich, M.-H., Maraschi, L., \& Urry, C.~M. 1997, Annual Review of Astronomy
  and Astrophysics, 35, 445, aDS Bibcode: 1997ARA\&A..35..445U

\bibitem[{Urry \& Padovani(1995)}]{urry_unified_1995}
Urry, C.~M. \& Padovani, P. 1995, Publications of the Astronomical Society of
  the Pacific, 107, 803, publisher: IOP Publishing

\bibitem[{Voges {et~al.}(2000)Voges, Aschenbach, Boller, Brauninger, Briel,
  Burkert, Dennerl, Englhauser, Gruber, Haberl, Hartner, Hasinger, Pfeffermann,
  Pietsch, Predehl, Schmitt, Trumper, \& Zimmermann}]{voges_rosat_2000}
Voges, W., Aschenbach, B., Boller, T., {et~al.} 2000, International
  Astronomical Union Circular, 7432, 3, aDS Bibcode: 2000IAUC.7432....3V

\bibitem[{Véron-Cetty \& Véron(2010)}]{veron-cetty_catalogue_2010}
Véron-Cetty, M.-P. \& Véron, P. 2010, Astronomy and Astrophysics, 518, A10

\bibitem[{Waddell {et~al.}(2024)Waddell, Buchner, Nandra, Salvato, Merloni,
  Gauger, Boller, Seppi, Wolf, Liu, Brusa, Comparat, Dwelly, Igo, \&
  Musiimenta}]{waddell_srgerosita_2024}
Waddell, S. G.~H., Buchner, J., Nandra, K., {et~al.} 2024, The {SRG}/{eROSITA}
  all-sky survey: {Hard} {X}-ray selected {Active} {Galactic} {Nuclei},
  arXiv:2401.17306 [astro-ph]

\bibitem[{Waddell {et~al.}(2023)Waddell, Nandra, Buchner, Wu, Shen, Arcodia,
  Merloni, Salvato, Dauser, Boller, Liu, Comparat, Wolf, Dwelly, Ricci,
  Brownstein, \& Brusa}]{waddell_erosita_2023}
Waddell, S. G.~H., Nandra, K., Buchner, J., {et~al.} 2023, The {eROSITA}
  {Final} {Equatorial} {Depth} {Survey} ({eFEDS}): {Complex} absorption and
  soft excesses in hard {X}-ray--selected active galactic nuclei,
  arXiv:2306.00961 [astro-ph]

\bibitem[{Wang {et~al.}(2021)Wang, Han, Li, Zhang, \& Cheng}]{wang_review_2021}
Wang, L., Han, M., Li, X., Zhang, N., \& Cheng, H. 2021, IEEE Access, 9, 64606,
  conference Name: IEEE Access

\bibitem[{Weigel {et~al.}(2017)Weigel, Schawinski, Caplar, Wong, Treister, \&
  Trakhtenbrot}]{weigel_agns_2017}
Weigel, A.~K., Schawinski, K., Caplar, N., {et~al.} 2017, The Astrophysical
  Journal, 845, 134

\bibitem[{Wolf {et~al.}(2020)Wolf, Salvato, Coffey, Merloni, Buchner, Arcodia,
  Baron, Carrera, Comparat, Schneider, \& Nandra}]{wolf_exploring_2020}
Wolf, J., Salvato, M., Coffey, D., {et~al.} 2020, Monthly Notices of the Royal
  Astronomical Society, 492, 3580

\bibitem[{Wright {et~al.}(2010)Wright, Eisenhardt, Mainzer, Ressler, Cutri,
  Jarrett, Kirkpatrick, Padgett, McMillan, Skrutskie, Stanford, Cohen, Walker,
  Mather, Leisawitz, Gautier, McLean, Benford, Lonsdale, Blain, Mendez, Irace,
  Duval, Liu, Royer, Heinrichsen, Howard, Shannon, Kendall, Walsh, Larsen,
  Cardon, Schick, Schwalm, Abid, Fabinsky, Naes, \&
  Tsai}]{wright_wide-field_2010}
Wright, E.~L., Eisenhardt, P. R.~M., Mainzer, A.~K., {et~al.} 2010, The
  Astronomical Journal, 140, 1868

\bibitem[{Yu \& Tremaine(2002)}]{yu_observational_2002}
Yu, Q. \& Tremaine, S. 2002, Monthly Notices of the Royal Astronomical Society,
  335, 965

\end{thebibliography}

\begin{appendix}
\section{Catalog column description} \label{appendix:catalogue column}

The result of the work presented in this paper has been compiled into a single catalog available online\footnote{\url{https://www.zeuthen.desy.de/nuastro/ML_reconstructed_AGN_catalogue/}}. This includes the \num{21050} reconstructed sources, with results from the obscuration classifier and estimation of $z$, $L_{\rm X}$, $L_{\rm Bol}$, $M_{\rm BH}$, $L_{\rm Edd}$, and $\lambda_{\rm Edd}$ with associated reconstruction uncertainties. \newline 

In addition, the \num{7613} SPIDERS sources used in the training sample are also included. A description of the catalog's columns is given below. Features providing X-ray, IR, and optical information were taken from the references listed in Table \ref{tab:inputs}.

\textit{Column 1}--- \textbf{X-ray detection}: Flag indicating whether the X-ray source was detected in the 2RXS or XMMSL2 survey \citep{boller_second_2016,saxton_first_2008}. \\
\textit{Column 2}--- \textbf{Name}: X-ray identifier from 2RXS or XMMSL2 survey. \\
\textit{Column 3-4}--- \textbf{RA, DEC}: Right ascension and declination of the X-ray detection (J2000) in degrees. \\
\textit{Column 5-6}--- \textbf{Flux, flux error}: X-ray flux and error converted to the 0.5-2 keV band in log$_{10}$(erg cm$^{-2}$s$^{-1}$). \\
\textit{Column 7}--- \textbf{ALLW\_ID}: WISE All-Sky Release catalog name \citep{cutri_vizier_2021} \\ 
\textit{Column 8-9}--- \textbf{ALLW\_RA, ALLW\_DEC}: J2000 AllWISE right ascension and declination, in degrees. \\
\textit{Column 10-13}--- \textbf{W[1234]}: AllWISE Vega magnitude in the W1, W2, W3, and W4 bands. \\
\textit{Column 14-17}--- \textbf{W[1234] error}: AllWISE Vega magnitude errors in the W1, W2, W3, and W4 bands. \\
\textit{Column 18-19}--- \textbf{Gaia mean flux, Gaia mean flux error}: \textit{Gaia} mean flux and error in units of e-s$^{−1}$.\\
\textit{Column 20}--- \textbf{CLASS}: Broad spectral classification computed by the SDSS-DR16 spectroscopic pipeline.\\
\textit{Column 21}--- \textbf{SUBCLASS}: Detailed spectral classification computed by the SDSS-DR16 spectroscopic pipeline.\\
\textit{Column 22-26}--- \textbf{psfMag\_[ugriz]}: Point spread function magnitude of the optical counterpart to the IR source in the \textit{ugriz} band (mag, AB).\\
\textit{Column 27-31}--- \textbf{psfMagErr\_[ugriz]}: Uncertainties on the PSF magnitude in the \textit{ugriz} band (mag, AB).\\
\textit{Column 32}--- \textbf{z}: Redshift of the source, and uncertainty (in log$_{10}$).\\
\textit{Column 33}--- \textbf{z error}: Uncertainty on the redshift.\\
\textit{Column 34}--- \textbf{Luminosity}: X-ray luminosity in the 0.5-2 keV band in log$_{10}$(erg.s$^{-1}$).\\
\textit{Column 35}--- \textbf{Luminosity error}: Uncertainty on the X-ray luminosity in the 0.5-2 keV band in log$_{10}$(erg.s$^{-1}$).\\
\textit{Column 36}--- \textbf{Bolometric luminosity}: Bolometric luminosity in log$_{10}$(erg.s$^{-1}$).\\
\textit{Column 37}--- \textbf{Bolometric luminosity error}: Uncertainty on the bolometric luminosity in log$_{10}$(erg.s$^{-1}$).\\
\textit{Column 38}--- \textbf{Black hole mass}: BH mass in log$_{10}$($M_{\rm \odot}$). \\
\textit{Column 39}--- \textbf{Black hole mass error}: Uncertainty on the BH mass in log$_{10}$($M_{\rm \odot}$). \\
\textit{Column 40}--- \textbf{Eddington luminosity}: Eddington luminosity in log$_{10}$(erg.s$^{-1}$).\\
\textit{Column 41}--- \textbf{Eddington luminosity error}: Uncertainty on the Eddington luminosity in log$_{10}$(erg.s$^{-1}$).\\
\textit{Column 42}--- \textbf{Eddington ratio}: Eddington ratio in log.\\
\textit{Column 43}--- \textbf{Eddington ratio error}: Uncertainty on the Eddington ratio in log.\\
\textit{Column 44}--- \textbf{Reconstructed}: Flag indicating whether the source and values from columns 32-43 come from this work's ML reconstruction (flag==1) or from the SPIDERS AGN spectroscopic catalog  (flag==0) \citep{coffey_sdss-ivspiders_2019}.\\
\textit{Column 45}--- \textbf{Known z}: Flag indicating whether the redshift values from column 32-33 come from this work's ML reconstruction (flag==0) or from the previous spectroscopic visual derived redshift (flag==1) \citep{dwelly_spiders_2017,veron-cetty_catalogue_2010}.\\
\textit{Column 46}--- \textbf{Obscuration}: Value between 0 and 1 indicating whether the source is obscured (obscuration $\sim$ 1) or not (obscuration $\sim$ 0), from the ML classifier presented in Sect. \ref{sec:classification}.\\
\textit{Column 47}--- \textbf{Obscuration error}: Uncertainty on the obscuration value.\\

\section{Feature selection of type 2 AGNs} \label{appendix:feature selection}
\begin{figure}
\centering
\includegraphics[width=0.35\textwidth]{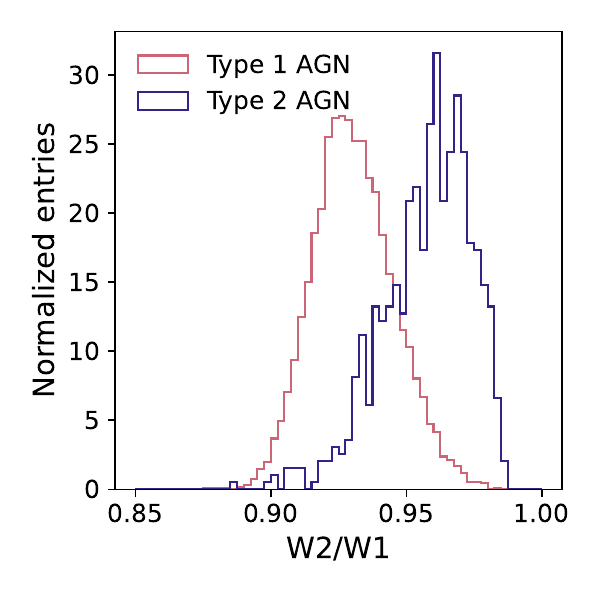}
\includegraphics[width=0.35\textwidth]{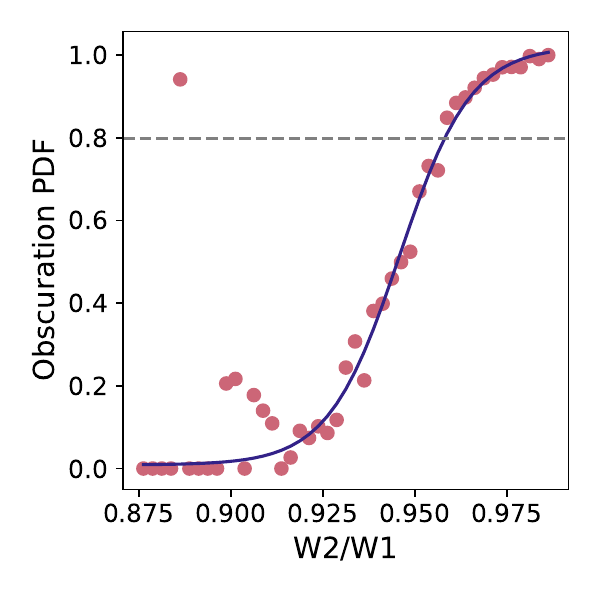}
\caption{Obscuration PDF definition. (\textit{Top}) distribution of the SDSS sources classified as Type 1 and Type 2 AGN as a function of W2/W1 magnitudes. (\textit{Bottom}) Obscuration PDF derived from the left figure and Eq. \ref{eq:obscuration_pdf} from a sigmoid fit to the points. The dashed gray line represents the cut threshold which defines whether an AGN is of Type 1 or 2. It was chosen by doing a grid search over Obscuration PDF values and choosing the value giving the best F1-score.}
\label{fig:feature_selection1}
\end{figure}

Using the known SDSS classification for the subsample of \num{9535} AGN, we can distinguish type 2 from type 1 galaxies in the W2-W1 space (see top panel of Fig. \ref{fig:feature_selection1}, following the method outlined in \cite{abbasi_search_2022}. We can use these two distributions to define an ``obscuration'' PDF as:
\begin{equation}
\centering
\rm Obscuration(W2-W1) = \frac{\mathcal{P}(type 2)}{\mathcal{P}(type 2) +  \mathcal{P}(type 1)}
\label{eq:obscuration_pdf}
\end{equation} 
where $\mathcal{P}(\rm type 1)$ and $\mathcal{P}(\rm type 2)$ are the probabilities of an AGN being of type 1 or of type 2, respectively, according to the normalized histograms of Fig. \ref{fig:feature_selection1}

\begin{figure}
\centering
\includegraphics[width=0.35\textwidth]{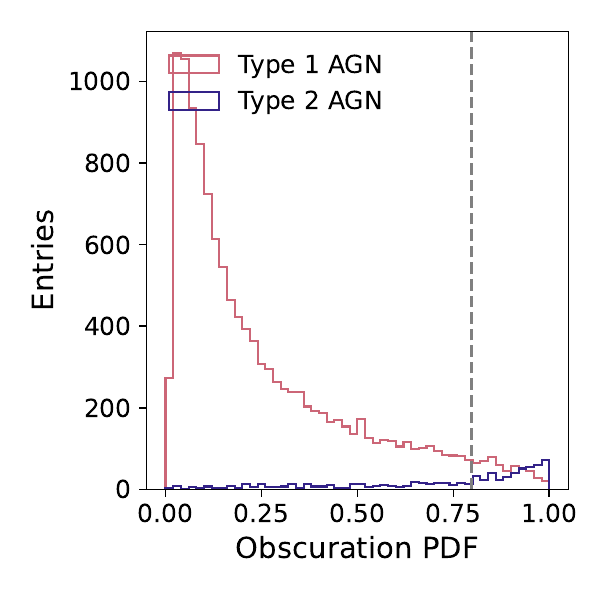}
\includegraphics[width=0.35\textwidth]{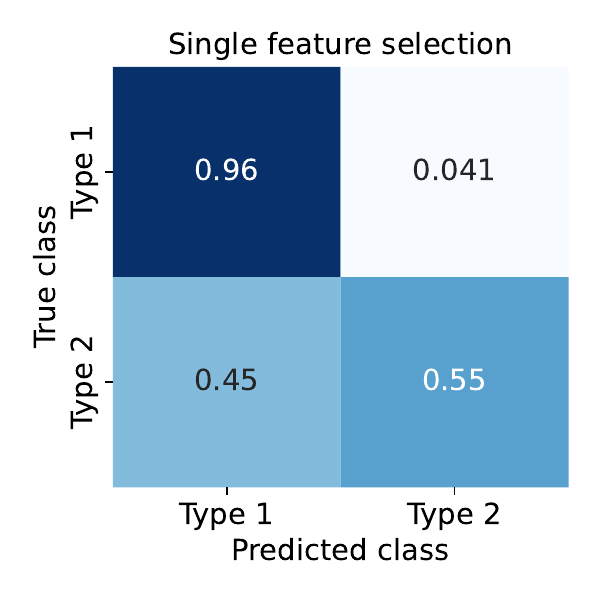}
\caption{Obscuration PDF precision recall curve definition. (\textit{Top}) Distribution of SDSS defined Type 1 and 2 AGN as a function of their derived Obscuration PDF. The vertical gray dashed line represents the threshold value giving the optimal classification perfomance. (\textit{Bottom}) Confusion matrix derived from the left distribution, by scanning through values between 0 and 1 and calculating the precision and recall.}
\label{fig:feature_selection2}
\end{figure}

By applying Equation \ref{eq:obscuration_pdf} to these two distributions, we obtain the obscuration PDF shown in the bottom panel of Fig. \ref{fig:feature_selection1}, with the blue line representing a sigmoid fit to the data points. Using this fitted function, we obtain the distribution of SDSS-classified type 1 and type 2 AGN as a function of the derived obscuration PDF (top panel of Fig. \ref{fig:feature_selection2}. By scanning through obscuration PDF values between 0 and 1, we can calculate the precision and recall for each threshold , based on the definitions given in Sect. \ref{sec:classification}. This then gives us confusion matrix presented in the bottom panel of Fig. \ref{fig:feature_selection2}. Using this single feature classification, we reach a type 1 TNR of 96\%, but a poor type 2 identification power with the TPR at 55\% only, for an optimized obscuration PDF threshold of 0.80.

\section{Handling of null entries} \label{appendix:null entries}
\begin{figure}
\centering
\includegraphics[width=0.35\textwidth]{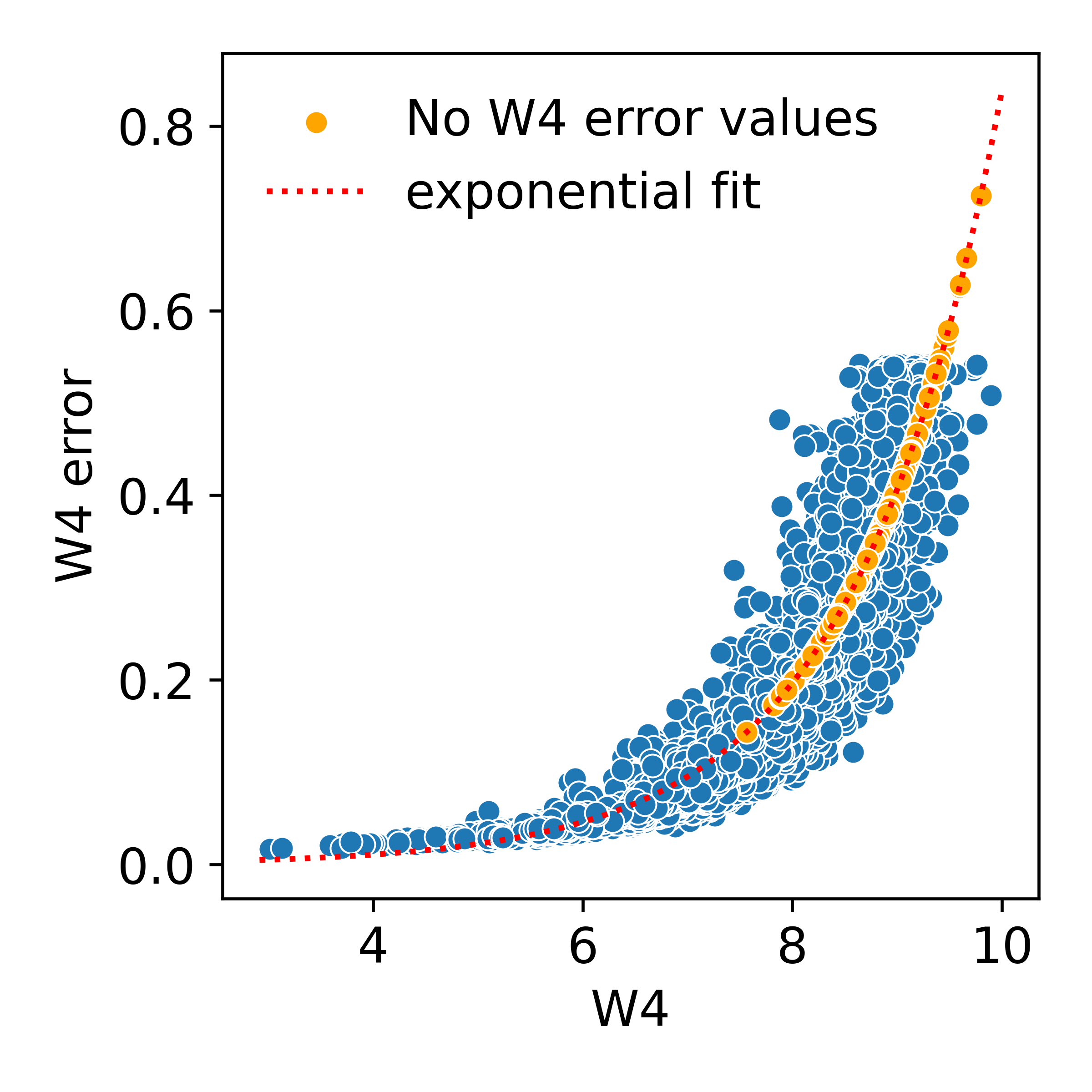}
\caption{Handling of null entries: W4 error values for points missing one are estimated (yellow points) using an exponential fit to the W4 error vs W4 plane.}
\label{fig:null_entries}
\end{figure}

The supervised ML algorithm cannot accept null entries for any of the features. This is true for the training sample, as well as the catalog that is to be reconstructed. As explained in Sect. \ref{subsec:optical}, demanding SDSS photometry observations for all AGN sources represents the greatest cut that can be applied to the data, however, some features still remain incomplete. Instead of indiscriminately removing these data points, we looked for correlations between fully complete features and partially incomplete ones: we fit the function which describes the relationship in that parameter space in order to derive dummy values. In the SPIDERS catalog, \num{685} (\num{278}) sources have AllWISE W3 (W4) magnitudes but are missing corresponding photometric error measurements. Similarly \num{278} SPIDERS have no entries for the Gaia mean flux and errors. In total, the training sample has \num{918} sources with null entries, while this number is \num{9337} for the reconstructed catalog. 
The top panel of Fig. \ref{fig:null_entries} shows the W4 magnitude as a function of the W4 error for AGNs with complete information in the SPIDERS sample. An exponential function is fit and W4 error are interpolated for sources which are missing entries (yellow points). The soft X-ray flux $F_{\rm 0.5-2 keV}$ is also used to derive a value for the Gaia mean flux using a log-log fit. The relationship between the Gaia mean flux and its associated error itself is then used to complete the error column. 

\begin{figure}
\centering
\includegraphics[width=0.35\textwidth]{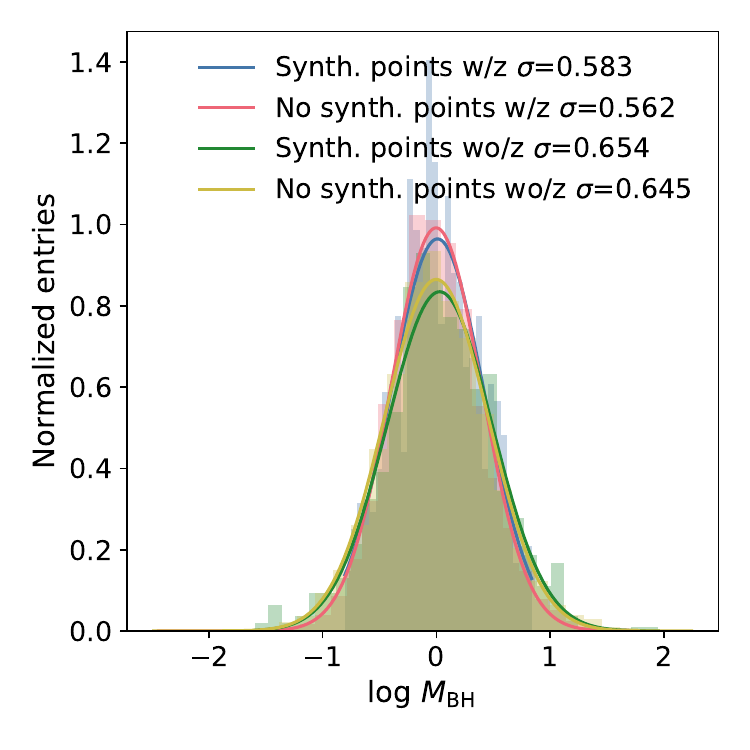}
\includegraphics[width=0.35\textwidth]{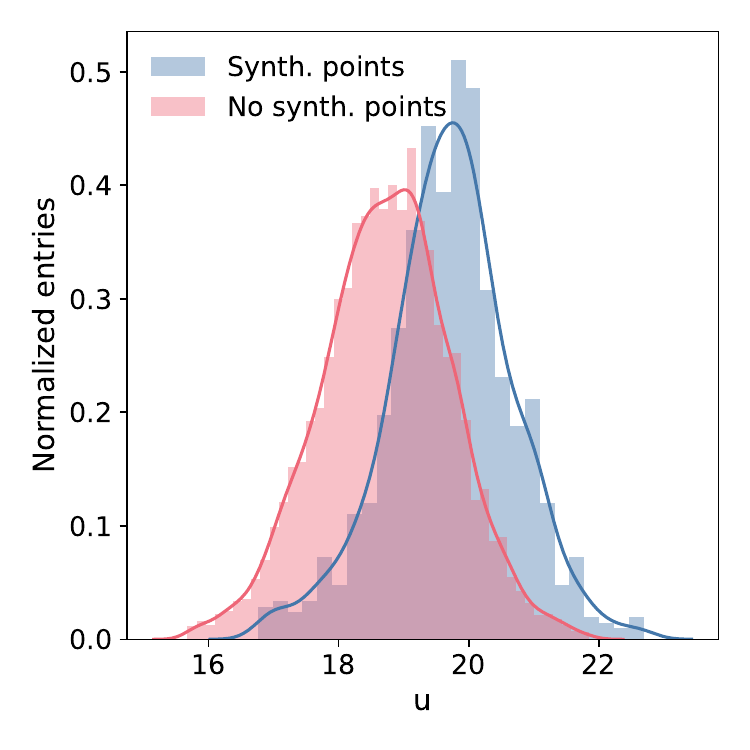}
\caption{Pull distribution with gaussian fits for $M_{\rm BH}$ (\textit{top}) help us study the effect of creating synthetic entries in the training sample. On average, the reconstruction resolution degrades by $\sim$10\% when synthetic points for IR errors and/or Gaia mean fluxes are included. AGN sources with synthetic entries lie on the fainter end of the multi-wavelength spectra (SDSS \textit{u}-mag shown on the \textit{bottom} panel).}
\label{fig:trainig_sample_synthetic_points}
\end{figure}

\begin{figure}
\centering
\includegraphics[width=0.35\textwidth]{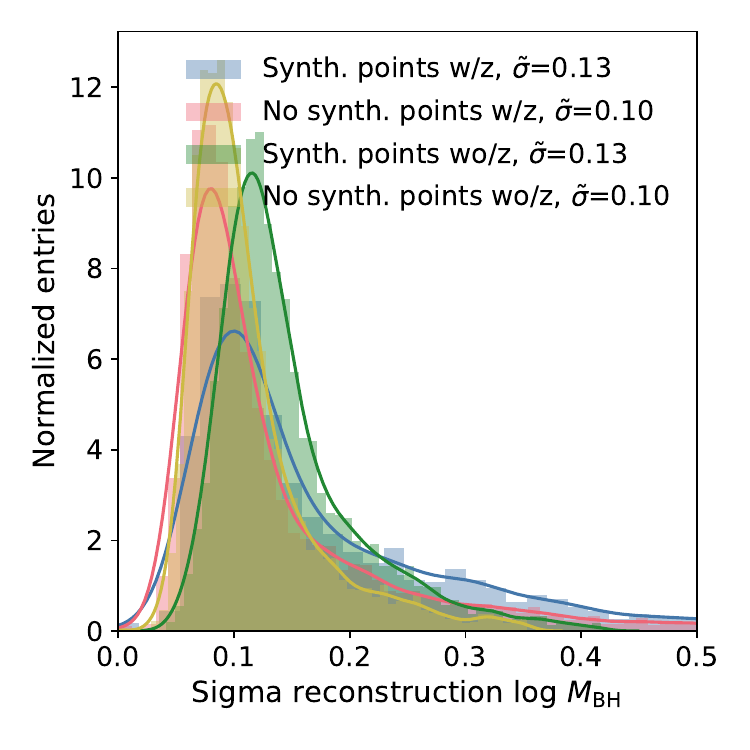}
\includegraphics[width=0.35\textwidth]{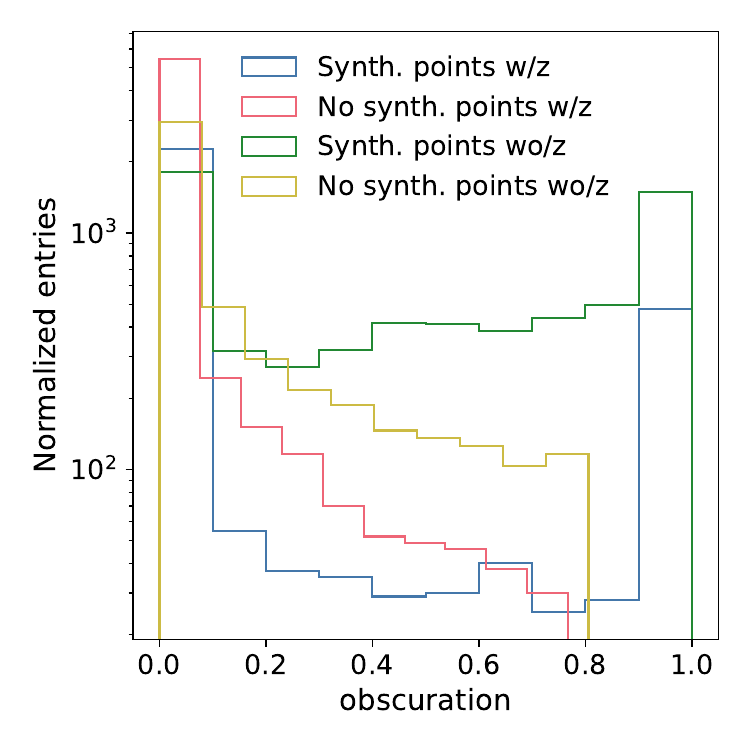}
\caption{Effect of synthetic entries generation for the reconstructed sample. \textit{Top}: error distribution for the reconstructed $M_{\rm BH}$. \textit{Bottom}: reconstruction obscuration level as a function of several subsamples. AGN sources with synthetic entries for IR errors and/or Gaia mean fluxes are the same dim sources that make up the Type 2 AGN sample from Fig.\ref{fig:mbh_2types_pred}.}
\label{fig:reco_sample_synthetic_points}
\end{figure}

We studied the effect of creating synthetic points for IR errors and Gaia mean fluxes and errors, in the simulation-based approach (see Sect. \ref{sec:pseudo-sets}) chosen. Fig. \ref{fig:trainig_sample_synthetic_points} shows pull distributions for $M_{\rm BH}$ and $\lambda_{\rm Edd}$ as a function of whether the AGN sources have any synthesized entries, for both $ML_{\rm w/z}$ and $ML_{\rm w/o z}$. Comparing the standard deviation $\sigma$ from the Gaussian, we note a degradation of $\sim$10\% in the fit resolution when synthesized points are added. However, points with missing entries are those which lie on the fainter end of the phase space (see right panel of Fig. \ref{fig:trainig_sample_synthetic_points}), in the tail of the training sample's distribution: these points are intrinsically harder to reconstruct with the smaller sample used to train to the ML regressor. Similarly, when looking at the uncertainties for the ML-reconstructed (unknown) catalog, AGN sources with synthetic entries have a larger $\sigma_{\rm err}$, as is shown for $M_{\rm BH}$ reconstruction on the left panel of Fig. \ref{fig:reco_sample_synthetic_points}. However, again, as the right panel of the figure indicates, the ratio of type 2 sources within the synthetic sample is far greater, so much so that the error PDF plot is equivalent to the one shown in Fig. \ref{fig:mbh_2types_pred}; namely the synthetic points have a slightly poorer reconstruction because they are faint.

\end{appendix}

\end{document}